\documentclass[letterpaper, 11pt]{article}

\usepackage{jheppub}
\usepackage{bm}
\usepackage{graphicx}
\usepackage{epstopdf}
\usepackage{amsmath, amssymb}
\usepackage{multirow}

\usepackage{amsfonts,amssymb,epsfig,amsmath,mathtools,tabu}
\usepackage{verbatim,booktabs}

\usepackage[inline]{enumitem}

\usepackage{hyperref,caption,subcaption}
\usepackage{xcolor,tikz,graphicx,afterpage}
\usetikzlibrary{shapes.geometric,positioning}
\newcommand{\be}{\begin{eqnarray}}
\newcommand{\ee}{\end{eqnarray}}
\newcommand{\nn}{\nonumber}
\newcommand{\bn}{\begin{enumerate}}
\newcommand{\en}{\end{enumerate}}

\def\IC{\mathbb{C}}

\def\IP{\mathbb{P}}

\def\IZ{\mathbb{Z}}

\def\CF{{\cal F}}

\def\CL{{\cal L}}

\def\CN{{\cal N}}
\def\CO{{\cal O}}

\def\CQ{{\cal Q}}

\def\CZ{{\cal Z}}

\def\a{\alpha}

\def\e{\epsilon}

\def\k{\kappa}

\def\s{\sigma}

\def\w{\omega}

\def\D{\Delta}

\def\half{\frac{1}{2}}

\def\Tr{{\rm Tr}}

\title{Instantons from Blow-up}

\author[a]{Joonho Kim,}
\author[b]{Sung-Soo Kim,}
\author[c]{Ki-Hong Lee,}
\author[a]{Kimyeong Lee,}
\author[a]{and Jaewon Song}
\affiliation[a]{School of Physics, Korea Institute for Advanced Study, Seoul 02455, Korea}
\affiliation[b]{School of Physics, University of Electronic Science and Technology of China,\\ No.4, Section 2, North Jianshe Road, Chengdu, Sichuan 610054, China}
\affiliation[c]{Department of Physics and Astronomy \& Center for Theoretical Physics\\ Seoul National University, Seoul 08826, Korea}
\emailAdd{joonhokim@kias.re.kr}
\emailAdd{sungsoo.kim@uestc.edu.cn}
\emailAdd{khlee11812@gmail.com}
\emailAdd{klee@kias.re.kr}
\emailAdd{jsong@kias.re.kr}

\abstract{
We generalize Nakajima-Yoshioka blowup equations to arbitrary gauge group with hypermultiplets in arbitrary representations. Using our blowup equations, we compute the instanton partition functions for 4d $\CN = 2$ and 5d $\CN = 1$ gauge theories for arbitrary gauge theory with a large class of matter representations, without knowing explicit construction of the instanton moduli space. Our examples include exceptional gauge theories with fundamentals, $SO(N)$ gauge theories with spinors, and $SU(6)$ gauge theories with rank-3 antisymmetric hypers.  Remarkably, the instanton partition function is completely determined by the perturbative part.

}

\preprint{KIAS-P19047, SNUTP19-004}

\begin{document}
\maketitle

\section{Introduction} \label{sec:intro}

The Seiberg-Witten prepotential provides a complete description for the low energy dynamics of 4d $\mathcal{N}=2$ or 5d $\mathcal{N}=1$ gauge theory in its Coulomb branch \cite{Seiberg:1994rs,Seiberg:1994aj}. It is a function of the vacuum expectation value (VEV) of the scalar in the vector multiplet that parameterizes the Coulomb branch moduli space. Quantum correction to the prepotential is known to be one-loop exact, while there also exist non-perturbative corrections coming from Yang-Mills instantons. 

An efficient way to compute the fully quantum corrected prepotential $\mathcal{F}$ is to study the Nekrasov partition function $\mathcal{Z}$ on $\Omega$-deformed $\mathbb{C}^2$ or $\mathbb{C}^2 \times S^1$. 
It can be written as the product of the classical, one-loop, and instanton contributions,
\begin{align}
  \mathcal{Z}(\vec{a}, \vec{{m}}, \e_1, \e_2, {q}) = {Z}_{\textrm{class}}(\vec{a}, \e_1, \e_2, {q}) \ {Z}_{\textrm{1-loop}} (\vec{a}, \vec{{m}}, \e_1, \e_2) \ {Z}_{\textrm{inst}}(\vec{a}, \vec{{m}}, \e_1, \e_2, {q}),
\end{align}
where the instanton piece is the fugacity sum over all multi-instanton contributions:
\begin{align}
  {Z}_{\textrm{inst}}(\vec{a}, \vec{{m}}, \e_1, \e_2, {q}) = 1 + \sum_{n=1}^\infty {q}^n {Z}_n(\vec{a}, \vec{{m}}, \e_1, \e_2).
\end{align}
Once the Nekrasov partition function is known, one can extract the Seiberg-Witten prepotential via taking the $\e_1, \e_2 \to 0$ limit as $\mathcal{F} = \lim_{\epsilon_{1, 2}\rightarrow 0} \epsilon_1 \epsilon_2 \log{\CZ}$ \cite{Nekrasov:2002qd,Nekrasov:2003rj,Nakajima:2003pg, Braverman:2004cr}.

The instanton part of the partition function in the $\Omega$-background can be computed once we know the appropriate instanton moduli space. For the classical gauge groups, the ADHM construction of the moduli space \cite{Atiyah:1978ri} provides a direct way to compute the instanton partition function. The ADHM construction can be understood as the quantum mechanics describing the D$p$-D$(p+4)$ system. The Higgs branch moduli space of the D$p$ system gives the desired moduli space. Matter fields can be also introduced by including more branes, for instance,
by considering the world-volume theory on the D0-branes of the D0-D4-D8 system. 
By using the localization on the 1d system on the D0-branes or its dimensional reduction \cite{Moore:1997dj, Bruzzo:2002xf}, the contour integral formula of the partition function has been obtained for the case of classical gauge groups with a particular choice of matter representations \cite{Nekrasov:2004vw, Marino:2004cn, Fucito:2004gi, Hollands:2010xa, Hollands:2011zc}. 
The precise choice of the contour of the ADHM integral has been derived in \cite{Hwang:2014uwa, Cordova:2014oxa,Hori:2014tda} following the Jeffrey-Kirwan residue formula in 2d elliptic genus \cite{Benini:2013xpa,Benini:2013nda}. 

However, there is no ADHM type construction for the exceptional gauge groups or generic type of matter fields even for the classical group. The string-theoretic picture implies that they require strong-coupling dynamics or non-Lagrangian field theories to realize instanton moduli space of exceptional theories as a vacuum moduli space. Even though there has been a number results regarding the exceptional instantons that we review later in the beginning of Section \ref{sec:example}, a complete way for general instanton counting is still lacking. 

To this end, we generalize the blowup formula of Nakajima-Yoshioka (NY) \cite{Nakajima:2003pg,Nakajima:2003uh,Nakajima:2005fg, Gottsche:2006bm, Nakajima:2009qjc, Gottsche:2010ig}.\footnote{This is an equivariant generalization of the blow-up formula of Donaldson invariants \cite{FintushelStern}, which is also derived and generalized in the context of Seiberg-Witten theory in \cite{Moore:1997pc, Marino:1998bm}.}
 In \cite{Keller:2012da}, the blow-up formula was used to compute the instanton partition function for exceptional gauge group without matter by extrapolating the NY blowup equation to arbitrary gauge group. This is tested against the superconformal index of 4d SCFT where the Higgs branch is given by the instanton moduli space \cite{Gaiotto:2012uq}. Since the Nekrasov partition function computes the topological string partition function for certain toric Calabi-Yau spaces, a similar blowup formula for topological string theory is expected. Indeed such formulae are found and developed in \cite{Grassi:2016nnt, Gu:2017ccq, Huang:2017mis, Gu:2018gmy,Gu:2019dan}. Especially in \cite{Gu:2018gmy,Gu:2019dan}, non-perturbative partition functions for 6d SCFTs are obtained using the blowup equation. 
We generalize the Nakajima-Yoshioka (NY) blowup equations \cite{Nakajima:2003pg,Nakajima:2003uh,Nakajima:2005fg, Gottsche:2006bm, Nakajima:2009qjc, Gottsche:2010ig} to arbitrary gauge group (with a possible 5d Chern-Simons term or discrete theta angle) with hypermultiplets in arbitrary representations. We propose a blow-up formula for a general gauge theory with arbitrary matter representations, under the condition that the matter representation is not `too large' as we discuss shortly. This enables us to compute the instanton partition functions for numerous gauge theories that have not been known before, without relying on the explicit construction of the moduli space. 

The basic idea is as follows: Let us consider a one-point blow-up $\hat{\IC}^2$ of the flat space $\IC^2$. The \emph{full} partition function on $\hat{\IC}^2$ can be written in terms of the products of the \emph{full} partition function of $\IC^2$. But at the same time, the partition function on the blowup is identical to that of the flat space since we can smoothly blow-down $\hat{\IC}^2$ to $\IC^2$. On $\hat{\IC}^2$, we can insert certain topological operator associated with the 2-cycle that turns out to be trivial via selection rule as long as the matter representation is not `too large'. This provides us functional relations for the partition function, that we call the blowup equations of the form
\begin{align}
 \CZ = \sum_{\vec{k} \in \Delta} \CZ^{(N), d}(\vec{k}) \CZ^{(S), d}(\vec{k}) \qquad \textrm{ for }\quad 0 \le d \le d_{\text{max}} \ , 
\end{align}
for some value of $d_\text{max}$ that depends on the gauge group and matter content. Here, $\CZ^{(N/S), d} (\vec{k})$ is given entirely in terms of the flat space partition function $\CZ$ and the sum is over the co-root vectors of the gauge group. 
It turns out that this equation is sufficient to determine the instanton partition function itself as long as $d_\text{max} \ge 2$. Remarkably, this blowup equation leads to a set of recursion relations which can completely determine the instanton contribution from the perturbative part of the partition function. Therefore we arrive at a surprising conclusion: 
\begin{quote}
\centering
The \emph{perturbative} physics determines the \emph{non-perturbative} physics! 	
\end{quote}
Sometimes in the resurgence analysis, the perturbative partition function constrains or even determine the non-perturbative part. This is not necessarily the case, especially for the case of 4d $\CN=2$ and 5d $\CN=1$ gauge theory that we consider \cite{Honda:2016mvg}. In our case, we demand the consistency of the partition function as we change the spacetime smoothly, without assuming any analytic property, which turns out to be sufficient to determine the full partition function from the perturbative part. In fact, it was noticed several decades ago in \cite{Edelstein:1998sp, Edelstein:1999xk} that the instanton part of the  prepotential can be determined recursively via perturbative part. What we find here is that the same statement holds at the level of partition function in Omega background as well.

Using the blowup equation we find, we obtain the following universal expression of the 1-instanton partition function for arbitrary gauge group and matter (5d version):
\begin{align}
\boxed{
  Z_1=\ \frac{e^{-\frac{b}{2}(\e_1 + \e_2)}\prod_{l}e^{\frac{m^\text{tw}_l I_2(\mathbf{R}_l)}{2}}}{(1-e^{-\e_1})(1-e^{-\e_2})} \sum_{\vec{k}\in\Delta_\ell}\frac{e^{\frac{\kappa_{\text{eff}}}{2}(\vec{a}\cdot\vec{k}-d_{ijk}a^ik^jk^k)}\prod_{\w \in \mathbf{R}_l} \CL_{\vec{k}\cdot\vec{\w}}(\vec{a}\cdot \vec{\w} + m_{l}^\text{tw})}{(1-e^{-\e_1-\e_2-\vec{a}\cdot\vec{k}})(1-e^{\vec{a}\cdot\vec{k}})\prod_{\vec{\alpha}\cdot \vec{k}=-1}(1-e^{-\vec{a}\cdot\vec{\alpha}})}}.
\end{align} 
Various symbols in this expression will be explained later in section \ref{sec:blowup}. But we highlight here that this formula depends only on group-theoretic data such as the set of long roots $\Delta_\ell$ and weight vectors for representation $\mathbf{R}_l$ of each hypermultiplet labeled by $l$. 
We emphasize that even though this expression looks completely universal, this formula turns out to be valid only if the matter representations satisfy certain constraint. For example, it fails for the matters in the adjoint representation. We find our 1-instanton formula in 5d is valid if 
\begin{align}
  d_\text{max} = h^\vee -\frac{1}{2}\sum_l I_2(\mathbf{R}_l) \ge 2 \qquad \textrm{for } \quad G \neq SU(N) \textrm{ or } Sp(N) \ ,
\end{align}
where $h^\vee$ is the dual Coxeter number of the gauge group and $I_2(\mathbf{R})$ is the quadratic Dynkin index of the representation $\mathbf{R}$ and $l$ runs over all hypermultiplets that are charged under the gauge group $G$. We give analogous expressions for $G=SU(N)$ or $Sp(N)$ in section \ref{subsec:numd}. The complication arises because of possible Chern-Simons term and discrete theta angle. One can consider 4d version of the partition function as well. In this case, an analogous formula turns out to be valid for any gauge group and matters with $h^\vee - \half \sum_l I_2(\mathbf{R}_l) >1$ since we do not have a Chern-Simons coupling nor discrete theta angle in this case. 
We compute instanton partition functions for a large number of examples and test against known results, from which we build our confidence for the generalized blowup equation we find. 

The organization of this paper is as follows: In Section \ref{sec:blowup}, we give a physical derivation of the blowup equations. From this, we obtain a recursion formula to compute instanton partition function at any number of instantons. Especially, we derive a concise closed-form formula for one instanton partition function that works for a large number of theories. We give a precise condition for our formula to work. In the case of 4d Nekrasov partition function, we are able to derive the bound on the matter representation for which the formula to work. We suggest an analogous bound for 5d version, by extrapolating known results. In Section \ref{sec:example}, we test our formula against various known cases. Moreover, we also obtain many previously unknown partition functions. Some of them are expressed as the character expansion, whose form is explicitly given in Appendix \ref{sec:data}. We then conclude with several future directions.

%%%%%%%%%%%%%%%%%%%%%%%%%%%%%%%%%%%%%%%%%%%%%%%%%%%%%%%%%%%%%%%%

\section{Instanton Counting from Blow-up} \label{sec:blowup}

The essential idea of using the blow-up of ${\IC}^2$ for instanton counting is that the gauge theory partition function for a 4d $\CN=2$ (or 5d $\CN=1$) theory on the blow-up of a point $\hat{\IC}^2$ (or $\hat{\IC}^2 \times S^1$) can be written in two different ways. This will allow us to write a recursion relation for the instanton partition function that can be solved rather easily \cite{Nakajima:2003pg, Nakajima:2003uh,Nakajima:2005fg, Keller:2012da}.

\subsection{Blowup equation}

\paragraph{Localization on the blow-up $\hat{\IC}^2$}

One of the expressions for the partition function $\hat{\mathcal{Z}}$ on the blow-up $\hat{\IC}^2$ comes from the Coulomb branch localization, which results that $\hat{\mathcal{Z}}$ can be obtained by patching together the flat-space partition function $\mathcal Z$ \cite{Nekrasov:2003vi}.

The blow-up $\hat{\IC}^2$ of the complex plane is constructed from $\IC^2$ by replacing the origin with a compact 2-cycle $\IP^1$. In particular, the geometry is identical to the total space of the line bundle of degree $(-1)$ over $\IP^1$. One can parametrize $\mathcal{O}(-1)\rightarrow \IP^1$ using the homogeneous coordinates $(z_0, z_1, z_2)$, satisfying the projective condition $(z_0, z_1, z_2) \sim (\lambda^{-1}z_0, \lambda^1 z_1, \lambda^1 z_2)$ for any $\lambda \in \IC^*$, where the two-cycle  $\IP^1 \subset \hat{\IC}^2$ corresponds to the locus $z_0 = 0$. 
We are interested in the $U(1)^2$ equivariant partition function, with the $U(1)^2$ action $V$ rotating the complex coordinates $(z_0, z_1, z_2)$ as follows:
\begin{align}
  (z_0, z_1, z_2) \mapsto (z_0, e^{\e_1}z_1, e^{\e_2}z_2).
\end{align}
Instantons are located at two fixed points of the $U(1)^2$ action, \textit{i.e.}, the north/south poles of the $\mathbb{P}^1$, whose coordinates are 
$(z_0, z_1, z_2)  = (0,1,0)$ and $(0,0,1)$. Around these fixed points, ($\IC^*$-invariant) local coordinates are given by $(z_0 z_1, z_2/z_1)$ and $(z_0 z_2, z_1/z_2)$ respectively. The local weights under the $U(1)^2$ action $V$ near the fixed points are:
\begin{align}
\begin{split}
  (z_0 z_1,\,z_2/z_1) \mapsto  (e^{\e_1}z_0 z_1, \,e^{\e_2 - \e_1}z_2/z_1) & \qquad \qquad \text{(near the north pole)}\\
  (z_0 z_2,\,z_1/z_2) \mapsto (e^{\e_2}z_0 z_2, \,e^{\e_1 - \e_2}z_1/z_2) &  \qquad \qquad \text{(near the south pole)}
\end{split}
\end{align}

The full partition function $\hat{\mathcal Z}$ on $\hat{\IC}^2$, which includes both the perturbative and instanton contributions, can be obtained by performing the localization on the Coulomb branch. On the Coulomb branch, the gauge group is generically broken to $U(1)^r$ where $r$ is the rank of the gauge group. The $U(1)^r$ equivariant parameters $\vec{a}$ naturally appear in the partition function. One needs to sum over all distinct field configurations with zero-sized instantons located at the north and south poles. All the inequivalent configurations are labeled by the $r$-dimensional vector $\vec{k}$ of the first Chern numbers, corresponding to different flux configurations on the two-cycle $\mathbb{P}^1$. When the gauge group has $U(1)$ factor, we can turn on the external flux that can be supported on the $\IP^1$. We assume there is no such a factor in the gauge group. 
Summing up, $\hat{\mathcal Z}$ can be expressed in terms of the partition function $\CZ$ on $\IC^2$ as \cite{Nekrasov:2003vi, Gottsche:2006bm, Gottsche:2006tn, Gasparim:2008ri, Bonelli:2012ny}
\begin{align} \label{eq:blowup}
  \hat{\mathcal Z}(\vec{a}, \e_1, \e_2, q, \vec{m}) = \sum_{\vec{k} \in \Lambda} \CZ^{(N)}(\vec{k}) \CZ^{(S)}(\vec{k}) \ , 
\end{align}
where the flux sum is taken over the co-root lattice $\Lambda$ of the gauge algebra. Each factor represents the partition function localized at the $U(1)^2$ fixed points (north/south-poles of the $\IP^1 \subset \hat{\IC}^2$) given as 
\begin{align}
\begin{split}
 \CZ^{(N)}(\vec{k}) &\equiv \CZ (\vec{a}+ \vec{k} \e_1, \e_1, \e_2 - \e_1, q, \vec{m}-\half{\e_1}) \ , \\
 \CZ^{(S)}(\vec{k}) & \equiv  \CZ (\vec{a}+\vec{k} \e_2, \e_1 - \e_2, \e_2, q, \vec{m}-\half{\e_2})  \ . 
\end{split}
\end{align}
 In addition to the Coulomb branch parameters, the partition function depends on the Omega deformation parameters $\e_1, \e_2$ and also mass parameters $\vec{m}$. The instanton fugacity $q$ takes the following form: For a 4d theory, it is given as $q = e^{2\pi i \tau} = \Lambda^{b_0} $ where $\tau$ is the complexified gauge coupling and $\Lambda$ being the dynamical scale of the gauge theory. The exponent $b_0$ is the 1-loop beta function coefficient. For a 5d theory, it is also given by the exponentiated gauge coupling as $q=e^{-\frac{1}{g^2}} \equiv e^{-m_0}$. 
Notice that the Coulomb parameter $\vec{a}$ gets an appropriate shift at each fixed point $p$, induced by the non-trivial magnetic flux $\vec{k}$ on the blown-up $\IP^1$, with the proportionality constant $H|_p$. The values of the moment map $H$ for the $U(1)^2$ action $V$,  \textit{i.e.}, $dH = \iota_V \omega$, at the north and south poles are given as
\begin{align}
  H|_\text{NP} = \epsilon_1 \text{ and } H|_\text{SP} = \epsilon_2.
\end{align}
The mass parameters also get shifted since the hypermultiplet mass is twisted by $SU(2)_R$, which makes the combination $m - \frac{\e_1+\e_2}{2}$ invariant at the fixed points.\footnote{One can instead use the shifted mass to simplify the formula involving mass. We use unshifted mass to match with the existing formulae in the literature.} 

\paragraph{Partition function on $\hat{\IC}^2$ vs $\IC^2$}
Another important fact for the partition function $\hat{\CZ}$ on the blow-up $\hat{\IC}^2$ is that $\hat{\CZ}$ is actually identical to the flat-space partition function $\CZ$  \cite{Nakajima:2003pg,Nakajima:2003uh,Nakajima:2005fg, Gottsche:2006bm, Nakajima:2009qjc, Gottsche:2010ig}.
The blow-up $\hat{\IC}^2$ is identical to $\IC^2$ except for the origin, which is replaced by the blown-up sphere $\IP^1$. 
Since the Nekrasov partition function gets contributions only from the small instantons localized at the fixed points of the $U(1)^2$ equivariant action $V$, the size of the divisor should not affect the partition function as we smoothly shrink it. So we expect that $\hat{\CZ} = \CZ$. This implies the following relation: 
\begin{align} 
  \label{eq:blowup-generic}
  \CZ = \hat{\CZ} = \sum_{\vec{k} \in \Lambda} \CZ^{(N)}(\vec{k}) \CZ^{(S)}(\vec{k}).
\end{align}

This blow-up identity can be thought of as a special case of more generalized orbifold partition functions \cite{Sasaki:2006vq,Bonelli:2012ny,Ito:2013kpa, Bruzzo:2013daa, Bruzzo:2014jza}. 
For example, the Nekrasov partition function on the orbifold $\IC^2/\IZ_2$ can be computed in two different ways, one is via formula analogous to \eqref{eq:blowup} by combining the contributions from two fixed points of the blown-up geometry $\CO(-2) \to \IP^1$. The other way is to compute the partition function at the orbifold point using the ADHM construction for the orbifolds. The Nekrasov partition function still remains the same as we blow up or down the singular point.\footnote{This simple picture does not necessarily hold when there are too many hypermultiplets, due to some subtle scheme dependence related to the wall-crossing \cite{Ito:2013kpa}.} 
The only difference in our case is that we blow-up or down a non-singular point instead of a singular point. 

\paragraph{Correlation functions in 4d}

The equation \eqref{eq:blowup-generic} itself is not enough to fix the partition function completely, since there are 3 unknown functions and only one relation. It turns out the necessary additional relations can be found from the insertion of non-trivial $\mathcal{Q}$-closed operators \cite{Nakajima:2003pg,Nakajima:2005fg} associated to the two-cycle on the blow-up. 

In the 4d Donaldson-twisted theory, the $\mathcal{Q}$-invariant observable $\CO_2$ associated to a two-cycle can be constructed by applying the topological descent procedure twice to the Casimir invariant $\CO_0 = \text{Tr}(\Phi^2)$ as \cite{Witten:1988ze}
\begin{align}
\begin{split}
0 &= \{ \CQ, \CO_0 \}, 
~ d\CO_0 = \{\CQ, \CO_1 \}, 
~ d\CO_1 = \{\CQ, \CO_2 \}, \\
~ d\CO_2 &= \{\CQ, \CO_3 \}, 
~ d\CO_3 = \{\CQ, \CO_4 \}, 
~ d\CO_4 = 0 \ . 
\end{split}
\end{align} 
In our case, we consider a $U(1)^2$-equivariant version of the topological descent procedure, that is to choose $\CQ$ so that $\CQ^2 = \CL_V$ and also change $d \to D \equiv d + \iota_V$ to obtain the operator associated to the two-cycle. In terms of the component fields, it can be written as \cite{Bershtein:2015xfa}
\begin{align} \label{eq:muC}
  {\CO}_{\IP^1} = \int_{\IP^1} \CO_2 = \int_{M_4} \left\{ \omega \wedge \text{Tr}\Big(\Phi F + \half \psi \wedge \psi\Big) + H\, \text{Tr}\Big( F \wedge F \Big) \right\}  \ .
\end{align} 
Here $\omega$ and $H$ are the K\"ahler two-form on the $\IP^1$ and the moment map $\iota_V \omega = dH$, respectively. $M_4$ denotes the spacetime. The first part of \eqref{eq:muC} without $H$ is the non-equivariant version of the topological operator associated to two-cycle. 
It is convenient to study the generating function $\langle e^{t \, {\CO}_{\IP^1}}\rangle $ of the correlators $\langle {\CO}_{\IP^1} \ldots {\CO}_{\IP^1} \rangle$. 
This causes a shift of the instanton parameter by $q \to q \exp(t H)$ at the fixed points of the blow-up $\hat{\IC}^2$ \cite{Nakajima:2003pg,Nakajima:2003uh, Nakajima:2005fg}.  
The expectation value of the generating function can be written as
\begin{align} \label{eq:corgen}
  \hat{\CZ}^{t} \equiv \langle e^{t  {\CO}_{\IP^1}}\rangle = \sum_{\vec{k} \in \Lambda}  \CZ^{(N),t}(\vec{k}) \cdot \CZ^{(S),t}(\vec{k})\ ,  
\end{align}
where
\begin{align}
  \label{eq:param-np-sp}
\begin{split}
  \CZ^{(N),t}(\vec{k}) &\equiv \CZ(\vec{a}+\vec{k} \e_1, \e_1, \e_2-\e_1, q \exp(t \e_1), \vec{m} -\tfrac{1}{2}\e_1) \ , \\
  \CZ^{(S),t}(\vec{k}) &\equiv \CZ(\vec{a}+\vec{k} \e_2, \e_1 - \e_2, \e_2, q \exp(t \e_2), \vec{m} -\tfrac{1}{2}\e_2)\ . \\
\end{split}
\end{align}

Now, as we shrink the two-cycle $\IP^1$ to recover the flat $\IC^2$, the effect of inserting $({\CO}_{\IP^1})^d$ turns out to give a vanishing contribution for small $d$ due to the selection rule. We recall that the instanton breaks the $U(1)_R$ symmetry to the discrete subgroup $\IZ_{2b_0}$ with $b_0 = 2h^\vee - \sum_l I_2(\mathbf{R}_l)$ where the sum is over all hypermultiplets, and $h^\vee$ is the dual Coxeter number of the gauge group and $\mathbf{R}_l$ denotes the representation of the $l$-th hypermultiplet and $I_2(\mathbf{R})$ being the quadratic Dynkin index.\footnote{We normalize it so that $I_2(\mathbf{F})= 1$ for the fundamental representation $\mathbf{F}$.} The first term of the operator ${\CO}_{\IP^1}$ (the two-form piece) carries $R$-charge $+2$, which is the familiar non-equivariant version. This discrete $R$-charge is sometimes called as a ghost number. The correlation functions vanish unless the $R$-charges add up to zero, modulo $2b_0 = 4h^\vee - 2\sum_l I_2(\mathbf{R}_l)$. 
Therefore, expanding \eqref{eq:corgen} in powers of $t$, we find
\begin{align}
  \label{eq:donaldson-4d}
  \boxed{  \langle e^{t {\CO}_{\IP^1}}\rangle = \CZ + \CO \left(t^{2h^\vee -  \sum_l I_2(\mathbf{R}_l)}\right) } \ . 
\end{align}
This is our blowup equation. 
To show this, notice that each term at order $t^m$ carries pieces with $R$-charge between $0$ and $2m$. When $m < b_0$, the only possible non-trivial contribution comes from the $R=0$ piece $ \int H F\wedge F$ at zero instanton sector.  This piece vanishes for zero instanton sector (at the north/south poles). For $n$-instanton sector, one should have $R=2b_0 n$, which is the condition to absorb the fermionic zero modes. For $m \ge b_0$, we always have a term that absorbs all the fermionic zero modes (or the term that has $R \equiv 0 \textrm{ mod } 2b_0 n)$ so they do not vanish. 

We see that as long as the hypermultiplet representation is not too large, \emph{i.e.}, when $b_0 = 2h^\vee - \sum_l I_2(\mathbf{R}_l) > 2$, this allows us to write 3 independent relations for the 3 unknown variables. One can expand $\langle e^{t \, \CO_{\IP_1}}\rangle$ to order $t^2$, $\mathcal{O}(t^2)$ and then recursively solve for $\CZ$ at each instanton number. So the instanton part of the partition function will be completely determined from the perturbative partition function. An explicit form of the recursion relation will be studied in Section~\ref{subsec:recursion}.

\paragraph{Correlation functions in 5d}
We now turn to 5d $\CN=1$ gauge theory wrapped on $S^1$. The Casimir invariant $\Tr (\Phi^2)$ and its descendants are no longer considered as well-defined observables. Instead, there are two types of $\CQ$-invariant observables \cite{Baulieu:1997nj}. The first type of observables are constructed from the 5d Wilson loop on the $S^1$ by applying the descent procedure. The second type of observables introduce the 3d (K\"ahler) Chern-Simons term, which can be written as \cite{Losev:1995cr,Baulieu:1997nj}
\begin{align} 
  \mathcal{O}_{\IP^1} = \exp\bigg[
    \int_{S^1 \times M_4} \bigg( & \omega \wedge \text{Tr}\Big(A \wedge dA + \frac{2}{3}A\wedge A \wedge A \Big) \\ \nonumber
     & + \omega \wedge \Big( \phi \, F + \frac{1}{2}\psi \wedge \psi \Big)\wedge dt + H\, \text{Tr}\Big( F \wedge F \Big) \wedge dt \bigg)
     \bigg] \ .
\end{align}
It can be viewed as the natural $S^1$ uplift of \eqref{eq:muC} via exponentiation. The correlation function is now given by
\begin{align}
  \label{eq:cor5d}
  \hat{\CZ}^{d} \equiv \langle (\mathcal{O}_{\IP^1})^{d}\rangle = \sum_{\vec{k} \in \Lambda}  \CZ^{(N),d}(\vec{k}) \cdot \CZ^{(S),d}(\vec{k}) \ , 
\end{align} 
where
\begin{align}
  \label{eq:param-np-sp5d}
\begin{split}
  \CZ^{(N),d}(\vec{k}) &\equiv \CZ(\vec{a}+\vec{k} \e_1, \e_1, \e_2-\e_1, q \exp \left((d - \tfrac{b}{2})\e_1 \right), \vec{m} -\tfrac{1}{2}\e_1) \ , \\
  \CZ^{(S),d}(\vec{k}) &\equiv \CZ(\vec{a}+\vec{k} \e_2, \e_1 - \e_2, \e_2, q \exp \left( (d - \tfrac{b}{2})\e_2\right), \vec{m} -\tfrac{1}{2}\e_2)\ . 
\end{split}
\end{align}
Here the quantity $b$ is given as 
\begin{align}
 b \equiv h^\vee - \frac{1}{2} \sum_i I_2(\mathbf{R}_i) - \kappa_{\textrm{eff}} \ , \quad \kappa_{\textrm{eff}} = \kappa - \half \sum_i I_3(\mathbf{R}_i) \ , 
\end{align}
where $I_2(\mathbf{R})$ and $I_3(\mathbf{R})$ are quadratic and cubic Casimir invariants respectively. 
We note that $d$ appearing in the exponential in \eqref{eq:param-np-sp5d} has to be an integer to be gauge-invariant.

The reason that the instanton parameter is further shifted by $\exp(\frac{b}{2} H|_p) $ is that the instanton mass parameter is twisted by $SU(2)_R$ as in the case of the hypermultiplet mass. The $SU(2)_R$ twisted mass of the instanton soliton is given by $m_\text{inst} \equiv m_{0, \textrm{eff}} - \kappa_\text{eff}\,\e_+$. The effective Chern-Simons coupling $\k_{\textrm{eff}}$ also induces an electric charge to the instanton, contributing to its ground state energy as $E_0 = m_\text{inst} - \vec{a} \cdot \vec{\Pi}$, where $\vec{\Pi}$ is the $U(1)^r \subset G$ electric charge.\footnote{This agrees with the supersymmetric Casimir energy of the ADHM quantum mechanics.} To keep the effective instanton mass $m_\text{inst}$ invariant at a fixed point $p$ of the blow-up $\hat{\IC}^2$, we require the shifted gauge coupling $m_0|_p$  to be 
\begin{align}
  \label{eq:coupling-shift}
  m_0|_p = m_0 + \frac{b}{2}H|_p \quad \text{ with }\quad
  b \equiv h^{\vee}-\sum_i\frac{I_2(\mathbf{R}_i)}{2} - \kappa_\text{eff}.
\end{align}

For the case of 5d pure $\mathcal{N}=1$ SYM, the correlation function turns out to be  
\begin{align}
  \label{eq:cor5d-unity}
\boxed{  \langle (\mathcal{O}_{\IP^1})^{d}\rangle = \CZ \quad \text{ for }\quad 0\leq d \leq d_\text{max} } \ , 
\end{align}
where $d_\text{max} = h^\vee$.\footnote{This was shown in \cite{Nakajima:2005fg} for the case of $G=SU(N)$.} We call \eqref{eq:cor5d-unity} as the blowup equation. The value of $d_{\text{max}}$ depends on the matter content and gauge group. For $d_\text{max} \geq 2$, there are a sufficient number of algebraic relations to determine the instanton partition function recursively in increasing order of instantons. This fact was utilized in \cite{Keller:2012da} to compute instanton partition function for the gauge theories with exceptional gauge groups, for which the ADHM construction of instanton moduli space is unknown. 

In this paper, we aim at developing the relation \eqref{eq:cor5d-unity} for various 5d 
$\mathcal{N}=1$ gauge theories with hypermultiplets in various representations, so as to compute the instanton partition function. We will identify a certain bound on $d$ in Section~\ref{subsec:numd} as the \emph{necessary} condition for \eqref{eq:cor5d-unity} for a large number of theories. 
We conjecture that the bound on $d$ we obtain is actually sufficient to obtain the blowup equation \eqref{eq:cor5d-unity}. While we do not attempt to prove this sufficiency, we compute $n$-instanton partition function $Z_n$, based on the recursion formula that will be derived shortly from \eqref{eq:cor5d-unity}, and confirm the agreement with the known result obtained from an alternative method.

We find a universal expression for the bound on $d$ when the gauge group is neither $SU(N)$ nor $Sp(N)$:
\begin{align}
  \label{eq:cor5d-dmax}
 d_{\textrm{max}} =  h^\vee -\frac{1}{2}\sum_l I_2(\mathbf{R}_l) \quad \textrm{for } G \neq SU(N) \textrm{ or } Sp(N).
\end{align}
This is essentially identical condition as in $4d$ $\CN=2$ gauge theory. But in 5d, some new effects come into play. 
For the $SU(N)$ case, we can have a Chern-Simons term generated at 1-loop, which alters the bound on $d$. When there is neither bare nor effective Chern-Simons coupling, the same bound holds for the $SU(N)$ case as well. The detailed condition will be given in section \ref{subsec:numd}. For the case of $Sp(N)$, one can turn on the discrete $\theta$-parameter and it turns out the bound on $d$ depends on this parameter.

\subsection{Recursion formula for 5d instanton partition function}
\label{subsec:recursion}

The blowup equation \eqref{eq:cor5d-unity} can be translated to a recursion formula on the (5d) $n$-instanton contribution $Z_n$ to the full partition function $\CZ$.
To derive this, we decompose the partition function $\CZ$ in terms of the classical, one-loop, and instanton pieces: 
\begin{align}
  \CZ(\vec{a}, \e_1, \e_2, q, \vec{m}) = Z_{\textrm{class}}(\vec{a}, \e_1, \e_2, q, \vec{m}) \cdot  Z_{\textrm{1-loop}} (\vec{a}, \e_1, \e_2, \vec{m}) \cdot Z_{\textrm{inst}}(\vec{a}, \e_1, \e_2, q, \vec{m}),
 \end{align}
where $Z_\text{inst}$ can be further expanded in terms of the instanton fugacity $q$ as\footnote{Sometimes the instanton partition function is expanded in powers of the shifted instanton mass $q \exp (-b \frac{\e_1+\e_2}{2})$ instead of $q$ \cite{Nakajima:2005fg, Gottsche:2006bm, Keller:2012da}. We expand it with the true instanton fugacity, which makes the symmetry property $\e_{1, 2} \to - \e_{1, 2}$ of $Z_n$ manifest. This is the one that we obtain using the ADHM quantum mechanics.}
\begin{align}
 Z_{\textrm{inst}} (\vec{a}, \e_1, \e_2, q, \vec{m}) = \sum_{n \ge 0} q^n Z_n (\vec{a}, \e_1, \e_2, m) \ . 
\end{align}
Then the blowup equation \eqref{eq:cor5d-unity} can be written as
\begin{align}
  \label{eq:recur-inst}
  \begin{split}
 Z_{\textrm{inst}} &= \sum_{\vec{k}} \left[ \frac{Z^{(N), d}_{\textrm{class}}(\vec{k}) \, Z^{(S), d}_{\textrm{class}}(\vec{k})}{Z_{\textrm{class}}}  \frac{Z^{(N), d}_{\textrm{1-loop}}(\vec{k})\, Z^{(S), d}_{\textrm{1-loop}}(\vec{k})}{Z_{\textrm{1-loop}}} \right] Z^{(N), d}_{\textrm{inst}}(\vec{k})\, Z^{(S), d}_{\textrm{inst}}(\vec{k}) \\
  &\equiv  \sum_{\vec{k}} f_d(\vec{k}) Z^{(N), d}_{\textrm{inst}}(\vec{k})\, Z^{(S), d}_{\textrm{inst}}(\vec{k}) \ , 
  \end{split}
\end{align}
where the superscript $(N/S),d$ denotes the appropriate shift of the parameters, specified in \eqref{eq:param-np-sp}. The function $f_d(\vec{k})$ is determined only via the perturbative part of the partition function. 

We recall the known expressions for the classical and 1-loop partition function (in 5d) \cite{Intriligator:1997pq,Nekrasov:2002qd,Shadchin:2005mx}:\footnote{There exists an ambiguity in writing the perturbative partition function, which depends on a choice of the $\IC^2$ boundary condition at infinity. The equations \eqref{eq:1-loop-vec} and \eqref{eq:1-loop-hyp} are fixed upon a specific choice. 
The `Casimir part' of $Z_{\text{1-loop}}$ is included here to make $f_d(\vec{k})_\text{1-loop}$ and thus the whole blow-up equations respect the charge conjugation, regardless of the ambiguity. We thank Hee-Cheol Kim for the related comment.}
\begin{align}
   { Z_{\textrm{class}}} &= \exp \Bigg[\frac{1}{\epsilon_1\epsilon_2}\left(\frac{1}{2}m_0 \, h_{ij}a_i a_j +\frac{\kappa}{6}d_{ijk} a^{i}a^j a^k\right) \Bigg],\\
   \begin{split}  \label{eq:1-loop-vec}
   { Z_{\textrm{1-loop}}^\text{vec}} &= \exp \Bigg[\frac{1}{\epsilon_1\epsilon_2} \,  \sum_{\vec{\alpha}\in\Delta}\Big(\frac{(\vec{a}\cdot\vec{\alpha}+\e_+)^3}{12}-\frac{\e_1^2+\e_2^2+24}{48}\,(\vec{a}\cdot\vec{\alpha}+\e_+)+1\Big)\Bigg]\\ & ~~\times \text{PE}\ \Bigg[- \frac{p_1 p_2}{(1-p_1)(1-p_2) }\sum_{\vec{\a} \in \D} e^{- \vec{a} \cdot \vec{\a}}  \Bigg]  \qquad \text{ for the vector multiplet}
   \end{split} \\
   \begin{split}  \label{eq:1-loop-hyp}
   { Z_{\textrm{1-loop}}^\text{hyp,$l$}} &= \exp \Bigg[-\frac{1}{\epsilon_1\epsilon_2} \sum_{\vec{\omega}\in\mathbf{R}_l}\Big(\frac{(\vec{a}\cdot\vec{\omega}+m_l)^3}{12}-\frac{\e_1^2+\e_2^2+24}{48}\,(\vec{a}\cdot\vec{\omega}+m_l)+1\Big)\Bigg]\\ &
   ~~\times \text{PE}\ \Bigg[+ \frac{(p_1 p_2)^{\frac{1}{2}} \cdot  y_\ell } {(1-p_1)(1-p_2) }\sum_{\vec{\omega}\in\mathbf{R}_l}e^{-\vec{a} \cdot \vec{\w}}\Bigg] \qquad \text{ for the $l$'th hypermultiplet}
   \end{split}
\end{align}
where $p_1 \equiv e^{-\e_1},\, p_2 \equiv e^{-\e_2},\, y_l \equiv e^{-m_{l}}, q \equiv e^{-m_0}$.\footnote{We assume a particular Weyl chamber in the Coulomb branch, i.e., $ 0< a_i < \e_+ < m$ for all $i\in \{1,\cdots, r\}$.} Also $\Delta$ is the set of all roots and $\vec{\omega}	$ runs over all weight vectors in representation $\mathbf{R}_\ell$. 
Here, PE represents the Plethystic exponential
\begin{align}
  \label{eq:PE}
  \text{PE}\left[f(\vec{a},\e_1,\e_2, m_0 ,\vec{m})\right] \equiv \exp\left(\sum_{n=1}^\infty\frac{1}{n} f(n\vec{a},n\e_1,n\e_2,n m_0,n\vec{m})\right).
\end{align}
We also set the radius of ${S}^1$ as $\beta = 1$. Also, the symbols $h_{ij}$ and $d_{ijk}$ are defined as 
\begin{align}
 h_{ij} = \Tr (T_i T_j ) \ , \quad
 d_{ijk} = \half \Tr T_i \{T_j,  T_k\} \ , 
\end{align}
where $T_i$ are the generators of the gauge algebra. 
They satisfy the relations
\begin{align}
\begin{split}
   \sum_{\vec{\omega}\in\mathbf{R}}(\vec{a}\cdot\vec{\omega})(\vec{b}\cdot\vec{\omega})(\vec{c}\cdot\vec{\omega})=&\,I_3(\mathbf{R})\,d_{ijk}\,a^ib^jc^k, \\
   \sum_{\vec{\omega}\in\mathbf{R}}(\vec{a}\cdot\vec{\omega})(\vec{b}\cdot\vec{\omega})=&\,I_2(\mathbf{R})\,h_{ij}\,a^ib^j,\\
   \sum_{\vec{\omega}\in\mathbf{R}}(\vec{a}\cdot\vec{\omega})=&\,0 , 
\end{split}
\end{align}
where $I_2(\mathbf{R})$ and $I_3(\mathbf{R})$ are the quadratic and cubic Dynkin indices. 

Substituting them to \eqref{eq:recur-inst}, we obtain the ratio of three different $Z$'s given as
\begin{align}
 f_d(\vec{k})_{\text{class}} &=  q^{\frac{\vec{k}\cdot\vec{k}}{2} } \, 
  (p_1p_2)^{(\frac{b}{2}-d) (\frac{\vec{k}\cdot\vec{k}}{2}) +\frac{\k}{6}d_{ijk}\,k^ik^jk^k} \times  e^{-(\frac{b}{2}-d)(\vec{a}\cdot\vec{k})}\,
  e^{-\frac{\kappa}{2}d_{ijk}\,a^ik^jk^k} \ , \\
  f_d(\vec{k})_{\text{1-loop}}^{\text{vec}}&= e^{\frac{h^\vee}{2}(\vec{a}\cdot\vec{k})}  \prod_{\a \in \D} \CL_{\vec{k} \cdot \vec{\a}} (\vec{a}\cdot\vec{\alpha}, \epsilon_1, \epsilon_2)^{-1} \ , \\
 \begin{split}
  f_d(\vec{k})_{\text{1-loop}}^{\text{hyp}} &= e^{-\frac{I_2(\mathbf{R}_l)}{4}(\vec{a}\cdot\vec{k})+\frac{I_3(\mathbf{R}_l)}{4}d_{ijk}\,a^ik^jk^k} (p_1p_2)^{\frac{I_2(\mathbf{R}_\ell)}{8}(\vec{k}\cdot\vec{k})-\frac{I_3(\mathbf{R}_l)}{12}d_{ijk}\,k^ik^jk^k} \\
   &\qquad \times   y_\ell^{-\frac{I_2(\mathbf{R}_\ell)}{4}(\vec{k}\cdot\vec{k})} \prod_{\w \in \mathbf{R}_l} \CL_{\vec{k}\cdot\vec{\w}}(\vec{a}\cdot \vec{\w} + m_{\text{tw},l}, \e_1,\e_2) \ , 
\end{split}
\end{align}
where we split the $f_d(\vec{k})$ into classical and 1-loop pieces for vector and hypermultiplet. 
Here we used $I_2(\textbf{adj}) = 2h^\vee$, $I_3(\textbf{adj})=0$, and also 
the fact $h_{ij}$ and $d_{ijk}$ are totally symmetric. We also define $m_{\text{tw}} \equiv m - \e_+$. The function $\CL_k (x,\e_1,\e_2)$ is introduced to denote concisely the combination of the PE parts:
\begin{align}
  \CL_k (x,\e_1,\e_2) \equiv \text{PE}\left[ e^{-x} \left(\frac{p_1^k \, p_2}{(1-p_1)(1-\frac{p_2}{p_1})}+\frac{p_1\, p_2^k}{(1-\frac{p_1}{p_2})(1-p_2)}-\frac{p_1p_2}{(1-p_1)(1-p_2)}\right) \right].
\end{align}
One can easily check that the expression inside the PE vanishes at $k=0, 1$. After some work, it is not difficult to find that 
\begin{align}
  \CL_k (x,\e_1,\e_2) =
   \begin{dcases}
   \prod_{m+n \le k-2}(1-p_1^{m+1} p_2^{n+1}e^{-x}) & \text{ for } k\geq +2 \\
   \prod_{m+n \le -k-1}(1-p_1^{-m} p_2^{-n}e^{-x}) & \text{ for } k\leq-1 \\
   1 & \text{ for } k=0, 1.
   \end{dcases} . 
\end{align}
Combining them all together, the recursion formula on the $n$-instanton piece $Z_n$ can be written as
\begin{align}
  \label{eq:recursion}
  Z_n = &\sum_{\frac{1}{2}\vec{k}\cdot\vec{k} + \ell + m = n}  
  \Bigg( (p_1p_2)^{(\frac{b}{2} -d) (\frac{\vec{k}\cdot\vec{k}}{2}) +\frac{\k_\text{eff}}{6}d_{ijk}\,k^ik^jk^k} e^{(d+\frac{\k_\text{eff}}{2})(\vec{a}\cdot\vec{k})}\,
 e^{-\frac{\k_\text{eff}}{2}d_{ijk}\,a^ik^jk^k} \\& 
 \times    \frac{\prod_{l} y_{\text{tw}, l}^{-I_2(\mathbf{R}_l)(\frac{\vec{k}\cdot\vec{k}}{4})} \prod_{\w \in \mathbf{R}_l} \CL_{\vec{k}\cdot\vec{\w}}(\vec{a}\cdot \vec{\w} + m_{\text{tw},l}, \e_1,\e_2)}{ \prod_{\a \in \D} \CL_{\vec{k} \cdot \vec{\a}} (\vec{a}\cdot\vec{\alpha}, \epsilon_1, \epsilon_2)}
 \cdot p_1^{(\frac{b}{2}-d)\ell}p_2^{(\frac{b}{2}-d)m}Z^{(N)}_{\ell}(\vec{k})  Z^{(S)}_m (\vec{k}) \Bigg) \nn,
 \end{align}
where $y_{\text{tw},l} \equiv e^{-m_{\text{tw},l}} = y_l / \sqrt{p_1p_2}$ and $l$ runs over all hypermultiplets in the theory. This is a generalization of the recursion formula found for the pure SYM case \cite{Nakajima:2005fg,Gottsche:2006bm}. 

\paragraph{Solving the recursion formulae}
The recursion relation \eqref{eq:recursion} can be rewritten as 
\begin{align}
  \label{eq:kinst}
  Z_n = p_1^{n(\frac{b}{2}-d)}Z^{(N)}_{n}  + p_2^{n(\frac{b}{2}-d)}  Z^{(S)}_n  + I_n^{(d)} \qquad \text{with an allowed range of $d$},
\end{align}
where $I_{n}^{(d)}$ is defined as
\begin{align} \label{eq:Ind}
  I_{n}^{(d)} = &\sum_{\stackrel{\frac{1}{2}\vec{k}\cdot\vec{k} + \ell + m = n}{\ell, m \neq n}}  
  \Bigg( (p_1p_2)^{(\frac{b}{2} -d) (\frac{\vec{k}\cdot\vec{k}}{2}) +\frac{\k_\text{eff}}{6}d_{ijk}\,k^ik^jk^k} e^{(d+\frac{\k_\text{eff}}{2})(\vec{a}\cdot\vec{k})}\,
 e^{-\frac{\k_\text{eff}}{2}d_{ijk}\,a^ik^jk^k}  \\& 
 \times    \frac{\prod_{l} y_{\text{tw}, l}^{-I_2(\mathbf{R}_l)(\vec{k}\cdot\vec{k}/4)} \prod_{\w \in \mathbf{R}_l} \CL_{\vec{k}\cdot\vec{\w}}(\vec{a}\cdot \vec{\w} + m_{\text{tw},l}, \e_1,\e_2)}{ \prod_{\a \in \D} \CL_{\vec{k} \cdot \vec{\a}} (\vec{a}\cdot\vec{\alpha}, \epsilon_1, \epsilon_2)}
 \cdot p_1^{(\frac{b}{2}-d)\ell}p_2^{(\frac{b}{2}-d)m} Z^{(N)}_{\ell}(\vec{k})  Z^{(S)}_m (\vec{k}) \Bigg) . \nn
\end{align}
Notice that we have a set of equations labeled by the parameter $d$. If the blowup equation holds for at least 3 values of $d$, we can solve it for $Z_n$. 
The $n$-instanton partition function $Z_n$ is given as the solution to the three linear equations \eqref{eq:kinst} with consecutive integers $\{d_0, d_0+1, d_0+2\}$,
\begin{align}
  Z_n(\vec{a},\e_1,\e_2, \vec{m})=\frac{p_1^n p_2^n \,I_{n}^{(d_0+2)}-(p_1^n+p_2^n)\,I_{n}^{(d_0+1)}+I_{n}^{(d_0)}}{(1-p_1^n)(1-p_2^n)}.
\end{align}
Since $I_n^{(d)}$ only involves low-order instanton corrections, the $n$-instanton partition function $Z_n$ can be constructed from $Z_{m<n}$, allowing us to obtain the full non-perturbative part $Z_\text{inst}$ in a recursive manner starting from $Z_0 = 1$. 

Therefore we arrive at a remarkable conclusion. The \emph{non-perturbative} partition function $Z_\text{inst}$ is completely fixed by the \emph{perturbative} partition function! We note that we do not reach this conclusion by requiring the perturbative series to be well-behaved, as is often done in the resurgence analysis. Instead, we demand consistency upon smooth deformation of the spacetime $\IC^2$ or $\IC^2 \times S^1$. Such consistency condition requires non-perturbative parts to exist and even enough to fix the instanton partition function (at least for a large number of examples).

Now, let us write the solution for 1-instanton explicitly. At one instanton level, the formula \eqref{eq:Ind} can be written as
 \begin{align}
  \begin{split}   \label{eq:1inst-i}
  I_1^{(d)} = &\sum_{\vec{k} \in \Delta_\ell}  \Bigg( 
     (p_1p_2)^{(\frac{b}{2}-d ) } e^{(d+\frac{\k_\text{eff}}{2})(\vec{a}\cdot\vec{k})}\,
    e^{-\frac{\k_\text{eff}}{2}d_{ijk}\,a^ik^jk^k}  \\& 
  \times \frac{\prod_{l} y_{\text{tw},l}^{-I_2(\mathbf{R}_l)/2} \prod_{\w \in \mathbf{R}_l} \CL_{\vec{k}\cdot\vec{\w}}(\vec{a}\cdot \vec{\w} + m_{\text{tw},l}, \e_1,\e_2)}{ (1-p_1p_2 e^{-\vec{a} \cdot \vec{k} })(1-p_1^{-1} e^{\vec{a} \cdot \vec{k} })(1-p_2^{-1} e^{\vec{a} \cdot \vec{k} })(1-e^{\vec{a} \cdot \vec{k} })\prod_{\vec{\a} \cdot \vec{k} = -1}(1-e^{-\vec{a}\cdot\vec{\alpha}})}
    \Bigg) , 
  \end{split}
\end{align}
where $\Delta_\ell$ is the set of long roots ($\vec{k}\cdot \vec{k} = 2$) and we used $Z_0 = 1$. 
It turns out to be more convenient to express $Z_1$ by decomposing $I_1^{(d)}$ into the flux sum, \emph{i.e.}, $I_1^{(d)} \equiv \sum_{\vec{k}\in\Delta_{\ell}} i_1^{(d)}(\vec{k})$, where
\begin{align}
  \label{eq:1inst-i-flux}
  \begin{split}
  i_{1}^{(d)}(\vec{k})\equiv &\ (p_1p_2)^{(\frac{b}{2}-d)}\,e^{(d+\frac{\kappa_{\text{eff}}}{2})(\vec{a}\cdot\vec{k})}e^{-\frac{\kappa_{\text{eff}}}{2}d_{ijk}a^ik^jk^k}\\
  &\,\times\frac{\prod_{l}(y_{l}^{\text{tw}})^{-{I_2(\mathbf{R}_l)}/{2}}\prod_{\w \in \mathbf{R}_l} \CL_{\vec{k}\cdot\vec{\w}}(\vec{a}\cdot \vec{\w} + m_{\text{tw},l}, \e_1,\e_2)}{(1-p_1p_2e^{-\vec{a}\cdot\vec{k}})(1-e^{\vec{a}\cdot\vec{k}})(1-p_1^{-1}e^{\vec{a}\cdot\vec{k}})(1-p_2^{-1}e^{\vec{a}\cdot\vec{k}})\prod_{\vec{\alpha}\cdot \vec{k}=-1}(1-e^{-\vec{a}\cdot\vec{\alpha}})}.
  \end{split}
\end{align}
Using the property $i_{1}^{(d_0+\aleph)}(\vec{k}) /i_{1}^{(d_0)}(\vec{k})=(p_1p_2)^{-\aleph}\,e^{\aleph(\vec{a}\cdot\vec{k})}$, the one-instanton partition function $Z_1$ can be written as
\begin{align}
  \label{eq:1inst-formula}
  Z_1=&\ \sum_{\vec{k}\in\Delta_\ell}\frac{(1-p_1^{-1}e^{\vec{a}\cdot\vec{k}})(1-p_2^{-1}e^{\vec{a}\cdot\vec{k}})}{(1-p_1)(1-p_2)}
  \cdot i_1^{(d_0)}(\vec{k}) 
   \\=&\ \frac{(p_1p_2)^{(\frac{b}{2}-d_0)}\prod_{l}(y_{l}^{\text{tw}})^{-{\frac{I_2(\mathbf{R}_l)}{2}}}}{(1-p_1)(1-p_2)} \sum_{\vec{k}\in\Delta_l}\frac{e^{(d_0+\frac{\kappa_{\text{eff}}}{2})(\vec{a}\cdot\vec{k})-\frac{\kappa_{\text{eff}}}{2}d_{ijk}a^ik^jk^k}\prod_{\w \in \mathbf{R}_l} \CL_{\vec{k}\cdot\vec{\w}}(\vec{a}\cdot \vec{\w} + m_{l}^\text{tw})}{(1-p_1p_2e^{-\vec{a}\cdot\vec{k}})(1-e^{\vec{a}\cdot\vec{k}})\prod_{\vec{\alpha}\cdot \vec{k}=-1}(1-e^{-\vec{a}\cdot\vec{\alpha}})} . \nn
\end{align}
Notice that there are multiple options for choosing $d_0$. However, we find that \eqref{eq:1inst-formula} is independent of a specific choice of $d_0$. Once we choose $d_0 = 0$, for instance, which works in most cases,\footnote{A numerical value of $d_0$ should be a half-integer for theories with $G = Sp(N)_{\theta=\pi}$.} \eqref{eq:1inst-formula} becomes
\begin{align}
\boxed{
  Z_1=\ \frac{(p_1p_2)^{\frac{b}{2}}\prod_{l}(y_{l}^{\text{tw}})^{-\frac{I_2(\mathbf{R}_l)}{2}}}{(1-p_1)(1-p_2)} \sum_{\vec{k}\in\Delta_\ell}\frac{e^{\frac{\kappa_{\text{eff}}}{2}(\vec{a}\cdot\vec{k}-d_{ijk}a^ik^jk^k)}\prod_{\w \in \mathbf{R}_l} \CL_{\vec{k}\cdot\vec{\w}}(\vec{a}\cdot \vec{\w} + m_{l}^\text{tw})}{(1-p_1p_2e^{-\vec{a}\cdot\vec{k}})(1-e^{\vec{a}\cdot\vec{k}})\prod_{\vec{\alpha}\cdot \vec{k}=-1}(1-e^{-\vec{a}\cdot\vec{\alpha}})}}\ .
\end{align} 
When the hypermultiplets are in the representations with $|\vec{k} \cdot \vec{w}|\le 1$ for all $\vec{w} \in \mathbf{R}$, we have
\begin{align}
  \label{eq:weight-expansion-non-sym}
  \prod_{\w \in \mathbf{R}} \CL_{\vec{k}\cdot\vec{\w}}(\vec{a}\cdot \vec{\w} + m_{\text{tw}}, \e_1,\e_2)  &=  \prod_{\vec{k}\cdot \vec{\w} = -1}(1-y_\text{tw} e^{-\vec{a}\cdot \vec{\w}}) .
\end{align}
The formula \eqref{eq:1inst-formula} indeed reduces to the pure YM partition function derived in \cite{Keller:2011ek, Keller:2012da} upon removing hypermultiplets and Chern-Simons levels up to the overall factor $(p_1 p_2)^{\frac{b}{2}} = e^{-\frac{h^\vee}{2} (\e_1 + \e_2)}$ that accounts for the shift of instanton fugacity. 

We claim that \eqref{eq:1inst-formula} is the closed-form expression for the one-instanton partition function, which holds \emph{universally for any gauge theory} with $d_{\text{max}}>2$.  
In section \ref{subsec:numd}, we study the structure of the blowup equations to bound the number of possible independent equations.

%%%%%%%%%%%%%%%%%%%%%%%%%%%%%%%%%%%%%%%%%
\subsection{Number of independent blowup equations} \label{subsec:numd}

We are mainly interested in 4d $\CN=2$ and 5d $\mathcal{N}=1$ gauge theories which are UV-complete. The UV-complete set of 4d $\CN=2$ gauge theories are classified in \cite{Bhardwaj:2013qia}. For 5d gauge theories that are UV complete as 5d SCFTs,  possible matter representations are restricted to \cite{Jefferson:2017ahm}:\footnote{A gauge group is always assumed to be simple in the current paper.}
\begin{itemize}
  \item fundamental representation for $SU(N)$, $SO(N)$, $Sp(N)$, $G_2$, $F_4$, $E_6$, $E_7$
  \item antisymmetric representation for $SU(N)$, $Sp(N)$
  % \item symmetric representation for $SU(N)$
  \item spinor representation for $SO(N)$ with $7 \leq N \leq 14$
  \item rank-3 antisymmetric representation for $Sp(3)$, $Sp(4)$, $SU(6)$, $SU(7)$
  \item symmetric representation for $SU(N)$.
\end{itemize}
In the case of 4d, we can also have the following additional cases: 
\begin{itemize}
	\item adjoint representation for arbitrary group
	\item rank-3 antisymmetric for $SU(8)$
	\item \textbf{16} for $Sp(2)$ (half-hypermultiplet)
\end{itemize}
We note that though our blow-up formula is applicable to a large number of 5d theories with various matter representations, we are not able to apply our formula for some cases including the one with adjoint hypermultiplet since the number of independent blowup equations is smaller than 3.

The formula \eqref{eq:recursion} is valid only for a certain range of $d$, for which $\langle (\mathcal{O}_{\IP^1})^{d}\rangle = \CZ$. We want to narrow down the valid range of $d$ by performing a simple sanity check on the blowup equation for the one-instanton partition function:
\begin{align}
  \label{eq:1inst}
  Z_1 = p_1^{\frac{b}{2}-d}Z^{(N)}_1  + p_2^{\frac{b}{2}-d}  Z^{(S)}_1  + I_1^{(d)} \qquad \text{with an allowed range of $d$} \ .
\end{align}
Specifically, we want to examine the expansion of each term in \eqref{eq:1inst} in powers of $p_1p_2 \ll 1$. The  leading exponent of each term behaves as {\allowdisplaybreaks
\begin{align}
\begin{split}
  I_1^{(d)} &\sim \begin{dcases}
    g_0(\vec{a},  \vec{m}_{\text{tw}})\cdot  (p_1p_2)^{\frac{b}{2}-d + 1} + \cdots   &  \text{for }N_\text{sym} = 0\\
    g_0(\vec{a},  \vec{m}_{\text{tw}})\cdot (p_1p_2)^{\frac{b}{2}-d } + \cdots    &  \text{for }N_\text{sym} = 1
\end{dcases}
   \\
  Z_1 &\sim   \ \ \, g_1(\vec{a},  \vec{m}_{\text{tw}})\cdot (p_1p_2)^{\frac{s}{2}}  + \cdots   \\
  p_1^{\frac{b}{2}-d}Z_1^{(N)}\sim p_2^{\frac{b}{2}-d}  Z^{(S)}_1 &\sim  \ \ \,  g_2(\vec{a},  \vec{m}_{\text{tw}})\cdot  (p_1p_2)^{\frac{b}{4}-\frac{d}{2}+\frac{s}{4}}   + \cdots   ,
  \end{split}
\end{align}
where $g_{0, 1, 2}(\vec{a}, \vec{m}_{\text{tw}})$ are functions independent of $p_{1, 2}$ and $N_{\text{sym}}$ denotes the number of symmetric representation. The numerical value of $s$ will be obtained shortly for a variety of gauge theories for which ADHM-like construction is available.
Notice that for the equation \eqref{eq:1inst} to be true, some terms on the right-hand side should have the leading exponent less than or equal to that of $Z_1$. Therefore, the condition $d - \frac{b}{2} \geq -\frac{s}{2}$ is naturally imposed, setting a lower bound on $d$. 

Similarly, an upper bound on $d$ can be found from an expansion of \eqref{eq:1inst} with respect to $ 1/p_1p_2 \ll 1$.\footnote{This is equivalent to assuming a different parameter regime $ 0 < a_i < -\e_+ < m$ for all $1\leq i \leq r$. In general, an explicit form of the 1-loop partition function \eqref{eq:1-loop-vec}--\eqref{eq:1-loop-hyp} can change depending on a parameter regime, thus affecting \eqref{eq:recursion}. However, all the above expressions remain valid under flipping a sign of $\epsilon_+$, such that we can simply study the expansion of the single terms in \eqref{eq:1inst} with respect to $ 1/p_1p_2 \ll 1$.} 
Each term in \eqref{eq:1inst} can be written as
\begin{align}
\begin{split}
  I_1^{(d)} &\sim \begin{dcases}
    h_0(\vec{a},  \vec{m}_{\text{tw}})\cdot  (1/p_1p_2)^{d-\frac{b}{2} + 1} + \cdots   &  \text{for }N_\text{sym} = 0\\
    h_0(\vec{a},  \vec{m}_{\text{tw}})\cdot (1/p_1p_2)^{d-\frac{b}{2} } + \cdots    &  \text{for }N_\text{sym} = 1
\end{dcases}
   \\
  Z_1 &\sim  \ \ \,  h_1(\vec{a},  \vec{m}_{\text{tw}})\cdot (1/p_1p_2)^{\frac{s'}{2}}  + \cdots   \\
  p_1^{\frac{b}{2}-d}Z_1^{(N)} \sim  p_2^{\frac{b}{2}-d}  Z^{(S)}_1&\sim \ \ \,  h_2(\vec{a},  \vec{m}_{\text{tw}})\cdot  (1/p_1p_2)^{\frac{d}{2}-\frac{b}{4}+\frac{s'}{4}}   + \cdots   .
\end{split}
\end{align}
Again, for \eqref{eq:1inst} to be consistent, the leading exponent of $Z_1$ should be greater than or equal to those of the terms on the right-hand side. Such a requirement imposes an upper bound on $d$, namely  $\frac{s'}{2} \geq  d - \frac{b}{2}$.
Combining the two inequalities, one can identify the following range
\begin{align}
  \label{eq:bound}
  -\frac{s}{2} + \frac{b}{2} \leq d \leq \frac{s'}{2} + \frac{b}{2},
\end{align}
as a necessary condition for \eqref{eq:1inst}. 
We explicitly checked that the $n$-instanton partition function $Z_n$ actually satisfies all the $(\frac{s+s'}{2})$ recursion relations up to a certain value of $n > 1$ for numerous examples whose $Z_n$ is already known from alternative methods. 
This is true even though the bound \eqref{eq:bound} itself is merely a \emph{necessary} condition found from one-instanton analysis. 
Based on this empirical observation, we claim that the 5d recursion formulae \eqref{eq:recursion} within the above range of $d$ is true at all instanton orders.

Another remarkable thing is that a numerical value of  $(s, s')$ exhibits the very simple pattern across a broad range of theories whose gauge group is not $SU(N)_\k$. %(See Table~\ref{tbl:list}.)
\begin{align}
  \label{eq:n-np}
 s = s' &= h^\vee - \frac{1}{2} \sum_{l}I_2(\mathbf{R}_l)  && \text{ for } \quad   G\neq SU(N)_\kappa \text{ nor } Sp(N) \nn \\
 s = s' - 2\left\{\frac{N_f}{2} \right\} &= h^\vee - \frac{1}{2} \sum_{l}I_2(\mathbf{R}_l)
 &&   \text{ for }  \quad G = Sp(N)_{\theta=0}\\
 s = s' + 2 \left\{\frac{N_f}{2} \right\} &= h^\vee - \frac{1}{2} \sum_{l}I_2(\mathbf{R}_l) + 1 &&   \text{ for }  \quad G = Sp(N)_{\theta=\pi} \nn
\end{align}
where $\{ x\} \equiv x - \lfloor x \rfloor $ denote the non-integer part of $x$.
As the above numerical pattern \eqref{eq:n-np} emerges for all $G \neq SU(N)_\k$ examples that we studied, we conjecture that \eqref{eq:n-np} is generally true, thereby taking the recursion formulae \eqref{eq:recursion} with
\begin{alignat}{2}
 \label{eq:d-range-nonsu}
  0 & \leq d \leq h^\vee - \frac{1}{2} \sum_{l}I_2(\mathbf{R}_l)  & \text{ for } \quad G \neq SU(N)_\kappa \text{ nor } Sp(N),  \nn \\
  0 &\leq  d \leq h^\vee - \frac{1}{2} \sum_{l}I_2(\mathbf{R}_l) + \left\{\frac{N_f}{2} \right\} & \qquad \text{ for }\quad G =Sp(N)_{\theta=0}, \\
  -\frac{1}{2} &\leq  d \leq h^\vee - \frac{1}{2} \sum_{l}I_2(\mathbf{R}_l) + \frac{1}{2} - \left\{\frac{N_f}{2} \right\} & \qquad \text{ for }\quad G =Sp(N)_{\theta=\pi}, \nn
\end{alignat}
as a basic assumption to obtain the partition function $\mathcal{Z}$ for any $G \neq SU(N)_\k$ gauge theory. 
It would be desirable to understand from the first principle the range  \eqref{eq:d-range-nonsu} of $d$ for which \eqref{eq:recursion} holds true.

It turns out to be more difficult to characterize a general pattern behind $(s, s')$ for $SU(N)_\kappa$ gauge theories, due to extra complication caused by the 5d Chern-Simons level $\kappa$. Here we consider two particular classes of $SU(N)_\kappa$ gauge theories for illustration.  For $SU(N)_\kappa + N_f\mathbf{F}$ gauge theory ($N_f$ fundamental hypermultiplets) with $N_f + 2|\kappa| \leq 2N$, we find that 
\begin{align}
  \label{eq:sun-nf}
  s &= \begin{dcases}
   \frac{N_f}{2}
  &  \quad \text{ if }  \kappa_\text{eff} = N - N_f, \\
  N - \frac{1}{2} \sum_{l}I_2(\mathbf{F}) + |\k_\text{eff}| 
  &  \quad  \textstyle\text{ otherwise},
  \end{dcases}\\
  s' &=
    \begin{dcases}
      \frac{N_f}{2}
     & \quad \textstyle\text{ if }  \kappa_\text{eff} = -N, \\
     N - \frac{1}{2} \sum_{l}I_2(\mathbf{F}) + \left|\bar{\k}_\text{eff}\right| 
     & \quad \textstyle\text{ otherwise} ,
    \end{dcases}  
\end{align}
where $\bar{\k}_\text{eff} \equiv \k + \half \sum_l I_3 (\mathbf{F}) $. 
Plugging in these values to \eqref{eq:bound}, we find the range of $d$ to be
\begin{align}
  \label{eq:sun-nf-range}
  0 \leq &\ d \leq N &\text{ if } \kappa &= -N + \frac{N_f}{2}, \nn \\
  0 \leq &\ d \leq N-\frac{N_f}{2}-\kappa &\text{ if } \kappa & \in \left(-N +\frac{N_f}{2}, -\frac{N_f}{2} \right], \nn \\
  0 \leq &\ d \leq N &\text{ if } \kappa &\in \left[-\frac{N_f}{2}, +\frac{N_f}{2}\right], \\
  \frac{N_f}{2}-\kappa \leq &\ d \leq N & \text{ if } \kappa &\in \left[\frac{N_f}{2}, N -\frac{N_f}{2}\right),  \nn \\
  0 \leq &\ d \leq N & \text{ if } \kappa &= N -\frac{N_f}{2},  \nn 
\end{align}
which always includes the range $0 \leq d \leq N$. Thus the recursion formula \eqref{eq:recursion} holds for at least $3$ values of $d$, which is enough to determine the partition function $Z_{\text{inst}}$ completely.

For the $SU(N)_\kappa + N_f\mathbf{F} + 1\mathbf{AS}$ theory ($N_f$ fundamentals and 1 anti-symmetric tensor) with $N_f + 2|\kappa| \leq N+4$, we find
\begin{align}
  \label{eq:sun-nf-na}
  \begin{split}
  s &= \min  
  \left(N - \frac{1}{2} \sum_{l}I_2(\mathbf{R}_l) - (\k_\text{eff}-2),\ N - \frac{1}{2} \sum_{l}I_2(\mathbf{R}_l) + 2 \left\{\frac{\k_\text{eff}}{2}\right\}\right) \\
  s' &= \min  
  \left(N - \frac{1}{2} \sum_{l}I_2(\mathbf{R}_l) + \Big(\k_\text{eff}+\sum_{l}I_3(\mathbf{R}_l)+2\Big),\right. \\& \left. \qquad \qquad \  N - \frac{1}{2} \sum_{l}I_2(\mathbf{R}_l) + 2\Big\{\!\!-\frac{\k_\text{eff}}{2}+\frac{1}{2}\sum_{l}I_3(\mathbf{R}_l) \Big\}\right)
  \end{split} 
\end{align} for most cases except for
\begin{align}
s=&\,\frac{N}{2}+2 \qquad N\in2\mathbb{Z},\,N_f=0,\,\kappa=\frac{N}{2}+1,\nn\\
s'=&\,\frac{N}{2}+2 \qquad N\in2\mathbb{Z},\,N_f=0,\,\kappa=-\frac{N}{2}-1,
\end{align}
from which one can identify the valid range of $d$ via \eqref{eq:bound}. 
As long as there exist at least three distinct allowed values for $d$ for given $(N, \kappa)$, the corresponding partition function $Z_{\text{inst}}$ can be solved from the recursion formula \eqref{eq:recursion}. 

We also consider $SU(6)_\kappa  + 1 \mathbf{TAS}$ theory (one rank-3 antisymmetric tensor) with $|\kappa| \leq 3$ in Section~\ref{sec:example}. This model can be Higgsed to two disjoint copies of $SU(3)_\kappa$ theory without a bifundamental hypermultiplet \cite{Hayashi:2019yxj}. At the level of the partition function, Higgsing is realized by turning off $m_\text{tw} = 0$ and imposing the $SU(3)$ traceless conditions. As neither of them modifies $s$ nor $s'$, the numerical value of $(s,s')$ must be identical to that of $SU(3)_\kappa$ gauge theory, the blowup equation always holds for the range $0\leq d \leq 3$. Therefore, the recursion formula \eqref{eq:recursion} is enough to determine the instanton partition function $Z_{\text{inst}}$ for $SU(6)_\kappa  + 1 \mathbf{TAS}$ theory as well.

\begin{table}[t!]
  \centering
  \begin{tabular}{|c|c|c|c|c|}
    \hline
    $G$ & Hypermultiplets & Conditions for $d_{\text{max}}\ge 2$ &   $(s, s')$ & $d$ \\
    \hline \hline
    $SU(N)_\kappa$ & $N_f \mathbf{F}$ & Always  &    \eqref{eq:sun-nf} & \eqref{eq:sun-nf-range} \\
    \cline{1-5}
    \multirow{2}{*}{$SU(N)_\kappa$} & \multirow{2}{*}{$N_f \mathbf{F}+1\textbf{AS}$} & $N_f \le N-1$ &    \multirow{2}{*}{\eqref{eq:sun-nf-na}} & \multirow{2}{*}{\eqref{eq:bound}} \\\cline{3-3}
	& & $N_f=N,\; \kappa\equiv N+1\,(\text{mod } 2)$ & & \\
    \cline{1-5}
    $Sp(N)_{\theta=0}$ & $N_f\mathbf{F}+N_a \mathbf{AS}$ & $N_a(N-1)+\lfloor N_f/2\rfloor\leq  N-1$ &   \multirow{7}{*}{\eqref{eq:n-np}}& \multirow{7}{*}{\eqref{eq:d-range-nonsu}} \\
    \cline{1-3}
    $Sp(N)_{\theta=\pi}$ & $N_f\mathbf{F}+N_a\mathbf{AS}$ & $N_a(N-1)+\lceil N_f/2\rceil\le  N$ & & \\
    \cline{1-3}
    $SO(2N)$ & $N_v\mathbf{V}+N_s\mathbf{S}+N_c\mathbf{C}$ & $N_v+2^{N-4}(N_s+N_c) \le  2N-4$ &  & \\
    \cline{1-3}
    $SO(2N+1)$ & $N_v\mathbf{V}+N_s\mathbf{S}$ & $N_v+2^{N-3}N_s \le 2 N-3$ & &   \\
    \cline{1-3}
    $E_6$ & $N_f\mathbf{F}+N_{\bar{f}}\mathbf{\bar{F}}$ & $N_f+N_{\bar{f}} \le 3 $ & &   \\
    \cline{1-3}
    $E_7$ & $N_f\mathbf{F}$ & $N_f \le 2$ & &   \\
    \cline{1-3}
    $E_8$ & $\varnothing$ &  & &   \\
    \hline
    \end{tabular}
    \caption{List of 5d gauge theories whose partition function is determined via the blowup equations. The number of hypermultiplets are bounded so that there are at least 3 blowup equations. 
    For the case of $SU(N) + N_f \mathbf{F}$ theory, it turns out that the Young diagram formula \eqref{eq:SUn-young} always satisfy at least 3 blowup equations. When $N_f + 2|\kappa| > 2N$, however, this formula does not produce the correct partition function for the UV field theory as we discuss in the text.}
    \label{tbl:list}
\end{table}
We give the list of theories we consider in the current paper in Table \ref{tbl:list}.

%%%%%%%%%%%%%%%%%%%%%%%%%%%%%%%%%%%%%%%%%%%%%%%%%%%%%%%%%%%%%%%%

\section{Examples} \label{sec:example}

The recursion formula \eqref{eq:recursion} for the $n$-instanton partition function and also the general expression \eqref{eq:1inst-formula} at one-instanton order are widely applicable to 5d $\mathcal{N}=1$ (and also similarly to 4d $\CN=2$) gauge theory whose $(s,s')$ satisfies $\frac{s+s'}{2}\geq 2$. Combined with the observation that $(s,s')$ follows \eqref{eq:n-np} in most cases, they become a very efficient approach to obtaining the BPS partition function on $\IC^2 \times S^1$ (or $\IC^2$), unless the matter representation is `too large.' 

Conventionally, the instanton partition function can be computed by employing the ADHM construction of the instanton moduli space \cite{Atiyah:1978ri,Nekrasov:2002qd,Nekrasov:2003rj} or by applying the topological vertex formalism to the 5-brane web \cite{Aganagic:2003db,Iqbal:2007ii}. Both are based on a certain UV realization of 5d $\mathcal{N}=1$ gauge theory via geometric engineering in string theory. 
Even though IR 5d gauge theory sometimes can be obtained using more than one string theory realizations, the correct UV completion might be only achieved through specific string theory realizations. For instance, the $SU(2)$ gauge theory with $N_f$ fundamental hypermultiplets with $N_f\geq 5$ must be embedded into D4-D8-O8 brane system to be UV-completed as 5d $E_{N_f+1}$ Minahan-Nemeschansky SCFT \cite{Seiberg:1996bd, Minahan:1996fg, Minahan:1996cj}. Ordinary $(p,q)$ 5-brane web with colliding branes (without O-planes) indicate UV inconsistency \cite{Aharony:1997ju}. 
A sensible QFT observable can thus be obtained only through a proper embedding of the gauge theory into string theory. In some occasions, an extra factor dressing the true QFT observable may appear during the above instanton computation, which is sensitive to the choice of a string theory embedding.
Our blow-up formula \eqref{eq:recursion} does not explicitly specify a particular UV completion nor string theory embedding. However, we observe that the formula does prefer a particular string theory embedding of the gauge theory. For example, for the $SU(2)$ gauge theory with $N_f$ fundamental hypermultiplets, we find the partition function obtained from the blow-up formula agrees with the partition function obtained from the ordinary $(p,q)$ 5-brane webs. 

There are wide varieties of `exceptional' gauge theories (having exceptional gauge groups or exotic matter representations) whose UV completion is found as M-theory wrapped on a singular Calabi-Yau 3-fold \cite{Diaconescu:1998cn, Jefferson:2018irk, Bhardwaj:2018yhy, Apruzzi:2019opn}. 
As most exceptional theories lack the ADHM description \cite{Kim:2018gjo}, their instanton partition function $Z_\text{inst}$ has been studied in a case-by-case basis. Once the 5-brane web configuration engineering an exceptional theory is identified \cite{Zafrir:2015ftn, Hayashi:2018bkd, Hayashi:2019yxj}, the topological vertex formalism can be applied to compute the relevant partition function $\CZ$ \cite{Hollowood:2003cv, Iqbal:2007ii}. 
Alternatively, one can first construct the $\IC^2 \times T^2$ partition function for a related 6d gauge theory, based on its modularity and anomaly, then take the circle reduction to obtain the 5d partition function $\CZ$ \cite{DelZotto:2016pvm, DelZotto:2018tcj}. Several interesting exceptional theories have been studied so far, based on the above two approaches.
Sometimes, there exists auxiliary 4d $\CN=2$ SCFT \cite{Benini:2009gi} that realizes exceptional instanton moduli space as its Higgs branch.\footnote{Also 2d $\CN=(0, 4)$ version \cite{Putrov:2015jpa} for any 4d $\CN=2$ theory can be obtained upon twisted dimensional reduction, which allows us to compute the 6d instanton string partition function.} In this case, computing the superconformal index in the Higgs branch limit provides a way to compute the necessary instanton partition function for the exceptional gauge theory \cite{Gadde:2010te, Gadde:2011uv, Gaiotto:2012uq, Gadde:2015xta, Agarwal:2018ejn}. 
Likewise, 3d $\CN=4$ theory can realize exceptional instanton moduli space via its Coulomb branch \cite{Intriligator:1996ex}. Computing its Hilbert series (or the Coulomb branch limit of the superconformal index), one can compute the instanton partition function \cite{Cremonesi:2013lqa, Cremonesi:2014xha}. 
We will illustrate that bootstrapping the instanton partition function $Z_{\text{inst}}$ based on the recursion formula \eqref{eq:recursion} works well for those `exceptional' theories, providing their BPS spectrum efficiently.

\subsection{Theories with known ADHM description}
\label{subsec:ex-adhm}

Let us first consider the `standard' gauge theories with classical gauge groups, whose hypermultiplet admits UV realization as a perturbative string ending on D-branes. In these cases, the ADHM construction of the instanton moduli space is well-known \cite{Atiyah:1978ri,Nekrasov:2002qd,Shadchin:2005mx}. 
As for the $k$-instanton partition function $Z_k$, the Witten index of the relevant ADHM quantum mechanics can be computed by SUSY localization \cite{Kim:2011mv, Hwang:2014uwa,Hwang:2016gfw, Lee:2017lfw}, ending up collecting all Jeffrey-Kirwan residues of a multi-dimensional contour integral. We will examine whether the recursion formula \eqref{eq:recursion} actually produces  the same result as the localization computation.

\paragraph{SU(N)}

The ADHM construction for the $n$-instanton partition function, for $SU(N)_\kappa +N_f \mathbf{F}$ ($N_f$ fundamentals) theory with $N_f + 2|\kappa| \leq 2N$ is well-known. Its partition function can be written as a sum over Young diagrams as
\begin{align}
  \label{eq:SUn-young}
  Z_n^\text{ADHM} = \sum_{|\vec{Y}| = n}\prod_{i=1}^N\prod_{\s \in Y_i}\frac{e^{-\kappa \phi(s)} \prod_{l=1}^{N_f} 2\sinh{\frac{\phi(\s)+m_l}{2}}  }{\prod_{j=1}^N 2\sinh{\frac{E_{ij}}{2}}\,2\sinh{\frac{E_{ij}-2\e_+}{2}}} \ , 
\end{align}
where
\begin{align*}
  % \begin{split}
  E_{ij}(\s) &= a_i - a_j - \e_1 h_i(\s)+ \e_2(v_j(\s)+1)\\
  \varphi(\s)&= a_i  -\e_+-(n-1)\e_1-(m-1)\e_2
  % \end{split}
   \quad \qquad\text{ for }\quad \s = (m,n) \in Y_i \ . 
\end{align*}
Here $h_i(\s)$ denotes the distance from $\s$ to the right end of the diagram $Y_i$ by moving right and $v_j(\s)$ denotes the distance from $\s$ to the bottom of the diagram $Y_j$ by moving down.
We checked that the instanton partition functions $Z_1$ and $Z_2$ obtained from the recursion formula \eqref{eq:recursion} with \eqref{eq:sun-nf-range} and the 1-instanton expression \eqref{eq:1inst-formula} precisely agree with the above  $Z_{n=1,2}^\text{ADHM}$ for $N=2,3,4$. 

As we have said earlier, $Z_n^\text{ADHM}$ often contains an additional factor $Z_\text{extra}$ that captures the contribution from an extra branch of vacua of the ADHM quantum mechanics. It is sensitive to the string theory embedding (UV completion) of the gauge theory and can be regarded as spurious from the 5d QFT perspective. It is usually factorized from the true QFT partition function as 
\begin{align}
  \sum_{n=0}^\infty q^n \,Z_n^\text{ADHM}(\vec{a}, \e_{1},\e_2, \vec{m}) = Z_\text{QFT}(\vec{a}, \e_{1},\e_2, \vec{m},q) \cdot Z_\text{extra}(\e_{1},\e_2,  \vec{m},q).
\end{align}
A non-trivial $Z_\text{extra}\neq 1$ appears in the above expression \eqref{eq:SUn-young} if and only if $N_f + 2|\kappa| = 2N$. This factor can be identified as the contribution of D1-branes escaping from D5-branes which engineer the $SU(N)_\kappa + N_f\mathbf{F}$ gauge theory. Since $Z_n = Z_n^\text{ADHM}$, the same factor $Z_\text{extra}$ emerges from the recursion formula \eqref{eq:recursion} as well. The 5-brane web construction of the gauge theory is thus indirectly reflected in the recursion formula. 

A similar observation is that the 1-instanton expression \eqref{eq:1inst-formula} applied to $SU(2)_\kappa+N_f\mathbf{F}$ with $N_f \geq 5$ does not match the Witten index of the D0-D4-D8-O8${}^-$  quantum mechanics, which is the correct 1-instanton partition function.\footnote{The case with $SU(2) \simeq Sp(1
)$ is an exception, which allows $N_f \leq 7$ fundamental hypermultiplets \cite{Seiberg:1996bd}.} Instead, it coincides with the topological vertex computation applied to the 5-brane web with a colliding pair of branes, which engineers the $SU(2)$ gauge theory with $N_f \geq 5$ in the IR, but behaves badly in the UV. Again, this suggests that the recursion formula \eqref{eq:recursion} implicitly chooses a specific string theory construction of the gauge theory, i.e., the web of $(p,q)$ 5-branes. It would be interesting to figure out if there is a version of the recursion relation \eqref{eq:recursion} that allows us to choose the particular UV embedding of the gauge theory.

For the $SU(N)_\kappa + N_f\mathbf{F} + 1\mathbf{AS}$ theory ($N_f$ fundamental and 1 anti-symmetric hypermultiplets) with $N_f + 2|\kappa| \leq N+4$, the ADHM quantum mechanics is the worldvolume theory of D1-branes, probing the D5-NS5-D7-O7${}^{-}$ brane configuration that realizes the gauge theory. Let us compute the Witten index for 1 and 2 D1-branes, then compare with the blow-up computation based on the recursion formula \eqref{eq:recursion}. For instance, the Witten index for the single D1-brane can be written as
\begin{align}
\begin{split}
  &Z_1^\text{ADHM} = - \sum_{i=1}^N \frac{e^{-\k (a_i - \e_+)}}{2\sinh{\frac{\e_1}{2}}\,2\sinh{\frac{\e_2}{2}}} \frac{\prod_{l=1}^{N_f}2\sinh{\frac{-\e_+ + a_i + m_l}{2}}}{2\sinh{\frac{-3\e_+ + 2a_i + m_a}{2}}} \prod_{j \neq i}\frac{2\sinh{\frac{a_i + a_j + m_a - \e_+}{2}}}{2\sinh{ \frac{a_{i}-a_j}{2}}\, 2\sinh{\frac{2\e_+ - a_i + a_j}{2}}}  \\
  &- \frac{1}{2}\ \frac{e^{-\frac{\k}{2}(\e_+ - m_a)}}{2\sinh{\frac{\e_1}{2}}\,2\sinh{\frac{\e_2}{2}}} \left(\frac{\prod_{l=1}^{N_f} 2\sinh{\frac{\e_+ + 2m_l - m_a}{4}}}{\prod_{i=1}^N 2\sinh{\frac{3\e_+ - M - 2a_i}{4}}} - (-1)^{\k + \frac{N-N_f}{2}}\frac{\prod_{l=1}^{N_f} 2\cosh{\frac{\e_+ + 2m_l - m_a}{4}}}{\prod_{i=1}^N 2\cosh{\frac{3\e_+ - M - 2a_i}{4}}}\right).
\end{split}
\end{align}
Note that $Z_n^\text{ADHM}$ contains an extra factor $Z_\text{extra} \neq 1$ if $N_f + 2|\k| = N+4$, coming from the spectrum of D1-branes escaping from the D5-branes on which the gauge theory is supported. 
The appearance of $Z_\text{extra}\neq 1$ is an artifact of the string theory embedding, spurious from the 5d QFT perspective.
We checked that $Z_1^\text{ADHM}$ and the 1-instanton formula \eqref{eq:1inst-formula} agree for the $SU(3)$, $SU(4)$, $SU(5)$ theories whose $(n,n')$ satisfies $\frac{n+n'}{2} \geq 2$. 
We confirmed $Z_2 = Z_2^\text{ADHM}$ as well, where $Z_2$ is the solution of the recursion formulae \eqref{eq:recursion} with \eqref{eq:sun-nf-na}. The same spurious factor $Z_\text{extra}$ arises from the recursion formula, implying that our blowup equations are implicitly based on the D5-NS5-D7-O7${}^{-}$ brane realization of the gauge theory.\footnote{An exceptional case is the $SU(2)$ gauge theory, in which the antisymmetric hypermultiplet decouples and never affects the recursion formula. The corresponding $Z_n$ is the same as the Young diagram formula \eqref{eq:SUn-young}.}

\paragraph{Sp(N)}
The $n$-instanton partition function for $Sp(N)_\theta + N_f \mathbf{F}$ theory ($\theta$ being the discrete theta-angle for $Sp$ and $N_f$ fundamental hypermultiplets) with $N_f \leq 2N+4$ can be computed from the ADHM quantum mechanics of D1-D5-NS5-O5 branes, which engineers the gauge theory and its instantons. The Witten index for the D1-brane theory is written as
\begin{align}
  Z_1^\text{ADHM} &= \frac{1}{2}\,\frac{1}{2\sinh\frac{\e_1}{2}\,2\sinh\frac{\e_2}{2}}\Bigg(\frac{\prod_{l=1}^{N_f}\,2\sinh\frac{m_l}{2}}{\prod_{i=1}^{N}2\sinh\frac{\e_+\pm a_i}{2}}+e^{i\theta}\frac{\prod_{l=1}^{N_f}\,2\cosh\frac{m_l}{2}}{\prod_{i=1}^{N}2\cosh\frac{\e_+\pm a_i}{2}}\Bigg).
\end{align}
We checked that our 1-instanton formula \eqref{eq:1inst-formula} agrees with $Z_1^\text{ADHM}$ for the $Sp(2)$ and $Sp(3)$ gauge theories satisfying $N-1 \geq \lfloor \frac{N_f}{2} \rfloor$ (at $\theta = 0$) and $N \geq \lceil \frac{N_f}{2} \rceil$ (at $\theta = \pi$). 
We also confirmed that $Z_2^\text{ADHM} =  Z_2$, where $Z_2$ is the solution of the recursion formulae \eqref{eq:recursion} with \eqref{eq:d-range-nonsu}. Note that there is no spurious factor $Z_\text{extra}$ so that the ADHM and the blowup results agree $Z_n^\text{ADHM} = Z_n $ for these theories.

For the $Sp(N)_\theta + N_f \mathbf{F} + 1\mathbf{AS}$ theory ($N_f$ fundamental and 1 anti-symmetric hypermultiplets) with $N_f \leq 7$, the relevant ADHM quantum mechanics is the worldvolume gauge theory of D0-branes which probe the D4-D8-O8 brane configuration. It is well-known that the QFT on D4-branes exhibits an enhanced $E_{N_f + 1}$ flavor symmetry at the UV fixed point \cite{Seiberg:1996bd}. Let us consider the Witten index for one and two D0-branes \cite{Kim:2012gu,Hwang:2014uwa}. For a single D0-brane, we obtain the one instanton partition function to be 
\begin{align}
\begin{split}
&  Z_1^{\text{ADHM}}=\frac{1}{2} \, \frac{1}{2\sinh\frac{\e_1}{2}\,2\sinh\frac{\e_2}{2}\, 2\sinh\frac{m_a+ \e_+}{2} \, 2\sinh\frac{m_a-\e_+}{2}} \\
& \qquad \times\left(\,\frac{\prod_{i=1}^{N}2\sinh\frac{m_a\pm a_i}{2}\prod_{l=1}^{N_f}\,2\sinh\frac{m_l}{2}}{\prod_{i=1}^{N}2\sinh\frac{\e_+\pm a_i}{2}}
  +\frac{e^{i\theta}\prod_{i=1}^{N}2\cosh\frac{m_a\pm a_i}{2}\prod_{l=1}^{N_f}\,2\cosh\frac{m_l}{2}}{ \prod_{i=1}^{N}2\cosh\frac{\e_+\pm a_i}{2}}\right).
\end{split}
\end{align}
We find that $Z_1^{\text{ADHM}}$ itself is \emph{not} the same as the 1-instanton expression from the blowup \eqref{eq:1inst-formula} for the $Sp(2)_\theta$, $Sp(3)_\theta$ theories with $N_f \leq 1$ (at $\theta=0$) and $N_f \leq 2$ (at $\theta=\pi$). Instead, the difference between $Z_1$ and $Z_1^{\text{ADHM}}$ can be identified as the BPS index of D0-branes moving away from the D4-D8-O8 brane system \cite{Kim:2012gu,Hwang:2014uwa}. 
Similarly, we confirmed that the 2-instanton correction $Z_2$ captures the same 5d QFT spectrum as in $Z_2^\text{ADHM}$, upon subtracting the spurious contribution of escaping D0-branes. It is interesting that our blow-up formula does \emph{not} contain a spurious factor $Z_\text{extra}$.

\paragraph{SO(N)} One can compute the instanton partition function of $SO(N) + N_v \mathbf{V}$ theory ($N_v$ hypermultiplets in the vector representation) with $N_v \leq N-4$ using the ADHM quantum mechanics of the D1-D5-NS5-O5 brane system. For even $N$, the Witten index for a single D1-brane can be written as
\begin{align}
Z_1^\text{ADHM} = \sum_{i=1}^{N/2}\bigg(\frac{2\sinh(2\e_+-a_i)\,2\sinh(a_i-\e_+)\prod_{l=1}^{N_v}2\sinh\frac{m_l\pm(a_i-\e_+)}{2}}{2 \cdot 2\sinh\frac{\e_1}{2}\,2\sinh\frac{\e_2}{2}\prod_{j\neq i}2\sinh\frac{a_i\pm a_j}{2}\,2\sinh\frac{2\e_+-a_i\pm a_j}{2}}+(a_i\rightarrow -a_i)\bigg).
\end{align}
For odd $N$,
\begin{align}
  Z_1^\text{ADHM} =\sum_{i=1}^{\lfloor N/2\rfloor}\bigg(\frac{2\cosh\frac{2\e_+-a_i}{2}\,2\sinh(a_i-\e_+)\,\prod_{l=1}^{N_f}2\sinh\frac{m_l\pm(a_i-\e_+)}{2}}{2 \cdot 2\sinh\frac{\e_1}{2}\,2\sinh\frac{\e_2}{2}\, 2\sinh\frac{a_i}{2}\prod_{j\neq i}2\sinh\frac{a_i\pm a_j}{2}\,2\sinh\frac{2\e_+-a_i\pm a_j}{2}}+(a_i\rightarrow -a_i)\bigg).
\end{align}
The general 1-instanton expression \eqref{eq:1inst-formula} and the recursion formula \eqref{eq:recursion} are applicable for all  $N_v \leq N-4$. We explicitly verified that $Z_n^\text{ADHM} = Z_n$ for $n=1, 2$ and $4 \leq N \leq 9$, where $Z_1$ is written in \eqref{eq:1inst-formula} and $Z_2$ is the solution of the recursion formula \eqref{eq:recursion}.  We find that $Z_n^\text{ADHM} = Z_n$ involves a non-trivial extra factor $Z_\text{extra} \neq 1$ when $N_v = N-4$. This extra factor can be attributed to the D1-branes moving away from the D5-NS5-O5 brane system, where the 5d QFT lives. It implies that a specific UV realization of the gauge theory, i.e., type IIB string theory with D1-D5-NS5-O5, is implicit in our recursion formulae \eqref{eq:recursion} with \eqref{eq:d-range-nonsu}.

\subsection{Theories with spinor hypermultiplets}
\label{subsec:spinor}

So far, we have investigated the `standard' gauge theories that have certain D-brane set-ups in type IIA/IIB string theory to realize themselves and also their instantons. For the theory with a sufficient number of the blowup equations, the $n$-instanton partition function $Z_n$ can be determined as the solution of the blowup equations. We have found that this formula agrees with the instanton counting result using the ADHM construction, modulo possible extra factor $Z_\text{extra}$ that is sensitive to the string theory embedding of the gauge theory.

We take advantage of the universality of the blowup equation. Recall that the blow-up recursion formula \eqref{eq:recursion} holds for a certain range of $d$, i.e., the set of all integers between $0 \leq d \leq h^\vee - \frac{1}{2}\sum_l I(\mathbf{R}_l)$, when the gauge group $G$ is neither $ SU(N)_\k$ nor $Sp(N)_\theta$. In this case, there is no extra complication due to the Chern-Simons level $\kappa$ or the theta angle $\theta$.
One can solve the recursion formulae for the $n$-instanton correction $Z_n$ to the partition function, as long as $h^\vee - \frac{1}{2}\sum_l I(\mathbf{R}_l) \geq 2$, even for the exceptional gauge theories. We conjecture that $Z_n$ solved from the recursion formula would be the correct BPS data for UV-consistent 5d SCFTs, modulo an extra factor $Z_\text{extra}$ independent of the Coulomb VEV $\vec{a}$. This conjecture will be tested via comparison with \cite{Kim:2018gjo, DelZotto:2018tcj,Hayashi:2019yxj} which compute $\CZ$ for some exceptional cases.

In this section, we will focus on the $SO(N)$ gauge theories with spinor hypermultiplets.
We have a sufficient number of recursion formulae \eqref{eq:recursion} to determine the $n$-instanton partition function $Z_n$ of the $SO(N)$ gauge theory, if and only if  
\begin{align}
  \label{eq:spinor-range}
  \begin{split}
  N-4 &\geq N_{\bf v} + 2^{\frac{N-7}{2}}\cdot N_{\bf s} \qquad\qquad\qquad \text{ for odd $N$},\\
  N-4 &\geq N_{\bf v} + 2^{\frac{N-8}{2}}\cdot (N_{\bf s} + N_{\bf c})\qquad\quad \text{for even $N$},
  \end{split}
\end{align}
where $N_{\bf v}$, $N_{\bf s}$, and $N_{\bf c}$ denote the number of hypermultiplets in the vector, spinor and conjugate spinor representations, respectively. Our 1-instanton expression \eqref{eq:1inst-formula} is also applicable to the cases satisfying \eqref{eq:spinor-range}. 
We compare our formula against any known results for $SO(N)$ gauge theory with a number of spinor hypermultiplets  \cite{Kim:2018gjo, DelZotto:2018tcj}. We not only find perfect agreements for the case with the known results, but also  obtain partition functions for the previously unknown cases as well. 

\paragraph{SO(7)} The $n$-instanton contribution $Z_n$ of $SO(7) + N_{\bf s}\,\mathbf{S}$ theory can be obtained from the SUSY quantum mechanics proposed in \cite{Kim:2018gjo}, which can be summarized as the following $SU(4)$ Young diagram expression:
\begin{align}
\begin{split}
  Z_n^\text{YD} = \sum_{|\vec{Y}| = n} &\prod_{i=1}^4\prod_{s \in Y_i}\frac{2\sinh{(\phi(s))} \ 2\sinh{(\phi(s)-\e_+)} \ \prod_{l=1}^{N_\textbf{s}}2\sinh(\frac{m_l \pm \phi(s)}{2}) }{\prod_{j=1}^4 2\sinh{\frac{E_{ij}}{2}}\,2\sinh{\frac{E_{ij}-2\e_+}{2}}\,2\sinh{\frac{\e_+ - \phi(s) - a_j}{2}}} \\ \times &\prod_{i \leq j}^4 \prod_{\stackrel{s_{i,j} \in Y_{i,j}}{s_i < s_j}} \frac{2\sinh{\frac{\phi(s_i)+\phi(s_j)}{2}}\, 2\sinh{\frac{\phi(s_i)+\phi(s_j)-2\e_+}{2}}}{ 2\sinh{\frac{\e_1 - \phi(s_i)-\phi(s_j)}{2}} \, 2\sinh{\frac{\e_2 - \phi(s_i)-\phi(s_j)}{2}}}.
\end{split}
\end{align}
We verified that $Z_1^\text{YD}$ and the 1-instanton formula $Z_1$ in \eqref{eq:1inst-formula} agree for $N_{\bf s} \leq 3$. We further confirmed at two instanton order for $N_{\bf s} \leq 3$ that $Z_2^\text{YD} = Z_2$, where $Z_2$ is  the solution of the recursion formula \eqref{eq:recursion} with \eqref{eq:d-range-nonsu}. Such explicit comparison implies that the blow-up recursion formula \eqref{eq:recursion} indeed works for the $SO(7) + N_{\bf s}\,\mathbf{S}$ theory. 

The 1-instanton partition function of $SO(7) + 4\mathbf{S} + 1\mathbf{V}$ theory is given in (H.15) of \cite{DelZotto:2018tcj}.
From this expression, we can obtain the 1-instanton correction of $SO(7) + N_{\bf s}\,\mathbf{S} + N_{\bf v}\mathbf{V}$ theory with $( N_{\bf s}, N_{\bf v}) \leq ( 2,1)$ by integrating out hypermultiplets or equivalently taking some flavor chemical potentials to infinity. We confirmed that the result agrees with our general 1-instanton expression \eqref{eq:1inst-formula} up to order $(p_1p_2)^{13/2}$. Notice that our formula holds for any $N_{\bf v} + N_{\bf s} \le 2$ and can be used to compute arbitrary high orders in instanton number. 

\paragraph{SO(8)}
Our instanton formula should hold for $N_{\bf v} + N_{\bf s} + N_{\bf c} \le 4$. Let us compare it with known results. 

The 1-instanton result of $SO(8) + 1\mathbf{S} + 1\mathbf{C} + 1\mathbf{V}$ theory is found in (H.28) of \cite{DelZotto:2018tcj}.
It is expressed in terms of characters of irreducible representations $\chi_{\mathbf{R}}^S$, whose superscript $S \in \{G,v,s,c\}$ means either the gauge symmetry ($G$) or the flavor symmetry acting on the vector ($v$), spinor ($s$), or conjugate spinor ($c$) hypermultiplets. Their representation $\mathbf{R}$ is specified by the Dynkin label in the subscript. All irreducible characters for the flavor symmetry are assumed to be written in the orthogonal basis, to be compatible with our convention of mass parameters in \eqref{eq:1-loop-hyp}, \eqref{eq:recursion}, \eqref{eq:1inst-formula}. The mass parameters will be often distinguished by the superscript $S \in \{s,c,v\}$ according to the matter representation. The flavor symmetry is $Sp(N_{\bf v})_v \times Sp(N_{\bf s})_s \times Sp(N_{\bf c})_c$.

We can obtain the 1-instanton partition function of $SO(8) + N_{\bf s}\mathbf{S} + N_{\bf c}\mathbf{C} + N_{\bf v}\mathbf{V}$ theory with  $( N_{\bf s},  N_{\bf c}, N_{\bf v}) \leq ( 1,1,1)$ from (H.28) of \cite{DelZotto:2018tcj} by sending appropriate mass parameters to infinity.
All the results obtained in this way is consistent with our general 1-instanton expression \eqref{eq:1inst-formula} up to $t^{20}$ order, where $t \equiv \sqrt{p_1p_2}$. 
Furthermore, we are able to determine the unknown part of (H.28) of \cite{DelZotto:2018tcj} as 
\begin{align}
  \label{eq:SO8-v1s1c1}
  \begin{split}
 \tilde{Z}_1 &=  t^4 + \sum_{n=0}^\infty t^{5+2n} \chi_{(0n00)}^G \chi^v_{(1)}\chi^s_{(1)}\chi^c_{(1)} \\
 &\quad + \sum_{n=0}^\infty t^{6+2n}\left(\chi_{(1n00)}^G \chi^s_{(1)}\chi^c_{(1)}+ \chi_{(0n10)}^G \chi^s_{(1)}\chi^v_{(1)}
  +\chi_{(0n01)}^G \chi^c_{(1)}\chi^v_{(1)}\right)\\
 &\quad  + \sum_{n=0}^\infty t^{7+2n} \left(\chi^G_{(1n10)}\chi^s_{(1)} + \chi^G_{(1n01)}\chi^c_{(1)}+\chi^G_{(0n11)}\chi^v_{(1)}\right)- \sum_{n=0}^\infty t^{8+2n}\chi^G_{(1n11)} \ ,
 \end{split}
\end{align}
where $\tilde{Z}_1 \equiv (2\sinh{\frac{\e_{1}}{2}})(2\sinh{\frac{\e_{2}}{2}})  Z_1$ is the 1-instanton partition function with the center-of-mass factor removed.

Now we compare \eqref{eq:1inst-formula} with the 1-instanton partition function of $SO(8) + 2\mathbf{S}
 + 2\mathbf{C} + 2\mathbf{V}$ theory, written in (H.19) of \cite{DelZotto:2018tcj}. Our 1-instanton formula \eqref{eq:1inst-formula} applied to the $SO(8)$ theories having $( N_{\bf s},  N_{\bf c}, N_{\bf v}) \leq (2,1,1), (1,2,2), (1,1,2),(2,2,0),(2,0,2),(0,2,2)$ agree with (H.19) up to $t^{20}$ order, after suitably setting some mass parameters in (H.19) to infinity. We could further determine the unknown part of (H.19) of \cite{DelZotto:2018tcj} as 
 \begin{align}
  \label{eq:SO8-v2s2c2}
  \tilde{Z}_1 &= t^{-1} - t^3(\chi_{(01)}^v+\chi_{(01)}^s+\chi_{(01)}^c) 
  +t^5(\chi^G_{(1000)}\chi^s_{(10)}\chi^c_{(10)}+\chi^G_{(0010)}\chi^s_{(10)}\chi^v_{(10)}+\chi^G_{(0001)}\chi^c_{(10)}\chi^v_{(10)}) \nn \\&
  -t^6(\chi^G_{(1010)}\chi_{(10)}^s+\chi^G_{(1001)}\chi_{(10)}^c+\chi^G_{(0011)}\chi_{(10)}^v) + t^7 \chi_{(1011)}^G - \sum_{n=0}^\infty \Big(t^{5+2n}\chi_{(0n00)}^G \chi^s_{(10)}\chi^c_{(10)} \chi^v_{(10)}  \nn \\&
  + t^{6+2n}(\chi^G_{(1n00)}\chi^s_{(01)}\chi^c_{(01)}\chi^v_{(10)}+\chi^G_{(0n10)}\chi^s_{(01)}\chi^c_{(10)}\chi^v_{(01)}+\chi^G_{(0n01)}\chi^s_{(10)}\chi^c_{(01)}\chi^v_{(01)}) \nn \\&- t^{7+2n} 
  (\chi^G_{(1n10)}\chi^s_{(01)}\chi^c_{(10)}\chi^v_{(10)}+\chi^G_{(1n01)}\chi^s_{(10)}\chi^c_{(01)}\chi^v_{(10)}+\chi^G_{(0n11)}\chi^s_{(10)}\chi^c_{(10)}\chi^v_{(01)})  \\& +  t^{8+2n}
  (\chi^G_{(2n10)}\chi^s_{(01)}\chi^c_{(10)}+\chi^G_{(2n01)}\chi^s_{(10)}\chi^c_{(01)}+\chi^G_{(1n20)}\chi^s_{(01)}\chi^v_{(10)}+\chi^G_{(1n02)}\chi^c_{(01)}\chi^v_{(10)} \nn\\&
  +\chi^G_{(0n21)}\chi^s_{(10)}\chi^v_{(01)}+\chi^G_{(0n12)}\chi^c_{(10)}\chi^v_{(01)}) - t^{9+2n}(\chi^G_{(2n11)}\chi^s_{(10)}\chi^c_{(10)}+\chi^G_{(1n21)}\chi^s_{(10)}\chi^v_{(10)} \nn\\&
  +\chi^G_{(1n12)}\chi^c_{(10)}\chi^v_{(10)}) +  t^{10+2n} (\chi^G_{(2n21)}\chi^s_{(10)}+\chi^G_{(2n12)}\chi^c_{(10)} + \chi^G_{(1n22)}\chi^v_{(10)}) -  t^{11+2n} \chi^G_{(2n22)}\Big). \nn
\end{align}
Notice that \eqref{eq:SO8-v1s1c1} and \eqref{eq:SO8-v2s2c2} are manifestly invariant under the $SO(8)$ triality, transforming the $SO(8)$ representations as $(n_vn_an_cn_s) \rightarrow (n_sn_an_vn_c)$ 
along with $\chi^v_\mathbf{R} \rightarrow \chi^s_\mathbf{R} \rightarrow \chi^c_\mathbf{R} \rightarrow \chi^v_\mathbf{R}$.
It can be done by shuffling the Coulomb VEVs and renaming the flavor chemical potentials. We rearranged $Z_1$ in terms of the new variables $\vec{a}'$ or $\vec{a}''$, 
\begin{align}
  \label{eq:so8-triality}
  \begin{split}
  (a_1',a_2',a_3',a_4') &= \textstyle(\frac{-a_1+a_2+a_3-a_4}{2},\frac{-a_1+a_2+a_3-a_4}{2},\frac{-a_1+a_2+a_3-a_4}{2},\frac{-a_1+a_2+a_3-a_4}{2}) \\
  (a_1'',a_2'',a_3'',a_4'') &= \textstyle(\frac{+a_1-a_2-a_3-a_4}{2},\frac{-a_1+a_2-a_3-a_4}{2},\frac{-a_1-a_2+a_3-a_4}{2},\frac{+a_1+a_2+a_3-a_4}{2}),
  \end{split}
\end{align}
which exchanges the $SO(8)$ irreducible characters as
\begin{align}
  \chi_{(n_cn_an_sn_v)} (\vec{a}) &= \chi_{(n_vn_an_cn_s)} (\vec{a}')|_{\vec{a}' \rightarrow \vec{a}}, &
  \chi_{(n_sn_an_vn_c)} (\vec{a}) &= \chi_{(n_vn_an_cn_s)} (\vec{a}'')|_{\vec{a}'' \rightarrow \vec{a}}.
\end{align}
Dropping off primes from $Z_1 (\vec{a}', \epsilon_1,\epsilon_2; \vec{m}^s,\vec{m}^c,\vec{m}^v)$ or $Z_1 (\vec{a}'', \epsilon_1,\epsilon_2;\vec{m}^s,\vec{m}^c,\vec{m}^v) $, we indeed find
\begin{align}
  \begin{split}
  Z_1^{N_{\bf s}=N_{\bf c}= N_{\bf v}}\,(\vec{a}, \epsilon_1,\epsilon_2;\vec{m}^s,\vec{m}^c,\vec{m}^v) &= Z_1^{N_{\bf s}=N_{\bf c}= N_{\bf v}}\,(\vec{a}', \epsilon_1,\epsilon_2;\vec{m}^v,\vec{m}^s,\vec{m}^c)|_{\vec{a}' \rightarrow \vec{a}}\\
  Z_1^{N_{\bf s}=N_{\bf c}= N_{\bf v}}\,(\vec{a}, \epsilon_1,\epsilon_2;\vec{m}^s,\vec{m}^c,\vec{m}^v) &= Z_1^{N_{\bf s}=N_{\bf c}= N_{\bf v}}\,(\vec{a}'', \epsilon_1,\epsilon_2;\vec{m}^c,\vec{m}^v,\vec{m}^s)|_{\vec{a}'' \rightarrow \vec{a}} , 
  \end{split}
\end{align}
which is consistent with the triality. 

Similarly, we also found the 1-instanton formula \eqref{eq:1inst-formula} applied to $SO(8)$ theories with $( N_{\bf s},  N_{\bf c}, N_{\bf v}) \leq (4,0,0)$ or $(0,4,0)$ is compatible with the $SO(8)$ triality.
Starting with the 1-instanton result $Z_1^\text{ADHM} = Z_1^\text{ADHM} (\vec{a}, \epsilon_1,\epsilon_2, \vec{m}) $ obtained from the relevant ADHM quantum mechanics for $SO(8) +  N_{\bf v}\mathbf{V}$ theory with $ N_{\bf v} \leq 4$, we find 
\begin{align}
  \begin{split}
  Z_1^{N_{\bf c}, \, N_{\bf c}= N_{\bf v}= 0}\,(\vec{a}, \epsilon_1,\epsilon_2, \vec{m}) &= Z_1^{\text{ADHM}}(\vec{a}', \epsilon_1,\epsilon_2, \vec{m})|_{\vec{a}' \rightarrow \vec{a}}\\
  Z_1^{N_{\bf s}, \, N_{\bf s}= N_{\bf v}= 0}\,(\vec{a}, \epsilon_1,\epsilon_2, \vec{m}) &= Z_1^{\text{ADHM}}(\vec{a}'', \epsilon_1,\epsilon_2, \vec{m})|_{\vec{a}'' \rightarrow \vec{a}}.
  \end{split}
\end{align}

\paragraph{SO(9)}
For the $SO(9)$ theory with $N_{\bf s}$ spinor and $N_{\bf v}$ vector, our blowup formula is valid for $N_{\bf v} + 2 N_{\bf s} \le 5$. 
The 1-instanton formula \eqref{eq:1inst-formula} can be applied to $(N_\mathbf{s}, N_{\mathbf{v}}) \leq (1,3)$ or $(2,1)$, which has $Sp(N_\mathbf{s})_s \times Sp(N_\mathbf{v})_v$ flavor symmetry. It can be compared with the 1-instanton partition function of $SO(9) + 2\mathbf{S} + 3\mathbf{V}$ theory, which is written in (H.20) of \cite{DelZotto:2018tcj} up to $t^7$ order, after appropriately taking some mass parameters to infinity. We checked all their consistency up to the given order. For example, the character expansion of $\hat{Z}_{1}$ for $SO(9) + 2\mathbf{S} + 1\mathbf{V}$  can be written as  
\begin{align}
  \label{eq:SO9-v1s2}
  \begin{split}
  \tilde{Z}_1&=\, t^4\chi^v_{(1)}+t^5\chi^s_{(20)}-t^6\chi_{(0001)}^G\chi^s_{(10)}
   \\&\quad +\sum_{n=0}^{\infty}\Bigg(t^{6+2n}\chi^G_{(0n00)}\chi^s_{(02)}\chi^v_{(1)}
  -t^{7+2n}\left(\chi^G_{(1n00)}\chi^s_{(02)}+\chi^G_{(0n01)}\chi^s_{(11)}\chi^v_{(1)}\right) 
  \\&\qquad \quad+t^{8+2n}\left(\chi^G_{(1n01)}\chi^s_{(11)}+\chi^G_{(0n10)}\chi^s_{(20)}\chi^v_{(1)} 
  +\chi^G_{(0n02)}\chi^s_{(01)}\chi^v_{(1)}\right)
  \\&\qquad \quad-t^{9+2n}\left(\chi^G_{(1n10)}\chi^s_{(20)}+\chi^G_{(1n02)}\chi^s_{(01)}+\chi^G_{(0n11)}\chi^s_{(10)}\chi^v_{(1)}\right)\\
  &\qquad \quad+t^{10+2n}\left(\chi^G_{(1n11)}\chi^s_{(10)}+\chi^G_{(0n20)}\chi^v_{(1)}\right) -t^{11+2n}\chi^G_{(1n20)} \Bigg),
  \end{split}
  \end{align}
which is tested against the general formula \eqref{eq:1inst-formula} up to $t^{20}$ order. It is the same as (H.20) of \cite{DelZotto:2018tcj} after reducing the $Sp(3)_v$ characters by 
\begin{align}
  \chi^{v}_{(001)}\rightarrow \chi^{v}_{(1)},\qquad \chi^{v}_{(010)}\rightarrow 1,\qquad \chi^{v}_{(100)}\rightarrow 0,\qquad\chi^{v}_{(000)}\rightarrow 0.
\end{align}

\paragraph{SO(10)}
We apply our 1-instanton expression \eqref{eq:1inst-formula} to $SO(10) + N_{\mathbf s}\mathbf{S} + N_{\mathbf c}\mathbf{C} + N_{\mathbf v}\mathbf{V}$ theory with $( N_{\bf s},  N_{\bf c}, N_{\bf v}) \leq (2,0,2),(1,1,2), (0,2,2),(1,0,4),(0,1,4)$. The relevant flavor symmetry is $U(N_\mathbf{s} + N_\mathbf{c}) \times Sp(N_\mathbf{v})$ because  the $SO(10)$ (conjugate) spinor is a complex representation. Since the $SO(10)$ charge conjugation exchanges the spinor and conjugate spinor representations, 
\textit{i.e.}, $\chi_{(00001)}^G = (\chi_{(00010)}^G)^*$, the instanton partition function for $SO(10) + (N_{\mathbf s}\mp 1)\mathbf{S} + (N_{\mathbf c}\pm 1)\mathbf{C} + N_{\mathbf v}\mathbf{V}$  must be identified with that of $SO(10) + N_{\mathbf s}\mathbf{S} + N_{\mathbf c}\mathbf{C} + N_{\mathbf v}\mathbf{V}$ simply by flipping the sign of mass parameters for (conjugate) spinor hypermultiplets: 
\begin{align}
Z_{1}^{N_{\mathbf{s}},N_\mathbf{c},N_{\mathbf{v}}}(m^s_{1, \cdots, N_\mathbf{s}}; m^c_{1,\cdots,N_\mathbf{c}}) &= Z_{1}^{N_{\mathbf{s}}-1,N_\mathbf{c}+1,N_{\mathbf{v}}}(m^s_{1, \cdots, N_\mathbf{s}-1}; m^c_{1,\cdots,N_\mathbf{c}+1})\big|_{m^c_{N_\mathbf{c}+1} = -m^s_{N_\mathbf{s}}}\nn\\ 
  &= Z_{1}^{N_{\mathbf{s}}+1,N_\mathbf{c}-1,N_{\mathbf{v}}}(m^s_{1, \cdots, N_\mathbf{s}+1}; m^c_{1,\cdots,N_\mathbf{c}-1})\big|_{m^s_{N_\mathbf{s}+1} = -m^c_{N_\mathbf{c}}}.
\end{align}
This relation is explicitly confirmed in all above cases at 1-instanton order. 
We may want to compare \eqref{eq:1inst-formula} with the known 1-instanton partition function of $SO(10) + 1\mathbf{S} + 1\mathbf{C} + 4\mathbf{V}$ theory, written in (H.21) of \cite{DelZotto:2018tcj}, after taking relevant mass parameters  to infinity. However, (H.21) specifies $\tilde{Z}_1$ only up to $\mathcal{O}(t^5)$, which leaves nothing for comparison once we reduce the mass parameters. Thus the consistency between two expressions can be only weakly tested. For instance, $\tilde{Z}_{1}$ obtained from \eqref{eq:1inst-formula} for $SO(10) + N_\mathbf{s}\mathbf{S}+ N_\mathbf{c}\mathbf{C} + 4\mathbf{V}$ theory with $ N_\mathbf{s} 
+ N_\mathbf{c} = 2$ is displayed in \eqref{eq:so10-2s2v}, which turns out to be trivial upto $t^4$ order.

\paragraph{SO(12)}

The 1-instanton partition function of $SO(12) + 1\mathbf{S} + 6\mathbf{V}$ theory  is written in (H.22) of \cite{DelZotto:2018tcj}, up to $t^8$ order. It can be compared with our 1-instanton formula \eqref{eq:1inst-formula} applied to $SO(12) + N_\mathbf{s}\mathbf{S} + N_\mathbf{c}\mathbf{C} + N_\mathbf{v}\mathbf{V}$ theory with $(N_\mathbf{s},N_\mathbf{c}, N_{\mathbf{v}}) \leq (1,0,4)$ or $(0,1,4)$, whose flavor symmetry acting on matter multiplets is $SO(2N_\mathbf{s})_s \times SO(2N_\mathbf{c})_c \times Sp(N_\mathbf{v})_v$. For comparison, we need to appropriately decouple some mass parameters in (H.22) to infinity. It reduces the $Sp(6)_v$ characters in (H.22) to, e.g., the $Sp(4)_v$ irreducible characters as follows:
\begin{align}
\begin{split}
\begin{array}{llllll}
  \chi^{v}_{(000000)} &\rightarrow 0, & \quad \chi^{v}_{(100000)}&\rightarrow 0,& \quad \chi^{v}_{(010000)}&\rightarrow 1, \\ \chi^{v}_{(001000)}&\rightarrow \chi^{v}_{(1000)}, & \quad  \chi^{v}_{(000100)}&\rightarrow \chi^{v}_{(0100)}, & \quad \chi^{v}_{(000001)}&\rightarrow \chi^{v}_{(0001)}.
\end{array}
\end{split}
\end{align}
We explicitly confirmed that (H.22) and \eqref{eq:1inst-formula} agree up to the given order, for $(N_\mathbf{s},N_\mathbf{c}, N_{\mathbf{v}}) = (1,0,4)$. Moreover, we checked that the 1-instanton results ${Z}_1$ from \eqref{eq:1inst-formula} for $(N_\mathbf{s},N_\mathbf{c}, N_{\mathbf{v}})= (1,0,N_\mathbf{v})$ and $(0,1,N_\mathbf{v})$ could be interchanged as follows: 
\begin{align}
Z_{1}^{N_{\mathbf{s}}=1,N_\mathbf{c}=0,N_{\mathbf{v}}}(a_1,a_2,a_3,a_4,a_5,a_6) =Z_{1}^{N_{\mathbf{s}}=0,N_\mathbf{c}=1,N_{\mathbf{v}}}(a_1,a_2,a_3,a_4,a_5,-a_6).
\end{align}

\paragraph{Summary of new results}
We have compared so far the solution $Z_1$ of the recursion formulae \eqref{eq:recursion} with the known $1$-instanton partition function for various $SO(N)$ theories with spinor hypermultiplets. The comparison showed consistency for all the examples whose $Z_1$ had been computed \cite{Kim:2018gjo,DelZotto:2018tcj}. We also collect the character expansion of the 1-instanton partition function \eqref{eq:1inst-formula} in Appendix~\ref{sec:data} for novel $SO(N)$ theories with spinor matters. See Table~\ref{tbl:son-data} for the list of character expansions.

\begin{table}[h!]
  \centering
  \begin{tabular}{cc|c}
    \hline
    Gauge Group & Hypermultiplets & Equation No.\\
    \hline\hline
	$SO(8)$ & $1\mathbf{S}+1\mathbf{C}+1\mathbf{V}$ & \eqref{eq:SO8-v1s1c1}\\
	$SO(8)$ & $2\mathbf{S}+2\mathbf{C}+2\mathbf{V}$ & \eqref{eq:SO8-v2s2c2}\\
	$SO(8)$ & $3\mathbf{S}+1\mathbf{C}$ &  \eqref{eq:so8-s3c1}\\
	$SO(9)$ & $2\mathbf{S}+1\mathbf{V}$ & \eqref{eq:SO9-v1s2}\\
	$SO(10)$ & $2\mathbf{S}+2\mathbf{V}$ & \eqref{eq:so10-2s2v}\\
	$SO(10)$ & $3\mathbf{S}$ & \eqref{eq:so10-s3}\\
	$SO(11)$ & $1\mathbf{S}+3\mathbf{V}$ & \eqref{eq:so11-s1v3}\\
	$SO(12)$ & $2\mathbf{S}$ & \eqref{eq:so12-s2}\\
	$SO(12)$ & $1\mathbf{S}+1\mathbf{C}$ & \eqref{eq:so12-s1c1}\\
	$SO(13)$ & $1\mathbf{S}+1\mathbf{V}$ & \eqref{eq:so13-s1v1}\\
	$SO(14)$ & $1\mathbf{S}+2\mathbf{V}$ & \eqref{eq:so14-s1v2}\\\hline
    \end{tabular}
  \caption{Character expansion of $SO(N)$ theory with spinor hypermultiplets}
  \label{tbl:son-data}
\end{table}

\subsection{Theories with an exceptional gauge group}
\label{subsec:exceptional}

Let us continue to apply the recursion formulae \eqref{eq:recursion} and the general 1-instanton expression \eqref{eq:1inst-formula} to study the instanton partition function of exceptional gauge theories. One can find a sufficient number of recursion formulae \eqref{eq:recursion} to fix the $n$-instanton partition function $Z_n$, if and only if the gauge theory has the following number of fundamental hypermultiplets: 
\begin{align}
  \label{eq:exceptional-range}
  \begin{split}
    N_{f} \leq 2 & \qquad\text{ if }\quad G=G_2,\\
    N_{f} \leq 2 & \qquad\text{ if }\quad G=F_4,\\
    N_{f}+N_{\overline{f}} \leq 3 & \qquad\text{ if }\quad G=E_6,\\
    N_{f} \leq 2 & \qquad\text{ if }\quad G=E_7,\\
    \varnothing & \qquad\text{ if }\quad G=E_8.
  \end{split}
\end{align}
Notice that other representations do not appear in the recent classification of 4d $\CN=2$ SCFTs \cite{Bhardwaj:2013qia} nor 5d SCFTs \cite{Jefferson:2017ahm}.

\begin{table}[h!]
\centering
  \begin{tabular}{cc|c}
    \hline
    Gauge Group & Hypermultiplets & Equation No.\\
    \hline\hline
	$F_4$ & $2\mathbf{F}$ & \eqref{eq:F4F2}\\
	$E_6$ & $3\mathbf{F}$ & \eqref{eq:E6F3}\\
	$E_7$ & $2\mathbf{F}$ & \eqref{eq:E7F2}\\
	$E_8$ & $\emptyset$ &  \eqref{eq:E8}\\\hline
    \end{tabular}
  \caption{Character expansion of exceptional gauge theory with fundamental hypermultiplets}
  \label{tbl:exc-data}
\end{table}

We give explicit character expansion of the one instanton partition function in Appendix \ref{sec:data}. See Table \ref{tbl:exc-data} for the list of character expansions. 

\paragraph{$\bf G_2$}
A supersymmetric quantum mechanical model was proposed in  \cite{Kim:2018gjo}, whose Witten index corresponds to the $n$-instanton partition function of 
$G_2 + N_f \mathbf{F}$ theory with $N_f \leq 3$. Its index can be written as the following sum over $SU(3)$ colored Young diagrams:
\begin{align}
\begin{split}
  Z_n^\text{YD} = \sum_{|\vec{Y}| = n} &\prod_{i=1}^3\prod_{s \in Y_i}\frac{2\sinh{(\phi(s))} \ 2\sinh{(\e_+-\phi(s))} \ \prod_{l=1}^{N_f}2\sinh(\frac{m_l \pm \phi(s)}{2}) }{2\sinh{\frac{\e_+ - \phi(s)}{2}}\prod_{j=1}^3 2\sinh{\frac{E_{ij}}{2}}\,2\sinh{\frac{E_{ij}-2\e_+}{2}}\,2\sinh{\frac{\e_+ - \phi(s) - a_j}{2}}} \\ \times &\prod_{i \leq j}^3 \prod_{\stackrel{s_{i,j} \in Y_{i,j}}{s_i < s_j}} \frac{2\sinh{\frac{\phi(s_i)+\phi(s_j)}{2}}\, 2\sinh{\frac{\phi(s_i)+\phi(s_j)-2\e_+}{2}}}{ 2\sinh{\frac{\e_1 - \phi(s_i)-\phi(s_j)}{2}} \, 2\sinh{\frac{\e_2 - \phi(s_i)-\phi(s_j)}{2}}}.
\end{split}
\end{align}
Our 1-instanton formula \eqref{eq:1inst-formula} agrees with the above expression $Z_1^\text{YD}$ for all $N_f \leq 2$. Also at two instantons, we explicitly checked that $Z_2^\text{YD} = Z_2$, where $Z_2$ is the solution of the recursion formulae \eqref{eq:recursion} with \eqref{eq:d-range-nonsu}.

\paragraph{$\bf F_4$}
The 1-instanton partition function of $F_4 + 2 \mathbf{F}$ gauge theory is given in (H.31) of \cite{DelZotto:2018tcj}, which has $Sp(2)_f$ flavor symmetry. In terms of $F_4$ and $Sp(2)_f$ characters, 
\begin{align}
    \label{eq:F4F2}
\begin{split}
  & \tilde{Z}_1 =\,t^6\chi^f_{(01)}+t^7\chi^{f}_{(30)}-t^8\left(\chi^G_{(0001)}\chi^f_{(20)}+\chi^G_{(1000)}\right)+t^9\chi^G_{(0010)}\chi^f_{(10)}-t^{10}\chi^G_{(0100)}\\
  &+\sum_{n=0}^{\infty}\Bigg(t^{8+2n}\chi^G_{(n000)}\chi^f_{(03)}-t^{9+2n}\chi^G_{(n001)}\chi^f_{(12)}+t^{10+2n}\left(\chi^G_{(n010)}\chi^f_{(21)}+\chi^G_{(n002)}\chi^f_{(02)}\right)\\
  &\qquad\quad-t^{11+2n}\left(\chi^G_{(n100)}\chi^f_{(30)}+\chi^G_{(n011)}\chi^f_{(11)}\right) 
  +t^{12+2n}\left(\chi^G_{(n101)}\chi^f_{(20)}+\chi^G_{(n020)}\chi^f_{(01)}\right)\\
  &\qquad\quad-t^{13+2n}\chi^G_{(n110)}\chi^f_{(10)}+t^{14+2n}\chi^G_{(n200)}\Bigg).
\end{split}
\end{align}
We confirmed that our 1-instanton formula \eqref{eq:1inst-formula} agrees with the above expression up to $t^{15}$ order.

\paragraph{$\bf E_6$}

Let us apply our general 1-instanton expression \eqref{eq:1inst-formula} to $E_6 + N_f \mathbf{F} + N_{\bar{f}}\overline{\mathbf{F}}$ gauge theory with $(N_f,N_{\bar{f}}) \leq (3,0), (2,1), (1,2),(0,3)$ whose flavor symmetry is $U(N_f 
+ N_{\bar{f}})$. Since the fundamental and anti-fundamental representations are interchanged by the $E_6$ charge conjugation, their instanton partition functions should be identical upon inverting the sign of relevant mass parameters. We explicitly confirmed that \eqref{eq:1inst-formula} satisfies the relations
\begin{align}
\begin{split}
  Z_{1}^{N_f,N_{\bar{f}}}(m^f_{1, \cdots, N_f}; m^{\bar{f}}_{1,\cdots,N_{\bar{f}}}) &= Z_{1}^{N_f-1,N_{\bar{f}}+1}(m^f_{1, \cdots, N_f-1}; m^{\bar{f}}_{1,\cdots,N_{\bar{f}}+1})\big|_{m^{\bar{f}}_{N_{\bar{f}}+1} = -m^f_{N_f}}\\ 
  &= Z_{1}^{N_f+1,N_{\bar{f}}-1}(m^f_{1, \cdots, N_f+1}; m^{\bar{f}}_{1,\cdots,N_{\bar{f}}-1})\big|_{m^{f}_{N_{f}+1} = -m^{\bar{f}}_{N_{\bar{f}}}},
\end{split}
\end{align}
in all above cases. Furthermore, $Z_1$ at $(N_f, N_{\bar{f}}) = (3,0)$ can be compared with (H.35) of \cite{DelZotto:2018tcj} which displays the character expansion up to $t^{11}$ order. We checked their consistency except a sign mistake in the second term of (H.35). The full character expansion of $Z_1$ at $N_f=3$ and $N_{\bar{f}}=0$ is written in \eqref{eq:E6F3}, after turning off the $E_6$ Coulomb VEV $\vec{a} = 0$ for simplicity.

\paragraph{$\bf E_7$}

Our 1-instanton expression \eqref{eq:1inst-formula} is applicable to $E_7 + N_f\mathbf{F}$ gauge theory with $N_f \leq 2$, which has $SO(2N_f)$ flavor symmetry. We give the full character expansion of $Z_1$ at $N_f=2$ in \eqref{eq:E7F2} after setting $\vec{a}=0$ to shorten the expression. We also compared the result \eqref{eq:1inst-formula} applied to the $N_f = 1$ case with (H.40) of \cite{DelZotto:2018tcj} and found that they agree up to $t^{280}$ order.

\paragraph{$\bf E_8$}

The (centered) 1-instanton partition function of $E_8$ gauge theory can be written as 
\begin{align}
    \label{eq:E8}
  \tilde{Z}_1=\sum_{n=0}^{\infty}t^{29+2n}\chi^{E_8}_{(000000n0)}.
\end{align}
We confirmed that it agrees with our 1-instanton expression \eqref{eq:1inst-formula} up to $t^{520}$ order. 
It is actually proven in \cite{Keller:2011ek, Keller:2012da} that the (centered) 1-instanton formula \eqref{eq:1inst-formula} for any gauge group without matter can be written in terms of the character expression \cite{VinbergPopov, Garfinkle, Benvenuti:2010pq}
\begin{align}
 \tilde{Z}_1 = t^{h^\vee - 1} \sum_{n=0}^\infty t^{2n} \chi^G_{n \cdot \textbf{adj}} \ . 
\end{align}

%%%%%%%%%%%%%%%%%%%%%%%%%%%%%%%%%%%%%%%
\subsection{SU(6) theory with a rank-3 antisymmetric hypermultiplet}
\label{subsec:rank3}
Another non-trivial test of our blow-up recursion formulae \eqref{eq:recursion}  is the partition function for 5d $SU(6)$ theory with a hypermultiplet in the rank-3 antisymmetric representation ({\bf TAS}). This theory has can be Higgsed to a theory with $SU(3)\times SU(3)$ gauge symmetry that can be explicitly checked at the level of the partition function.

To have a UV fixed point, 5d $SU(6)$ theories can have up to 2 hypermultiplets in the rank-3 antisymmetric representation \cite{Jefferson:2017ahm}. Their type IIB 5-brane configurations were constructed in~\cite{Hayashi:2019yxj} with/without O5-planes. In particular, 5-brane web diagrams for $SU(6)+\frac12{\bf TAS}$ and $SU(6)+1{\bf TAS}$ do not contain orientifold planes, so that topological vertex method \cite{Aganagic:2003db, Iqbal:2007ii} can be straightforwardly applied to compute their partition functions. In \cite{Hayashi:2019yxj},  for instance, the partition function of $SU(6)_\frac52 + \frac12{\bf TAS}$ theory was computed up to two instantons using the topological vertex formalism.
%---------  <Figure>  ---------------%
\begin{figure}[t]
\centering
\includegraphics[width=12cm]{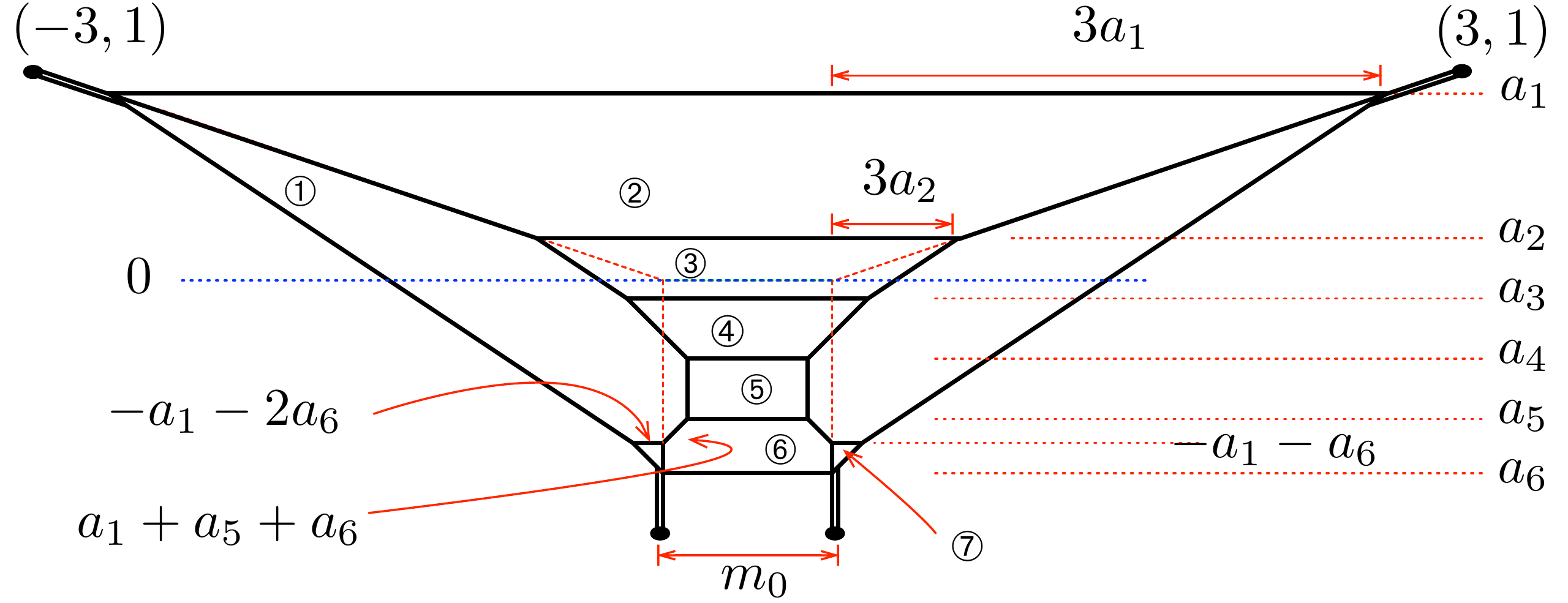}
\caption{A 5-brane web for $SU(6)_3$ theory with one massless hypermultiplet in the rank-3 antisymmetric representation.}
\label{fig:SU6-monopole}
\end{figure}
%----------- </Figure> ---------------%

Our blow-up equation \eqref{eq:cor5d-unity} demands all mass parameters to be generically turned on. In particular, we need a mass parameter for the rank-3 antisymmetric hypermultiplet. As one cannot introduce mass for a half-hypermultiplet, let us consider the $SU(6)_3$ theory with a full hypermultiplet in the rank-3 antisymmetric representation ($SU(6)_3+1{\bf TAS}$). 
An example for 5-brane web for $SU(6)_3+1{\bf TAS}$ is depicted in Figure~\ref{fig:SU6-monopole}. It is instructive to see if Figure~\ref{fig:SU6-monopole} is consistent with the expected prepotential.  The effective prepotential on the Coulomb branch of a 5d gauge theory with a gauge group $G$ and matter $f$ in a representation $\mathbf{R}_f$ is \cite{Intriligator:1997pq}
\begin{align}
\mathcal{F}(\phi) = \frac{m_0}{2}h_{ij}\phi_i\phi_j + \frac{\kappa}{6}d_{ijk}\phi_i\phi_j\phi_k + \frac{1}{12}\left(\sum_{\vec{\alpha}\in \D}\Big|\vec{\alpha} \cdot \vec{\phi}\,\Big|^3 - \sum_f\sum_{\vec{\w} \in \mathbf{R}_f}\Big|\vec{\w}\cdot \vec{\phi} + m_f\Big|^3\right). \label{prepotential}
\end{align}
Here, $m_0$ is the inverse of the gauge coupling squared, $\kappa$ is the Chern-Simons level and $m_f$ is a mass parameter for the matter $f$. $\vec{\alpha}$ is a root of the Lie algebra $\mathfrak{g}$ associated to $G$ and $\vec{\w}$ is a weight of the representation $\mathbf{R}_f$ of $\mathfrak{g}$. We also defined $h_{ij} = \text{Tr}(T_iT_j), d_{ijk} = \frac{1}{2}\text{Tr}\left(T_i\{T_j, T_k\}\right)$ where $T_i$ are the Cartan generators of the Lie algebra $\mathfrak{g}$. With the Coulomb branch moduli assigned in Figure \ref{fig:SU6-monopole} and the identification of Weyl chamber for the Coulomb VEV ($a_1\ge a_2\ge \cdots \ge a_{6}$, $\sum_{i=1}^{6}a_i=0$),
\begin{align}
	a_1= \phi_1,&& 
	a_2=\phi_2-\phi_1,&&
	a_3=\phi_3-\phi_2,&&
	a_4=\phi_4-\phi_3,&&
	a_5=\phi_5-\phi_4,&&
	a_6=-\phi_5, \label{orth2Dynkin}
\end{align}
one finds that the prepotential for $SU(6)_3$ with one \emph{massless} rank-3 antisymmetric matter takes the form of
\begin{align}
\begin{split}
	\mathcal{F}_{SU(6)_3+1{\bf TAS}} &= m_0 \big( \phi _1^2+\phi _2^2+\phi _3^2+\phi _4^2+\phi _5^2-\phi _1 \phi _2-\phi _2 \phi _3-\phi _3 \phi _4-\phi _4
   \phi _5\big) \cr
   &\quad + \frac{\phi _1^3}{3}+\frac{4 \phi _2^3}{3}+\frac{4
   \phi _3^3}{3}+\frac{4 \phi_4^3}{3}+\frac{4 \phi_5^3}{3}+4\phi_1^2 \phi_2 -5 \phi_1\phi_2^2  \cr
   &  \quad -2\phi_1 \big(\phi_3^2+\phi_4^2+\phi_5^2\big) +\phi_2^2 \phi_3-2 \phi_2 \phi_3^2-\phi_3 \phi_4^2-\phi _4^2 \phi_5 \cr 
   & \quad +2 \phi _1\phi _2 \phi _3 +2\phi_1\phi_3\phi_4+2\phi_1\phi_4\phi_5. \label{eq:prep+SU6-3}
\end{split}
\end{align}
One can easily see that the monopole string tensions $T_i=\partial{\mathcal{F}}/\partial{\phi_i}$ computed from the above prepotential \eqref{eq:prep+SU6-3} agree with the areas of the compact faces of the 5-brane web, i.e.,
\begin{align}
T_1=\textcircled{\scriptsize 1} + 2\times\textcircled{\scriptsize 2},  %\crcr
&& T_2=\textcircled{\scriptsize 3}, %\crcr
&&T_3=\textcircled{\scriptsize 4}, %\crcr
&&T_4=\textcircled{\scriptsize 5}, %\crcr
&&T_5=\textcircled{\scriptsize 6} 
	+ 2\times\textcircled{\scriptsize 7},
\end{align}
where the encircled numbers represent the area of apparent faces in Figure \ref{fig:SU6-monopole}. This shows that Figure~\ref{fig:SU6-monopole} is indeed consistent with the prepotential of $SU(6)_3+1{\bf TAS}$ gauge theory.

 %---- <Figure>  ---------------%
\begin{figure}[t]
\centering
\includegraphics[width=12cm]{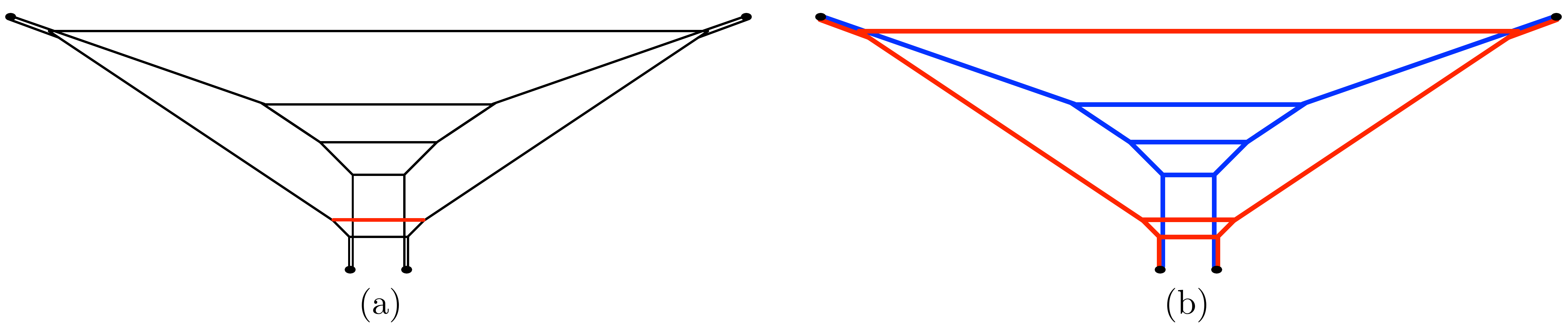}
\caption{(a) A Higgsing of $SU(6)_3+1{\bf TAS}$ into two $SU(3)_3$ theories by aligning internal D5-branes in red. (a) Two different $SU(3)_3$ theories are painted in blue and red, respectively. %As a result, a new Higgs branch emerges.
}
\label{fig:SU6-Higgsing}
\end{figure}
 %---- </Figure>  --------------%
 
Notice that this 5-brane web for $SU(6)_3+1{\bf TAS}$ suggests an intriguing Higgsing of the theory, which is the Higgsing of $SU(6)$ theory with one rank-3 antisymmetric hyper into two disjoint $SU(3)$ theories. It can be achieved by setting the Coulomb branch parameters as
 \begin{align}
 	a_5= - a_1-a_6, \qquad {\rm or~equivalently} \qquad \phi_4=\phi_1. \label{eq:HiggisingToSU3}
 \end{align}
 This tuning of the parameters, of course, reduces dimension of the Coulomb branch by one and also opens up a Higgs branch in such a way that the 5-brane web in Figure \ref{fig:SU6-monopole} becomes 5-brane web in Figure \ref{fig:SU6-Higgsing}(a) where the D5-branes on the upper edges of $\textcircled{\scriptsize 6}$ and $\textcircled{\scriptsize 7}$ are aligned and joint to become a single D5-brane denoted red in Figure~\ref{fig:SU6-Higgsing}(a). The resulting configuration is then a 5-brane configuration for two pure $SU(3)_3$ theories that are on top of each other, as shown in Figure~\ref{fig:SU6-Higgsing}(b). This is a 5-brane realization of Higgsing $SU(6)_3+1{\bf TAS}$ theory into two pure $SU(3)_3$ theories. It follows that 
 under this Higgsing, the prepotential for $SU(6)_3+1{\bf TAS}$ \eqref{eq:prep+SU6-3} theory reduces to a sum of prepotentials for two disjoint pure $SU(3)_3$ theories:
\begin{align}
\mathcal{F}_{SU(6)_{3}+1{\bf TAS}}{}\Big|_{a_1+a_5+a_6=0}	%\mathcal{F}(m_0,a_1,a_2,a_3,a_4,a_5,a_6)	
&\rightarrow \mathcal{F}_{SU(3)_3}(m_0,a_1,a_5,a_6)+\mathcal{F}_{SU(3)_3}(m_0,a_2,a_3,a_4).
\end{align}
This in turn implies that under this Higgsing, the partition function for $SU(6)_{3}+1{\bf TAS}$ should be expressed as a product of the partition functions of two pure $SU(3)_3$ theories:
\begin{align}
\CZ^{SU(6)_{3}+1{\bf TAS}}\big|_{\rm Higgsing}~\rightarrow ~~ \CZ^{SU(3)_{3}}\!(q,A_1, A_5,A_6)\,\CZ^{SU(3)_{3}}\!(q,A_2, A_3,A_4)\,Z_{\rm extra}(q)\,,
\label{eq:SU6-Higgs-SU3}	
\end{align}
where the parameters $q$ and $A_i$ are the K\"ahler parameters for instanton and Coulomb branch parameters, and  $Z_{\rm extra}(q)$ represents the overall extra terms that do not explicitly depend on the Coulomb branch moduli, which would correspond to a new decoupled mode appearing in Figure \ref{fig:SU6-Higgsing}.  In what follows, we explicitly compute the partition function for $SU(6)_3+1{\bf TAS}$ based on the 5-brane web and compare it with our general 1-instanton formula \eqref{eq:1inst-formula}. At two instantons, we will consider this Higgsing as a consistency check of our solution $Z_2$ obtained from the blowup recursion formulae \eqref{eq:recursion}.

%---------  Figure  ---------------%
\begin{figure}[t]
\centering
\includegraphics[width=12cm]{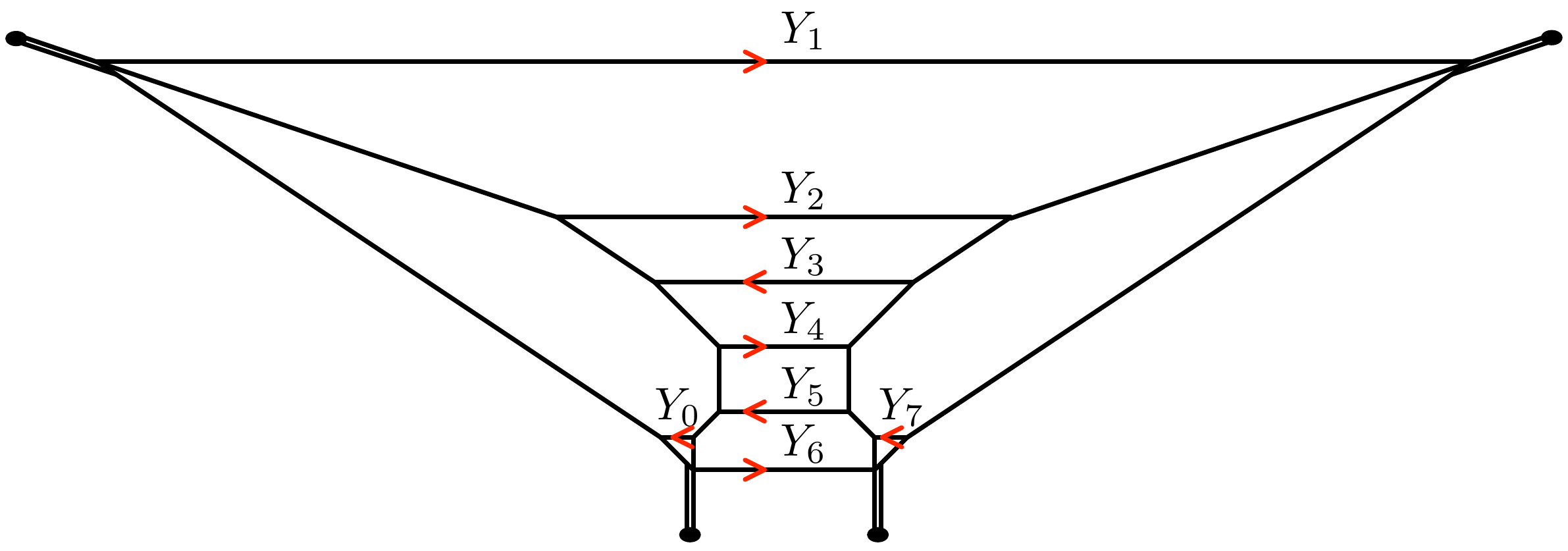}
\caption{A labeling of Young diagrams assigned to the horizontal edges of Figure \ref{fig:SU6-monopole}.}
\label{fig:SU6young}
\end{figure}
%----------------------------------%
To compute the instanton partition function based on the 5-brane web for $SU(6)_3+1{\bf TAS}$ given in Figure \ref{fig:SU6-monopole}, we assign the Young diagrams $Y_i$ to each horizontal edge of the web diagram as shown in Figure \ref{fig:SU6young} and use the topological vertex method. For convenience, we restrict ourselves to the unrefined case where $2\epsilon_+=\epsilon_1+\epsilon_2=0$. (See also a similar calculation done in \cite{Hayashi:2019yxj}.) As the web diagram in Figure \ref{fig:SU6-monopole} is left-right symmetric, it is convenient to split the web diagram to the left and right parts and glue them later to obtain the full partition function. Let us introduce the following fugacity variables to express the partition function.
\begin{align}
A_i \equiv e^{-a_i} \ \text{ for } \ i=1, \cdots, 6
%\quad \Big(\prod_{i=1}^{6}A_i =1\Big)\, 
% , \qquad q = e^{-\frac{8\pi^2}{g_0^2}}% = e^{-4\pi^2m_0}
, \qquad g\equiv \sqrt{p_1/p_2} = e^{-\epsilon_-},
\end{align}
in which the $SU(6)$ traceless condition $\prod_{i=1}^{6}A_i =1$ is assumed. Applying the topological vertex formalism \cite{Aganagic:2003db}, we find that 
\begin{align}
  \label{eq:znek-su6}
  \begin{split}
  \mathcal{Z} 
  =&\, \sum_{(Y_1, \cdots, Y_6)}q^{\sum_{i=1}^6|Y_i|} (-A_1^6)^{|Y_1|}(-A_2^6)^{|Y_2|}(-A_2^2A_3^4)^{|Y_3|}(-A_2^2A_3^2A_4^2)^{|Y_4| + |Y_5|}\\
  &\times f_{Y_1}(g)^5f_{Y_2}(g)^5f_{Y_3}(g)^3f_{Y_4}(g)f_{Y_5}(g)^{-1}f_{Y_6}(g)^{2}
  Z_{\text{left}}(\vec{Y})Z_{\text{right}}(\vec{Y}),
%  Z_{\text{half}}\,(Y_1, Y_2, Y_3, Y_4, Y_5, Y_6)^2. 
 \end{split}
\end{align}
where $\vec{Y}=(Y_1, Y_2, Y_3, Y_4, Y_5, Y_6)$. The left/right factor 
$Z_{\text{left}}(\vec{Y})$/$Z_{\text{right}}(\vec{Y})$ can be written as 
\begin{align}
\begin{split}
Z_{\text{left}}(\vec{Y}) = Z_{\text{right}}(\vec{Y})=& \,
\sum_{Y'} ( - A_1{}^{-1} A_6{}^{-2})^{|Y'|} 
g^{\frac{||Y'^t||^2+||Y'||^2}{2}} \tilde{Z}_{Y'}^2 f_{Y'}^2(g)
\prod_{i=1}^6 g^{\frac{||Y_i||^2}{2}} \tilde{Z}_{Y_i} 
\cr 
& 
\times 
R_{Y_1 Y_6^t}^{-1} (A_1 A_6{}^{-1})\,R_{Y' Y_6^t}^{-1} (A_1{}^{-1} A_6{}^{-2})\,R_{Y_1Y'^t }^{-1} (A_1^2A_6) \cr 
& 
\times  
 \prod_{2 \le i <  j \le 5} R_{Y_i Y_j^t}^{-1} (A_i A_j{}^{-1})
 \prod_{i=2}^5 R_{Y'^t Y_i} (A_1 A_i  A_6) 
\end{split}
\end{align}
in which the dummy variable $Y'$ should be interpreted as $Y_0$ for $Z_{\text{left}}(\vec{Y})$ and  $Y_7$ for $Z_{\text{right}}(\vec{Y})$. % summing over $Y_7$ instead of $Y_0$. %  \big|_{Y_0 \to Y_7}$.
Here, for a Young diagram $\lambda = (\lambda_1, \lambda_2, \cdots)$ and its transpose $\lambda^t$,  
\begin{align}
  |\lambda|=\sum_{i}\lambda_i, \qquad ||\lambda||^2 =\sum_{i}\lambda_i^2, \qquad  \tilde{Z}_{\lambda} 
&= \prod_{(i,j) \in \lambda} \frac{1}{1 - g^{\lambda_i + \lambda^t_j - i - j +1} }. 	
\end{align}
The framing factor $f_{\lambda}(g)$ is defined by
\begin{align}
f_\lambda(g) = (-1)^{|\lambda|}g^{\frac{1}{2}(g^{||\lambda^t||^2 - ||\lambda||^2})}.
\end{align}
And also, $R_{\lambda \mu } (Q)=R_{\mu \lambda} (Q)$ is defined by
\begin{align}
R_{\lambda \mu } (Q)%= \prod_{i.j=1}^{\infty} (1 - Q g^{i+j-\lambda_j - \mu_i -1}),\cr
=\text{PE} \left[ - \frac{g}{(1-g)^2} Q \right]
\times N_{\lambda^t \mu} (Q),
\end{align}
with PE representing the Plethystic exponential \eqref{eq:PE}
 and 
\begin{align}
N_{\lambda \mu} (Q) 
= \prod_{(i,j) \in \lambda} \left( 1 - Q g^{\lambda_i + \mu_j^t -i-j+1} \right)
\prod_{(i,j) \in \mu} \left( 1 - Q g^{-\lambda^t_j - \mu_i + i + j - 1} \right). 
\end{align}
Recall that the Nekrasov partition function is expressed as the following weighted sum: 
\begin{align}
\CZ = Z_{\text{pert}} \cdot \bigg(1 + \sum_{k=1}^{\infty}q^kZ_k\bigg) ,
\end{align}
where $Z_{\text{pert}}$ is the perturbative partition function, while $Z_k$ stands for the $k$-instanton partition function. 
The perturbative part of the partition function $Z_{\text{pert}}$ comes from the summand of \eqref{eq:znek-su6} at empty Young diagrams, i.e., %$(Y_1, Y_2, Y_3, Y_4, Y_5, Y_6) = (\varnothing,\varnothing,\varnothing,\varnothing,\varnothing,\varnothing)$. 
$(Y_1, Y_2, Y_3, Y_4, Y_5, Y_6) = (\o,\o,\o,\o,\o,\o)$.
It is given by
\begin{align}
\begin{split}
Z_{\rm pert} 
=& \,
Z_{\text{left}}(\o,\o,\o,\o,\o,\o)
Z_{\text{right}}(\o,\o,\o,\o,\o,\o)
\cr
=&\, 
\text{PE} \Biggl[ 
\frac{2g}{(1-g)^2} 
\Bigl( \frac{A_1}{A_6} + \frac{1}{A_1A_6^2} + A_1^2  A_6 
+ \sum_{2 \le i <  j \le 5} \frac{A_i}{A_j}
- \sum_{i=2}^5 A_1 A_i  A_6
\Bigr)
\Biggr]
\cr 
& 
\times \bigg(\sum_{Y'} ( - A_1{}^{-1} A_6{}^{-2})^{|Y'|} \ 
g^{\frac{\Vert Y'^t \Vert ^2+\Vert Y'\Vert ^2}{2}} \tilde{Z}_{Y'}(g)^2 f_{Y'}^2(g)
\cr 
& \qquad \qquad
\textstyle N_{Y'^t \o }^{-1} (A_1{}^{-1} A_6{}^{-2})
 N^{-1}_{Y' \o } (A_1^2  A_6)\prod_{i=2}^5 N_{Y' \o } (A_1 A_i  A_6)\bigg)^2 ,\label{eq:zpert-top}
\end{split}
\end{align}
where the last two lines can be  combined into the following closed-form expression:
\begin{align}
  \text{PE} \Biggl[ 
    \frac{2g}{(1-g)^2} 
    \Bigl(\sum_{i=2}^5  \frac{A_1}{A_i} +\sum_{i=2}^5  \frac{A_i}{A_6} - \frac{1}{A_1A_6^2} - A_1^2  A_6 
    - \sum_{2\leq i<j \leq 5} A_1 A_i  A_j + \mathcal{O}(A_1^6)
    \Bigr)
    \Biggr].
\end{align}
We note here that when performing the Young diagram sum over $Y'$ in \eqref{eq:zpert-top} to compute the $Z_{\rm pert}$, we expand \eqref{eq:zpert-top} in terms of $A_1$ and, by $\mathcal{O}(A_1^6)$, we mean that the obtained result is explicitly compared up to $\mathcal{O}(A_1^6)$. As it is very unlikely that there will be a new term which suddenly appears  in higher orders than 6 in $A_1$, we believe that there are no further terms for $\mathcal{O}(A_1^6)$.  
It is clear then that \eqref{eq:zpert-top} is manifestly consistent with the equivariant index  \cite{Shadchin:2005mx} for 5d $SU(6)$ gauge theory with a hypermultiplet in the rank-3 antisymmetric representation, %$\mathbf{20}$, 
i.e., 
\begin{align}
  \label{eq:SU6pert}
  Z_{\rm pert}  = \text{PE} \Biggl[ 
    \frac{2g}{(1-g)^2} 
    \Bigl(\sum_{1\leq i<j \leq 6}  \frac{A_i}{A_j} 
    - \sum_{2\leq i<j \leq 6} A_1 A_i  A_j %+ \mathcal{O}(A_1^6)
    \Bigr)
    \Biggr].
\end{align}
The 1-instanton partition function $Z_1$ can be obtained from the summands of \eqref{eq:znek-su6} at Young diagrams satisfying $\sum_{i=1}^6 |Y_i|=1$. There are 6 different profiles of Young diagrams. The configuration $|Y_i| = 1$ and %$Y_{j\neq i} = \varnothing$
$Y_{j\neq i} = \o$ contribute to $Z_{1}$ as
\begin{align}
  + \frac{g}{(1-g)^2} \frac{A_i^6}{\prod_{j \neq i} (A_i - A_j)^2} 
  \Bigl(-A_i \sum_{j\neq i} A_j +  \sum_{j\neq i}\frac{1}{A_j}  - \frac{1}{A_i} + A_i^2
 \Bigr)^2. 
\end{align}
Summing over all six contributions, one finds
\begin{align}
  Z_1 = \sum_{i=1}^6 \frac{g}{(1-g)^2} \frac{A_i^6}{\prod_{j \neq i} (A_i - A_j)^2} 
  \Bigl(-A_i \sum_{j\neq i} A_j +  \sum_{j\neq i}\frac{1}{A_j}  - \frac{1}{A_i} + A_i^2
 \Bigr)^2. 
\end{align}
which is in agreement with our general 1-instanton formula \eqref{eq:1inst-formula}.

We checked that upon imposing the Higgsing condition \eqref{eq:SU6-Higgs-SU3}, i.e., $a_1 + a_5 + a_6 = 0$ and $a_2 + a_3 + a_4 = 0$, the 1-loop contribution \eqref{eq:SU6pert} can be factorized into a product of two $SU(3)$ vector multiplet indices \eqref{eq:1-loop-vec}. We also confirmed that the instanton corrections $Z_1$ and $Z_2$ obtained from the blowup recursion formulae \eqref{eq:recursion} with \eqref{eq:sun-nf-range} become
\begin{align}
  \begin{split}
  Z_1^{SU(6)_{3}+1{\bf TAS}}\big|_{\rm Higgsing}~&\rightarrow ~~ Z_1^{SU(3)_{3}}
  \!(A_1, A_5,A_6)+Z_1^{SU(3)_{3}}\!(A_2, A_3,A_4),\\
  Z_2^{SU(6)_{3}+1{\bf TAS}}\big|_{\rm Higgsing}~&\rightarrow ~~ Z_2^{SU(3)_{3}}\!(A_1, A_5,A_6)+Z_2^{SU(3)_{3}}\!(A_2, A_3,A_4) \\
  &\quad+ Z_1^{SU(3)_{3}}\!(A_1, A_5,A_6)\cdot Z_1^{SU(3)_{3}}\!(A_2, A_3,A_4),
  \end{split}
\end{align}
which satisfy the expected Higgsing relation \eqref{eq:SU6-Higgs-SU3}. Here, $Z_n^{SU(3)_3}$ is the Young
diagram formula \eqref{eq:SUn-young} which includes the Coulomb VEV independent contribution $Z_{\rm extra}(q)$.

%%%%%%%%%%%%%%%%%%%%%%%%%%%%%%%%%%%%%%%%%%%%%%%%%%%%
\section{Conclusion} \label{sec:conclusion}

In this paper, we have found the blowup equations for the Nekrasov partition function that hold for a large set of 4d and 5d gauge theories. 
We listed the theories that are determined via the blowup equations in Table \ref{tbl:list}, and tested against various examples in Section \ref{sec:example}. In particular, the blowup formula enables us to compute instanton partition functions for `exceptional' theories whose ADHM description is not known. 
One of the remarkable aspects of the blowup formulae is that the instanton part of the partition function is completely determined via the perturbative part of the partition function. Let us make a couple of comments on future directions. 

First, we have not given a fully general condition for the blowup formula to hold in the case of 5d $SU(N)$ gauge theory. For the case of 4d $\CN=2$ theory, the general conditions for arbitrary gauge theory is given by the selection rule obtained from an unbroken subgroup of $U(1)_R$ symmetry. It would be desirable to find an analogous explanation for $d_{\text{max}}$ in 5d $SU(N)$ theories. 

Secondly, there must be a broader set of blow-up relations for the 5d Nekrasov partition function $\CZ$, similar to those recently found for topological string partition functions and 6d minimal SCFTs \cite{Grassi:2016nnt, Gu:2017ccq, Huang:2017mis, Gu:2018gmy,Gu:2019dan}. We expect that there exists recursion formulae, derived from the generalized blow-up equations, realize different string theory embeddings of the gauge theory. It would be very interesting if one can reveal the connection between the choice of UV embedding and the blow-up equations. In this way, it may be possible to determine the partition function even for the theories that we are not able to fix in the current paper. 

Finally, we remark that though our blow-up formula is applicable to a fairly large set of theories that contain hypermultiplets of various representations, it is not clear how to implement our blow-up formula to theories with half-hypermultiplets.  There exist many interesting gauge theories with half-hypermultiplets, such as trifundamentals in generalized $SU(2)$ quiver gauge theories \cite{Gaiotto:2009we} that appear in AGT correspondence \cite{Alday:2009aq} or bifundamentals in $SO-Sp$ quiver theories. To the best of our knowledge, instanton counting with half-hypermultiplet has not been studied except for \cite{Hollands:2010xa, Hollands:2011zc} some time ago, and there is recent progress in \cite{Coman:2019eex}. Our blowup formula is naturally written in terms of the representation of a full hypermultiplet, therefore it is not obvious how to incorporate half-hypermultiplet. It would be interesting to develop a way to do instanton counting for half-hypermultiplets as well.

\acknowledgments
We thank Babak Haghighat, Saebyeok Jeong, Hee-Cheol Kim, Seok Kim and Jaemo Park for discussion. 
JK and JS thank the hospitality of UESTC where part of this work is done. 
The work of KL and JS is partly done at the 2019 Pollica summer workshop, which was supported in part by the Simons Foundation (Simons Collaboration on the Non-perturbative Bootstrap) and in part by the INFN. KL and JS are grateful for this support.
SSK and KHL also thank APCTP for hosting the Focus program ``Strings, Branes and Gauge Theories 2019," where part of the work is done.
JK, KL, and JS also thank the Simons Center for Geometry and Physics during the Simons Summer Workshop 2019 where part of this work is done. 
This work is supported in part by the UESTC Research Grant A03017023801317 (SSK), the National Research Foundation of Korea (NRF) Grants 2017R1D1A1B06034369 (KL, JS), and 2018R1A2B6004914 (KHL).

%\pagebreak

%%%%%%%%%%%%%%%%%%%%%%%%%%%%%%%%%%%%%%%%%%%%%%%%%%%%%%%%
\appendix
\section{One-instanton partition functions}
\label{sec:data}

This appendix collects the character expansion of the 1-instanton partition function $Z_1$ for a variety of 5d $\mathcal{N}=1$ gauge theories. For simplicity, we display the $\tilde{Z}_1 \equiv (2\sinh{\frac{\e_{1,2}}{2}}) \cdot Z_1$ which takes off the center-of-mass factor. They are written in terms of irreducible characters $\chi_{\mathbf{R}}^S$, whose superscript $S \in \{G,v,s,c,f,\bar{f}\}$ indicates the gauge symmetry ($G$) or the flavor symmetry acting on the vector ($v$), spinor ($s$), conjugate spinor ($c$), fundamental ($f$), or anti-fundamental ($\bar{f}$) hypermultiplets. The representation $\mathbf{R}$ of an irreducible character $\chi_{\mathbf{R}}^S$ is specified by its Dynkin label.\footnote{In this paper, we follow the convention of  LieART \cite{Feger:2012bs} to denote the Dynkin label of a representation $\mathbf{R}$.} An irreducible character for the flavor symmetry is assumed to be in the orthogonal basis, such that it can be consistent with the mass parameters $m_\ell$ introduced in Section~\ref{sec:blowup}. We will often distinguish the mass parameters by the superscript $S \in \{s,c,v,f,\bar{f}\}$ according to the matter representation. 

\paragraph{SO(8)} The flavor symmetry acting on $N_{\bf s} \mathbf{S} + N_{\bf c} \mathbf{C} + N_{\bf v} \mathbf{V}$ matter multiplets is given by $Sp(N_{\bf s})_s \times Sp(N_{\bf c})_c \times Sp(N_{\bf v})_v$. For $(N_{\bf s}, N_{\bf c}, N_{\bf v}) = (3,1,0)$, the character expansion of the 1-instanton result $\tilde{Z}_1$ is 
\begin{align}
  \label{eq:so8-s3c1}
\tilde{Z}_1=\sum_{n=0}^{\infty}&\,\Big(t^{5+2n}\chi^G_{(0n00)}\chi^s_{(001)}\chi^v_{(1)}-t^{6+2n}(\chi^G_{(0n01)}\chi^s_{(010)}\chi^v_{(1)}+\chi^G_{(1n00)}\chi^s_{(001)})\nn\\
&+\,t^{7+2n}(\chi^G_{(1n01)}\chi^s_{(010)}+\chi^G_{(0n02)}\chi^s_{(100)}\chi^v_{(1)})\\
&-t^{8+2n}(\chi^G_{(1n02)}\chi^s_{(100)}+\chi^G_{(0n03)}\chi^v_{(1)})+t^{9+2n}\chi^G_{(1n03)}\Big),\nn
\end{align}
which was compared with the closed-form expression \eqref{eq:1inst-formula} up to $t^{20}$ order.
We checked that the 1-instanton partition functions ${Z}_1$ from \eqref{eq:1inst-formula} for $(N_\mathbf{s},N_\mathbf{c}, N_{\mathbf{v}})= (3,1,0)$ and $(1,3,0)$ could be interchanged as follows: 
\begin{align}
  Z_{1}^{(N_{\bf s}, N_{\bf c}, N_{\bf v}) = (1,3,0)}(a_1,a_2,a_3,a_4) = Z_{1}^{(N_{\bf s}, N_{\bf c}, N_{\bf v}) = (3,1,0)}(a_1,a_2,a_3,-a_4).
\end{align}
The $SO(8)$ triality \eqref{eq:so8-triality} was also confirmed as in Section~\ref{subsec:spinor}. Namely, we found that 
\begin{align}
  \begin{split}
  Z_1^{(N_{\bf s}, N_{\bf c}, N_{\bf v}) = (3,1,0)}\,(\vec{a}, \epsilon_1,\epsilon_2;\vec{m}^s,\vec{m}^c,0)  &= Z_1^{(N_{\bf s}, N_{\bf c}, N_{\bf v}) = (0,3,1)}\,(\vec{a}', \epsilon_1,\epsilon_2;0,\vec{m}^s,\vec{m}^c)|_{\vec{a}' \rightarrow \vec{a}}\\
  &= Z_1^{(N_{\bf s}, N_{\bf c}, N_{\bf v}) = (1,0,3)}\,(\vec{a}'', \epsilon_1,\epsilon_2;\vec{m}^c,0,\vec{m}^s)|_{\vec{a}'' \rightarrow \vec{a}},
  \end{split}\\
  \begin{split}
    Z_1^{(N_{\bf s}, N_{\bf c}, N_{\bf v}) = (1,3,0)}\,(\vec{a}, \epsilon_1,\epsilon_2;\vec{m}^s,\vec{m}^c,0)  &= Z_1^{(N_{\bf s}, N_{\bf c}, N_{\bf v}) = (0,1,3)}\,(\vec{a}', \epsilon_1,\epsilon_2;0,\vec{m}^s,\vec{m}^c)|_{\vec{a}' \rightarrow \vec{a}}\\
    &= Z_1^{(N_{\bf s}, N_{\bf c}, N_{\bf v}) = (3,0,1)}\,(\vec{a}'', \epsilon_1,\epsilon_2;\vec{m}^c,0,\vec{m}^s)|_{\vec{a}'' \rightarrow \vec{a}}.
    \end{split}
\end{align}
The character expansion for other $SO(8)$ theories with less number of hypermultiplets can be obtained from \eqref{eq:so8-s3c1} by decoupling some mass parameters to infinity. It was checked that the general 1-instanton formula \eqref{eq:1inst-formula} agrees with that.

\paragraph{SO(10)} The flavor symmetry acting on $N_{\bf s} \mathbf{S} + N_{\bf c} \mathbf{C} + N_{\bf v} \mathbf{V}$ hypermultiplets is $U(N_{\bf s} + N_{\bf c})_s \times Sp(N_{\bf v})_v$, reflecting  that the $SO(10)$ (conjugate) spinor representation is complex. For $N_{\bf s} + N_{\bf c} = 2$ and $N_{\bf v} = 2$, the character expansion of $\tilde{Z}_1$ is given by 
\begin{align}
  \label{eq:so10-2s2v}
\tilde{Z}_1&=\,t^5(\chi^s_{(2)_0}+\chi^v_{(01)})+t^6(\chi^s_{(2)_{-2}}+\chi^s_{(2)_2})-t^7(\chi^G_{(00001)}\chi^s_{(1)_{-1}}\chi^v_{(10)}+\chi^G_{(00010)}\chi^s_{(1)_1}\chi^v_{(10)}+\chi^G_{(01000)}\nn\\
&+\chi^G_{(10000)}(\chi^s_{(2)_{-2}}+\chi^s_{(2)_2}))+t^8(\chi^G_{(00100)}\chi^v_{(10)}+\chi^G_{(10001)}\chi^s_{(1)_{-1}}+\chi^G_{(10010)}\chi^s_{(1)_1})-t^9\chi^G_{(10100)}\nn\\
&\textstyle+\sum_{n=0}^{\infty}\Big(t^{7+2n}\chi^G_{(0n000)}(\chi^s_{(0)_{-4}}+\chi^s_{(0)_4}+\chi^s_{(4)_0})\chi^v_{(01)} -t^{8+2n}(\chi^G_{(0n001)}(\chi^s_{(1)_{-3}}+\chi^s_{(3)_1})\chi^v_{(01)} \nn\\&\qquad\quad\qquad\quad
+\chi^G_{(0n010)}(\chi^s_{(1)_3}+\chi^s_{(3)_{-1}})\chi^v_{(01)} +\chi^G_{(1n000)}(\chi^s_{(0)_{-4}}+\chi^s_{(0)_4}+\chi^s_{(4)_0})\chi^v_{(10)})\nn\\
&\qquad\quad+t^{9+2n}(\chi^G_{(0n100)}(\chi^s_{(2)_{-2}}+\chi^s_{(2)_2})\chi^v_{(01)}+\chi^G_{(0n002)}\chi^s_{(0)_{-2}}\chi^v_{(01)}+\chi^G_{(0n020)}\chi^s_{(0)_2}\chi^v_{(01)}\nn\\
&\qquad\qquad\qquad+\chi^G_{(0n011)}\chi^s_{(2)_0}\chi^v_{(01)}+\chi^G_{(1n001)}(\chi^s_{(1)_{-3}}+\chi^s_{(3)_1})\chi^v_{(10)}+\chi^G_{(1n010)}(\chi^s_{(1)_3}+\chi^s_{(3)_{-1}})\chi^v_{(10)}\nn\\
&\qquad\qquad\qquad+\chi^G_{(2n000)}(\chi^s_{(0)_{-4}}+\chi^s_{(0)_4}+\chi^s_{(4)_0}))\nn\\
&\qquad\quad-t^{10+2n}(\chi^G_{(0n101)}\chi^s_{(1)_{-1}}\chi^v_{(01)}+\chi^G_{(0n110)}\chi^s_{(1)_1}\chi^v_{(01)}+\chi^G_{(1n100)}(\chi^s_{(2)_{-2}}+\chi^s_{(2)_2})\chi^v_{(10)}\nn\\
&\qquad\qquad\qquad+\chi^G_{(1n002)}\chi^s_{(0)_{-2}}\chi^v_{(10)}+\chi^G_{(1n020)}\chi^s_{(0)_2}\chi^v_{(10)}+\chi^G_{(1n011)}\chi^s_{(2)_0}\chi^v_{(10)}\nn\\
&\qquad\qquad\qquad+\chi^G_{(2n001)}(\chi^s_{(1)_{-3}}+\chi^s_{(3)_1})+\chi^G_{(2n010)}(\chi^s_{(1)_3}+\chi^s_{(3)_{-1}})\nn\\
&\qquad\quad+t^{11+2n}(\chi^G_{(1n200)}\chi^v_{(01)}+\chi^G_{(1n101)}\chi^s_{(1)_{-1}}\chi^v_{(10)}+\chi^G_{(1n110)}\chi^s_{(1)_1}\chi^v_{(10)}+\chi^G_{(2n100)}(\chi^s_{(2)_{-2}}+\chi^s_{(2)_2})\nn\\
&\qquad\qquad\qquad+\chi^G_{(2n002)}\chi^s_{(0)_{-2}}+\chi^G_{(2n020)}\chi^s_{(0)_2}+\chi^G_{(2n011)}\chi^s_{(2)_0})\nn\\
&\qquad\quad-t^{12+2n}(\chi^G_{(1n200)}\chi^v_{(10)}+\chi^G_{(2n101)}\chi^s_{(1)_{-1}}+\chi^G_{(2n110)}\chi^s_{(1)_1})+t^{13+2n}\chi^G_{(2n200)}\Big).
\end{align}
where the $U(2)$ character $\chi^s_{(j)_b}$ is defined as (with $y_{s,i} \equiv e^{-m_i^s}$ and $y_{c,i} \equiv e^{-m_i^c}$ understood)
\begin{align}
  \chi^s_{(j)_b} = \begin{dcases}
    (y_{s,1}\,y_{s,2})^{b/2} \cdot \textstyle\sum_{a=0}^{j} \big({y_{s,1}}/{y_{s,2}}\big)^{-j/2+a}& \text{ for } \quad (N_\mathbf{s}, N_\mathbf{c})=(2,0),\\
    (y_{s,1}/y_{c,1})^{b/2} \cdot \textstyle\sum_{a=0}^{j} \big(y_{s,1}\,y_{c,1}\big)^{-j/2+a}& \text{ for } \quad (N_\mathbf{s}, N_\mathbf{c})=(1,1),\\
    (y_{c,1}\,y_{c,2})^{-b/2} \cdot \textstyle\sum_{a=0}^{j} \big(y_{c,1}/y_{c,2}\big)^{-j/2+a}& \text{ for } \quad (N_\mathbf{s}, N_\mathbf{c})=(0,2).
  \end{dcases}
\end{align}
Similarly, for $N_{\bf s} + N_{\bf c} = 3$ and $N_{\bf v} = 0$, the character expansion of $\tilde{Z}_1$ is given by 
\begin{align}
  \label{eq:so10-s3}
\tilde{Z}_1&=\,t^5(\chi^G_{(10000)}+\chi^s_{(02)_{-2}}+\chi^s_{(20)_2})-t^6
(\chi^s_{(01)_{-1}}+\chi^s_{(10)_1})+t^7\chi^G_{(00100)}\nn\\
&+\textstyle\sum_{n=0}^{\infty}\Big(t^{7+2n}(\chi^G_{(0n000)}(\chi^s_{(00)_{-6}}+\chi^s_{(00)_6}+\chi^s_{(40)_{-2}}+\chi^s_{(04)_2}))\nn\\
&\qquad \quad \ -t^{8+2n}(\chi^G_{(0n001)}(\chi^s_{(10)_{-5}}+\chi^s_{(03)_3}+\chi^s_{(31)_{-1}})+\chi^G_{(0n010)}(\chi^s_{(01)_5}+\chi^s_{(30)_{-3}}+\chi^s_{(13)_1}))\nn\\
&\qquad \quad \ +t^{9+2n}(\chi^G_{(0n100)}(\chi^s_{(20)_{-4}}+\chi^s_{(02)_4}+\chi^s_{(22)_0})+\chi^G_{(0n011)}(\chi^s_{(21)_{-2}}+\chi^s_{(12)_2})\nn\\
& \qquad \qquad\qquad+\chi^G_{(0n002)}(\chi^s_{(01)_{-4}}+\chi^s_{(30)_0})+\chi^G_{(0n020)}(\chi^s_{(10)_4}+\chi^s_{(03)_0}))\\
&\qquad \quad \ -t^{10+2n}(\chi^G_{(0n101)}(\chi^s_{(11)_{-3}}+\chi^s_{(21)_1})+\chi^G_{(0n110)}(\chi^s_{(11)_3}+\chi^s_{(12)_{-1}})\nn\\
& \qquad \qquad\qquad+\chi^G_{(0n003)}\chi^s_{(00)_{-3}}+\chi^G_{(0n030)}\chi^s_{(00)_3}+\chi^G_{(0n012)}\chi^s_{(20)_{-1}}+\chi^G_{(0n021)}\chi^s_{(02)_1})\nn\\
&\qquad \quad \ +t^{11+2n}(\chi^G_{(0n200)}(\chi^s_{(02)_{-2}}+\chi^s_{(20)_2})+\chi^G_{(0n102)}\chi^s_{(10)_{-2}}+\chi^G_{(0n120)}\chi^s_{(01)_2}+\chi^G_{(0n111)}\chi^s_{(11)_0})\nn\\
&\qquad \quad \ -t^{12+2n}(\chi^G_{(0n201)}\chi^s_{(01)_{-1}}+\chi^G_{(0n210)}\chi^s_{(10)_1})+t^{13+2n}\chi_{(0n300)}\Big),\nn
\end{align}
where the $U(3)$ character $\chi^s_{(mn)_c}$ is defined as
\begin{align}
  \label{eq:u3-flavor-char}
  \chi^s_{(mn)_c} =
    (w_1 w_2w_3)^{\frac{c-m+n}{3}}\bigg( \sum_{\substack{1\leq i_1\leq\cdots\leq i_m\leq3\\1\leq j_1\leq\cdots\leq j_n\leq 3}}\frac{w_{i_1}\cdots w_{i_m}}{w_{j_1}\cdots w_{j_n}}-\sum_{\substack{1\leq i_1\leq\cdots\leq i_{m-1}\leq3\\1\leq j_1\leq\cdots\leq j_{n-1}\leq 3}}\frac{w_{i_1}\cdots w_{i_{m-1}}}{w_{j_1}\cdots w_{j_{n-1}}}\bigg),
\end{align}
with
\begin{align}
(w_1,w_2,w_3)=
\begin{cases}
(y_{s,1},y_{s,2},y_{s,3}) & \text{for}\quad (N_\mathbf{s}, N_\mathbf{c})=(3,0),\\
(y_{s,1},y_{s,2},y_{c,1}^{-1}) & \text{for}\quad (N_\mathbf{s}, N_\mathbf{c})=(2,1),\\
(y_{s,1},y_{c,1}^{-1},y_{c,2}^{-1}) & \text{for}\quad (N_\mathbf{s}, N_\mathbf{c})=(1,2),\\
(y_{c,1}^{-1},y_{c,2}^{-1},y_{c,3}^{-1}) & \text{for}\quad (N_\mathbf{s}, N_\mathbf{c})=(0,3).
\end{cases}
\end{align}
Again, \eqref{eq:so10-2s2v} and \eqref{eq:so10-s3} was tested against the closed-form expression \eqref{eq:1inst-formula} up to $t^{20}$ order.

\paragraph{SO(11)} The flavor symmetry acting on $N_{\bf s} \mathbf{S} + N_{\bf v} \mathbf{V}$ hypermultiplets is $SO(2N_{\bf s})_s \times  Sp(N_{\bf v})_v$. For $N_{\bf s}=1$ and  $N_{\bf v} = 3$, the character expansion of  $\tilde{Z}_1$ can be written as
\begin{align}
  \label{eq:so11-s1v3}
\tilde{Z}_1&=\,t^5+t^6(\chi^v_{(001)}+(y_s^2+y_s^{-2})\chi^v_{(100)})+t^7((y_s^2+y_s^{-2}+1)\chi^v_{(010)}-(y_s^{2}+y_s^{-2})\chi^G_{(10000)})\nn\\
&\,-t^8((y_s+y_s^{_1})\chi^G_{(00001)}\chi^v_{(010)}+(y_s^2+y_s^{-2}+1)\chi^G_{(10000)}\chi^v_{(100)}+\chi^G_{(01000)}\chi^v_{(100)})\nn\\
&\,+t^9(\chi^G_{(00100)}\chi^v_{(010)}+\chi^G_{(10001)}(y_s+y_s^{-1})\chi^v_{(100)}+\chi^G_{(20000)}(y_s^2+y_s^{-2}+1)+\chi^G_{(11000)})\nn\\
&\,-t^{10}(\chi^G_{(10100)}\chi^v_{(100)}+\chi^G_{(20001)}(y_s+y_s^{-1}))+t^{11}\chi^G_{(20100)}\\
&\,+\sum_{n=0}^{\infty}\Big(t^{8+2n}(\chi^G_{(0n000)}(y_s^4+y_s^{-4}+1)\chi^v_{(001)})\nn\\
&\quad\qquad\ -t^{9+2n}(\chi^G_{(0n001)}(y_s^3+y_s^{-3})\chi^v_{(001)}+\chi^G_{(0n001)}(y_s+y_s^{-1})\chi^v_{(001)}+\chi^G_{(1n000)}(y_s^4+y_s^{-4}+1)\chi^v_{(010)})\nn\\
&\qquad\quad\ +t^{10+2n}(\chi^G_{(0n010)}(y_s^2+y_s^{-2})\chi^v_{(001)}+\chi^G_{(0n100)}(y_s^2+y_s^{-2}+1)\chi^v_{(001)}+\chi^G_{(0n002)}\chi^v_{(001)}\nn\\
&\qquad\qquad\qquad\ +\chi^G_{(1n001)}(y_s^3+y_s+y_s^{-1}+y_s^{-3})\chi^v_{(010)}+\chi^G_{(2n000)}(y_s^4+y_s^{-4}+1)\chi^v_{(100)})\nn\\
&\qquad\quad\ -t^{11+2n}(\chi^G_{(0n101)}(y_s+y_s^{-1})\chi^v_{(001)}+\chi^G_{(1n100)}(y_s^2+y_s^{-2}+1)\chi^v_{(010)}+\chi^G_{(1n010)}(y_s^2+y_s^{-2})\chi^v_{(010)}\nn\\
&\qquad\qquad\qquad\ +\chi^G_{(1n002)}\chi^v_{(010)}+\chi^G_{(2n001)}(y_s^3+y_s+y_s^{-1}+y_s^{-3})\chi^v_{(100)}+\chi^G_{(3n000)}(y_s^4+y_s^{-4}+1))\nn\\
&\qquad\quad\ +t^{12+2n}(\chi^G_{(0n200)}\chi^v_{(001)}+\chi^G_{(1n101)}(y_s+y_s^{-1})\chi^v_{(010)}+\chi^G_{(2n100)}(y_s^2+y_s^{-2}+1)\chi^v_{(100)}\nn\\
&\qquad\qquad\qquad\ +\chi^G_{(2n010)}(y_s^2+y_s^{-2})\chi^v_{(100)}+\chi^G_{(2n002)}\chi^v_{(100)}+\chi^G_{(3n001)}(y_s^3+y_s+y_s^{-1}+y_s^{-3}))\nn\\
&\qquad\quad\ -t^{13+2n}(\chi^G_{(1n200)}\chi^v_{(010)}+\chi^G_{(2n101)}(y_s+y_s^{-1})\chi^v_{(100)}+\chi^G_{(3n100)}(y_s^2+y_s^{-2}+1)\nn\\
&\qquad\qquad\qquad\ +\chi^G_{(3n010)}(y_s^2+y_s^{-2})+\chi^G_{(3n002)})\nn\\
&\qquad\quad\ +t^{14+2n}(\chi^G_{2n200)}\chi^v_{(100)}+\chi^G_{(3n101)}(y_s+y_s^{-1}))-t^{15+2n}\chi^G_{(3n200)}\Big). \nn
\end{align}
which was compared with the closed-form expression \eqref{eq:1inst-formula} up to $t^{20}$ order.

\paragraph{SO(12)} 
The flavor symmetry acting on $N_{\bf s} \mathbf{S} +N_{\bf c} \mathbf{C} + N_{\bf v} \mathbf{V}$ hypermultiplets is $SO(2N_{\bf s})_s \times SO(2N_{\bf c})_c \times  Sp(N_{\bf v})_v$. Here we turn off the Coulomb VEV $\vec{a} = 0$ for simplicity. The character expansion of  $\tilde{Z}_1$ at $(N_{\bf s},N_{\bf c}, N_{\bf v}) = (2,0,0)$ can be written as
\begin{align}
  \label{eq:so12-s2}
  \tilde{Z}_1&= \frac{t^{18}}{(1-t^2)^{18}}\Big(
  -96096\, (\chi^s_{(13)}+\chi^s_{(31)})\cdot {(7t^4+42t^2+72+42t^{-2}+7t^{-4})}\\
  &+10010\,(\chi^s_{(24)}+\chi^s_{(42)})\cdot {(9t^5+88t^3+243t+243t^{-1}+88t^{-3}+9t^{-5})}\nn\\
  &-352\,(\chi^s_{(15)}+\chi^s_{(51)})\cdot {(25t^6+474t^4+2169t^2+3504+2169t^{-2}+474t^{-4}+25t^{-6})}\nn\\
  &-2464\,(\chi^s_{(35)}+\chi^s_{(53)})\cdot {(2t^6+27t^4+108t^2+168+108t^{-2}+27t^{-4}+2t^{-6})}\nn\\
  &+11\,(\chi^s_{(06)}+\chi^s_{(60)})\cdot {(42t^7+1194t^5+8451t^3+21253t+21253t^{-1}+\cdots+42t^{-7})}\nn\\
  &+11\,(\chi^s_{(26)}+\chi^s_{(62)})\cdot (45t^7+1101t^5+6983t^3+16623t+16623t^{-1} + \cdots + 45t^{-7})\nn\\
  &-32\,(\chi^s_{(17)}+\chi^s_{(71)})\cdot (t^8+36t^6+336t^4+1176t^2+17641+1176t^{-2}+\cdots+t^{-8})\nn\\
  &+99\,\chi^s_{(22)}\cdot {(5t^9-90t^7+1623t^5+26743t^3+83103t+ (t\rightarrow t^{-1}))}\nn\\
  &+462\,(\chi^s_{(02)}+\chi^s_{(20)})\cdot {(t^{9}-18t^7+153t^5+4059t^3+13485t+ (t\rightarrow t^{-1}))}\nn\\
  &-32\,\chi^s_{(33)}\cdot {(t^{10}-18t^8+450t^6+13340t^4+66977t^2+110772+66977t^{-2}+\cdots+t^{-10})}\nn\\
  &+\chi^s_{(44)}\cdot {(t^{11}-18t^9+615t^7+26332t^5+187749t^3+466001t^{1}+466001t^{-1}+\cdots+t^{-11})}\nn\\
  &+(\chi^s_{(04)}+\chi^s_{(40)})\cdot (t^{13}-18t^{11}+153t^9-816t^7+58115t^5+730170t^{3}+2129595t^{1}+(t\rightarrow t^{-1}))\nn\\
  &-352\,\chi^s_{(11)}\cdot{(t^{10}-4t^8-99t^6+2496t^4+18246t^2+32976+18246t^{-2}+\cdots+t^{-10})}\nn\\
  &+(\chi^s_{(08)}+\chi^s_{(80)})\cdot {(t^9+48t^7+603t^5+2898t^3+6174t+(t\rightarrow t^{-1}))}\nn\\
  &+(t^{17}-18t^{15}+153t^{13}-739t^{11}+3753t^9-20195t^{7}+49881t^{5}+1203597t^{3}+4481279t^{1}+(t\rightarrow t^{-1}))\Big).\nn
\end{align}
It was explicitly checked that the 1-instanton partition function $Z_1$ at $(N_{\bf s},N_{\bf c}, N_{\bf v}) = (0,2,0)$ could be identified with the above as
\begin{align}
  {Z}_{1}^{(N_{\mathbf{s}},N_\mathbf{c},N_{\mathbf{v}}) = (0,2,0)}(a_1,a_2,a_3,a_4,a_5,a_6) = {Z}_{1}^{(N_{\mathbf{s}},N_\mathbf{c},N_{\mathbf{v}}) = (2,0,0)}(a_1,a_2,a_3,a_4,a_5,-a_6).
\end{align}
Similarly, the character expansion of $\tilde{Z}_1$ at $(N_{\bf s},N_{\bf c}, N_{\bf v}) = (1,1,0)$ can be displayed as follows:
\begin{align}
  \label{eq:so12-s1c1}
  \tilde{Z}_1&=\frac{t^{18}}{(1-t^2)^{18}}\sum_{\pm}\Big(
  -2462\,(y_s^{\pm1}y_c^{\pm4}+y_s^{\pm4}y_c^{\pm1})\cdot{(2t^6+27t^4+108t^2+168+108t^{-2}+27t^{-4}+2t^{-6})}\nn\\
  &+11\,(y_s^{\pm2}y_c^{\pm4}+y_s^{\pm4}y_c^{\pm2})\cdot{(45t^7+1101t^5+6983t^3+16623t+ (t\rightarrow t^{-1}))}\nn\\
  &+44\,(y_s^{\pm3}y_c^{\pm3})\cdot{(23t^7+587t^5+3925t^3+9609t+ (t\rightarrow t^{-1}))}\\
  &+44\,(y_s^{\pm1}y_c^{\pm3}+y_s^{\pm3}y_c^{\pm1})\cdot{(23t^7+2927t^5+26025t^3+70033t+(t\rightarrow t^{-1}))}\nn\\
  &-32\,(y_s^{\pm3}y_c^{\pm4}+y_s^{\pm4}y_c^{\pm3})\cdot{(t^8+36t^6+336t^4+1176t^2+1764+1176t^{-2}+\cdots+t^{-8})}\nn\\
  &-32\,(y_s^{\pm2}y_c^{\pm3}+y_s^{\pm3}y_c^{\pm2})\cdot{(t^8+465t^6+7629t^4+33351t^2+53244+33351t^{-2}+\cdots+t^{-8})}\nn\\
  &+(y_s^{\pm4}y_c^{\pm4})\cdot {(t^9+48t^7+603t^5+2898t^3+6174t+(t\rightarrow t^{-1}))}\nn\\
  &-32\,(y_s^{\pm1}y_c^{\pm2}+y_s^{\pm2}y_c^{\pm1})\cdot (t^{10}-17t^8+1069t^6+44069t^4+234770t^2+393168+234770t^{-2} +\cdots+t^{-10})\nn\\
  &-32\,(y_s^{\pm3}+y_c^{\pm3})\cdot{(t^{10}-17t^8+750t^6+17526t^4+83553t^2+136714+83358t^{-2}+\cdots+t^{-10})}\nn\\
  &-32\,(y_s^{\pm1}+y_c^{\pm1})\cdot (13t^{10}-79t^8+408t^6+97724t^4+587351t^2+1011546+587351t^{-2}+\cdots +13t^{-10}))\nn\\
  &+(y_s^{\pm4}+y_c^{\pm4})\cdot{(t^{11}-18t^9+615t^7+26332t^5+187749t^3+466001t+(t\rightarrow t^{-1}))}\nn\\
  &\,+(y_s^{\pm2}+y_c^{\pm2}) \cdot (t^{11}+477t^{9}-7305t^7+391411t^5+4750692t^3+13923764t^{1}+ (t\rightarrow t^{-1}))\nn\\
  &+4\,(y_s^{\pm1}y_c^{\pm1})\cdot{(3t^{11}+199t^9+132676t^7+1864041t^5+5630341t^3+5630341t^+(t\rightarrow t^{-1}))}\nn\\
  &+(y_s^{\pm2}y_c^{\pm2})(t^{13}-17t^{11}+1136t^9+804t^7+200385t^5+1971471t^{3}+5450836t+(t\rightarrow t^{-1}))\nn\\
  &+(t^{15}-17t^{13}+214t^{11}+1414t^9-33152t^7+704404t^{5}+11381979t^{3}+35592757t+(t\rightarrow t^{-1}))\Big),\nn
\end{align}
in which $\sum_{\pm}$ notation is understood as follows: $\sum_{\pm}x^{\pm1}y^{\pm1} = xy+xy^{-1}+x^{-1}y+x^{-1}y^{-1}$, $\sum_{\pm}x^{\pm1} = x+x^{-1}$, and $\sum_{\pm} 1 = 1$.

\paragraph{SO(13)} The flavor symmetry  on $N_{\bf s} \mathbf{S}+ N_{\bf v} \mathbf{V}$ matter multiplets is $SO(2N_{\bf s})_s \times  Sp(N_{\bf v})_v$. The character expansion of  $\tilde{Z}_1$ at $(N_{\bf s}, N_{\bf v}) = (1,1)$ is written follows, after setting the Coulomb VEV $\vec{a} = 0$ to keep the expression concise,
\begin{align}
  \label{eq:so13-s1v1}
  \tilde{Z}_1&=\frac{t^{20}}{(1-t^2)^{20}}\sum_{\pm}\Big(y_s^{\pm8}\chi^V_{(1)}\cdot (t^{10}+58 t^8+905 t^6+5580 t^4+15876 t^2+22344++15876 t^{-2} \cdots + t^{-10})\nn\\
  &-64\,y_s^{\pm7}\chi^V_{(1)} \cdot (t^9+45 t^7+540 t^5+2520 t^3+5292 t + (t\rightarrow t^{-1}))\nn\\
  &+26\,y_s^{\pm6}\chi^V_{(1)}\cdot (77 t^8+2541 t^6+22226 t^4+74811 t^2+110770 + 74811 t^{-2} + \cdots + 77 t^{-8})\nn\\
  &-5824\,y_s^{\pm5}\chi^V_{(1)} \cdot (7 t^7+154 t^5+924 t^3+2145 t + (t\rightarrow t^{-1}) )\nn\\
  &+y_s^{\pm4}\chi^V_{(1)} \cdot (t^{14}-19 t^{12}+170 t^{10}+766 t^8+576628 t^6 +7601283 t^4+29870761 t^2\nn\\
  &\qquad\qquad\qquad\qquad\qquad\qquad\qquad\qquad\qquad+46175700+ 29870761
  t^{-2} + 7601283 t^{-4}+\cdots +t^{-14})\nn\\
  &-64\,y_s^{\pm3}\chi^V_{(1)} \cdot (t^{11}-20 t^9+1256 t^7+83074 t^5+628311 t^3+1580032 t+ (t\rightarrow t^{-1}))\nn\\
  &+2002\,y_s^{\pm2}\chi^V_{(1)}\cdot (t^{10}-19 t^8+756 t^6+15006 t^4+66051 t^2+105146+66051 t^{-2} + \cdots + t^{-10})\nn\\
  &-64\,y_s^{\pm1}\chi^V_{(1)} \cdot (13 t^{11}-51 t^9-436 t^7+182670 t^5+1603925 t^3+4218449 t + (t\rightarrow t^{-1}))\nn\\
  &+\chi^V_{(1)} \cdot (t^{16}-19 t^{14}+274 t^{12}+3185 t^{10}-73808 t^8+1918679 t^6+46355974 t^4+212905247t^2\nn\\
  &\qquad\qquad\qquad\qquad\qquad\qquad\qquad\qquad\quad+342439014+212905247t^{-2}+46355974 t^{-4}+\cdots +t^{-16})\nn\\
  &-13\,y_s^{\pm8} \cdot (t^9+35 t^7+365 t^5+1575 t^3+3192 t+ (t\rightarrow t^{-1}))\nn\\
  &+256\,y_s^{\pm7}\cdot (3 t^8+80 t^6+630 t^4+2016
  t^2+2940+2016t^{-2}+\cdots+{3}t^{-8})\nn\\
  &-26\,y_s^{\pm6} \cdot (847 t^7+15989 t^5+89887 t^3+203357
  t+ (t\rightarrow t^{-1}) )\nn\\
  &-64\,y_s^{\pm5}\cdot (t^{10}-20 t^8-6180 t^6-75228 t^4-286725
  t^2-439416-{286725}t^{-2} +\cdots + t^{-10})\nn\\
  &+y_s^{\pm4}\cdot (t^{13}-20 t^{11}+2907 t^9-74785 t^7-4557934 t^5-33690015 t^3-83955034 t +  (t\rightarrow t^{-1}) )\nn\\
  &-64\,y_s^{\pm3}\cdot (t^{12}-19 t^{10}+807 t^8-24636 t^6-510121 t^4-2255129
  t^2-3592422\nn\\
  &\qquad\qquad\qquad\qquad\qquad\qquad\qquad\qquad\quad -{2255129}t^{-2}-510121 t^{-4}-24636 t^{-6} + \cdots + t^{-12})\nn\\
  &+2\,y_s^{\pm2}\cdot (7 t^{13}-140 t^{11}+3189 t^9+86972 t^7-7685485 t^5-71293018 t^3-190116261 t + (t\rightarrow t^{-1}))\nn\\
  &-64\,y_s^{\pm1} \cdot (t^{12}-84 t^{10}+2667 t^8-36526 t^6-1227485 t^4-5926190
  t^2-9643046
    \nn\\
  &\qquad\qquad\qquad\qquad\qquad\qquad\qquad\qquad\quad -{5926190}{t^{-2}}-1227485 t^{-4} -36526 t^{-6}+\cdots + t^{-12})\nn\\
  &+(t^{15}-20 t^{13}-602 t^{11}+5691 t^9+495005 t^7-22183672 t^5-225823570 t^3-617150913t+  (t\rightarrow t^{-1}) )\Big).
  \end{align}

  \paragraph{SO(14)} The classical flavor symmetry  on $N_{\bf s} \mathbf{S}+N_{\bf c} \mathbf{C}+ N_{\bf v} \mathbf{V}$ hypermultiplets is $U(N_{\bf s})_s \times U(N_{\bf c})_c  \times Sp(N_{\bf v})_v$. The character expansion of  $\tilde{Z}_1$ at $(N_{\bf s},N_{\bf c}, N_{\bf v}) = (1,0,2)$ is written as follows, after turning off the $SO(14)$ Coulomb VEV $\vec{a} = 0$,
  \begin{align}
  \label{eq:so14-s1v2}
  \tilde{Z}_1&=\frac{t^{22}}{(1-t^2)^{22}}\sum_{\pm}\Big(y_s^{\pm8}\chi^V_{(01)}\cdot (t^{11}+69 t^9+1309 t^7+10065 t^5+36828 t^3+69300
  t+  (t\rightarrow t^{-1}) )\nn\\
  &-64\,y_s^{\pm7}\chi^V_{(01)}\cdot (t^{10}+55 t^8+825 t^6+4950 t^4+13860
  t^2+19404+13860t^{-2} + \cdots + t^{-10})\nn\\
  &+26\,y_s^{\pm6}\chi^V_{(01)}\cdot (77 t^9+3234 t^7+36667 t^5+164401 t^3+338261
  t+  (t\rightarrow t^{-1}))\nn\\
  &-5824\,y_s^{\pm5}\chi^V_{(01)}\cdot (7 t^8+210 t^6+1694 t^4+5434
  t^2+7920+ 5434t^{-2} + \cdots + 7t^{-8}) \nn\\
  &+y_s^{\pm4}\chi^V_{(01)}\cdot (t^{15}-22 t^{13}+231 t^{11}-1540 t^9+614558 t^7\nn\\
  &\qquad\qquad\qquad\qquad\qquad\qquad\qquad\qquad\quad +11510191 t^5+62671224 t^3+139186397t+  (t\rightarrow t^{-1}))\nn\\
  &-832\,y_s^{\pm3}\chi^V_{(01)}\cdot (33 t^8+7744 t^6+83776 t^4+300104
  t^2+451192+300104t^{-2} + \cdots + 33t^{-8})\nn\\
  &+2002\,y_s^{\pm2}\chi^V_{(01)}\cdot (t^{11}-22 t^9+621 t^7+21262 t^5+134245 t^3+314181t+  (t\rightarrow t^{-1}) )\nn\\
  &-832\,y_s^{\pm1}\chi^V_{(01)}\cdot (t^{12}-t^{10}-231 t^8+15631 t^6+206987 t^4+790240t^2+1207976+ \nn\\
  &\qquad\qquad\qquad\qquad\qquad\qquad\qquad\qquad\quad +790240t^{-2} +206987 t^{-4} +  15631 t^{-6} + \cdots + t^{-12}) \nn\\
  &+\chi^V_{(01)}\cdot (t^{17}-22 t^{15}+335 t^{13}+3179 t^{11}-84595 t^9+1320011 t^7\nn\\
  &\qquad\qquad\qquad\qquad\qquad\qquad\qquad\qquad\quad +63966077 t^5+427850621 t^3+1020096033t+  (t\rightarrow t^{-1}) ) \nn\\
  &-14\,y_s^{\pm8}\chi^V_{(10)}\cdot  (t^{10}+42 t^8+539 t^6+2948 t^4+7854
  t^2+10824+ 7854t^{-2} + \cdots + t^{-10})\nn\\
  &+832\,y_s^{\pm7}\chi^V_{(10)}\cdot (t^9+33 t^7+330 t^5+1386 t^3+2772
  t + (t\rightarrow t^{-1}) )\nn\\
  &-2184 \, y_s^{\pm6}\chi^V_{(10)}\cdot (11 t^8+270 t^6+2002 t^4+6182t^2+8910+6182t^{-2} + \cdots + 11 t^{-8})\nn\\
  &+5824 \, y_s^{\pm5}\chi^V_{(10)}\cdot (77 t^7+1281 t^5+6677 t^3+14575
  t+ (t\rightarrow t^{-1}) )\nn\\
  &+52\, y_s^{\pm4}\chi^V_{(10)}\cdot (33 t^{10}-726 t^8-109153 t^6-1133396 t^4-3996580t^2-5980436 \nn\\
  &\qquad\qquad\qquad\qquad\qquad\qquad\qquad\qquad\quad -3996580t^{-2} -1133396 t^{-4} -109153 t^{-6}+ \cdots +  33 t^{-10})\nn\\
  &-64\, y_s^{\pm3}\chi^V_{(10)}\cdot (t^{13}-22 t^{11}+868 t^9-20559 t^7-726341 t^5-4583956 t^3-10718569t + (t\rightarrow t^{-1}) )\nn\\
  &+8008\, y_s^{\pm2}\chi^V_{(10)}\cdot (49 t^8-2102 t^6-29678 t^4-115094
  t^2-176638 -115094t^{-2} + \cdots + 49t^{-8})\nn\\
  &+4928\, y_s^{\pm 1}\chi^V_{(10)}\cdot (t^{11}-35 t^9+217 t^7+21505 t^5+152866 t^3+371316t+ (t\rightarrow t^{-1}))\nn\\
  &-8\,\chi^V_{(10)}\cdot t^{10}(112 t^{12}-189 t^{10}-104258 t^8+2855160 t^6+46213090 t^4+185620270t^2\nn\\
  &\qquad\qquad\qquad\qquad\qquad\qquad\qquad\qquad\quad +287407450 +185620270t^{-2}+46213090 t^{-4} + \cdots + 112t^{-12})\nn\\
  &+13\, y_s^{\pm8}\cdot (8 t^9+229 t^7+2101 t^5+8393 t^3+16401
  t+ (t\rightarrow t^{-1})) \nn\\
  &-5824\,y_s^{\pm7}\cdot (t^8+22 t^6+154 t^4+462t^{2}+660+{462}{t^{-2}} + \cdots +
   t^{-8})\nn\\
  &+26\,y_s^{\pm6}\cdot (6075 t^7+95425 t^5+483483 t^3+1042937
  t+ (t\rightarrow t^{-1}))\nn\\
  &-64\,y_s^{\pm5}\cdot (t^{12}-22 t^{10}+231 t^8+41580 t^6+427575 t^4+1498244
  t^2+2237312\nn\\
  &\qquad\qquad\qquad\qquad\qquad\qquad\qquad\qquad\quad +1498244t^{-2} +427575 t^{-4}+41580 t^{-6}+ \cdots + t^{-12})\nn\\
  &+91\,y_s^{\pm4}\cdot (11 t^{11}-473 t^9+7623 t^7+312675 t^5+2010490 t^3+4723994
  t+  (t\rightarrow t^{-1}))\nn\\
  &+5824\,y_s^{\pm3}\cdot (77 t^8-2046 t^6-32546 t^4-129768t^2-200508-129768t^{-2} + \cdots + 77 t^{-8})\nn\\
  &+2\,y_s^{\pm2}\cdot (7 t^{15}-154 t^{13}+2475 t^{11}-93720 t^9-257649 t^7 \nn\\
  &\qquad\qquad\qquad\qquad\qquad\qquad\qquad\qquad\quad +50128782 t^5+390072133 t^3+972422990t+ (t\rightarrow t^{-1}))\nn\\
  &-64\,y_s^{\pm1}\cdot (t^{14}-22 t^{12}+231 t^{10}-24927 t^8+317625 t^6+7227990 t^4+31070743t^2+48912688+ \nn\\
  &\qquad\qquad\qquad\qquad\qquad\qquad\qquad\qquad\quad +31070743t^{-2}+7227990 t^{-4}+317625 t^{-6}+ \cdots + t^{-14})\nn\\
  &+154\,(20 t^{11}-1740 t^9-16109 t^7+958563 t^5+8046291 t^3+20489955
  t+ (t\rightarrow t^{-1})).
  \end{align}
  We also confirmed that the 1-instanton partition function $Z_1$ for $(N_{\bf s},N_{\bf c}, N_{\bf v}) = (0,1,2)$ could be identified with the above as follows:
  \begin{align}
    {Z}_{1}^{(N_{\mathbf{s}},N_\mathbf{c},N_{\mathbf{v}}) = (0,1,2)}(\vec{a},\e_1,\e_2,m^c,\vec{m}^v)  = {Z}_{1}^{(N_{\mathbf{s}},N_\mathbf{c},N_{\mathbf{v}}) = (1,0,2)}(\vec{a},\e_1,\e_2,m^s,\vec{m}^v) |_{m^s \rightarrow -m^c }.
  \end{align}

\paragraph{$\bf E_6$}
The flavor symmetry on $N_f\mathbf{F} + N_{\bar{f}}\bar{\mathbf{F}}$ hypermultiplets is $U(N_f + N_{\bar{f}})$. The character expansion of $\tilde{Z}_1$ at $N_f+ N_{\bar{f}} = 3$ is written as follows:
  \begin{align}
    \label{eq:E6F3}
    \tilde{Z}_1&=\,\frac{t^{22}}{(1-t^2)^{22}}\Big((\chi^f_{(00)_{-9}}+\chi^f_{(00)_9})(t^{11}+56 t^9+945 t^7+6776 t^5+23815 t^3+43989 t+(t\rightarrow t^{-1}))\nn\\
&-27(\chi^f_{(10)_{-8}}+\chi^f_{(01)_8})(t^{10}+42 t^8+539 t^6+2948 t^4+7854 t^2+10824+7854t^{-2}+\cdots+t^{-10})\nn\\
&+351(\chi^f_{(20)_{-7}}+\chi^f_{(02)_7}) (t^9+28 t^7+253 t^5+1001 t^3+1947t+(t\rightarrow t^{-1}))\nn\\
&+351(\chi^f_{(01)_{-7}}+\chi^f_{(10)_7}) (t^9+33 t^7+330 t^5+1386 t^3+2772t+(t\rightarrow t^{-1}))\\
&+(\chi^f_{(30)_{-6}}+\chi^f_{(03)_6})(t^{12}-22 t^{10}-2694 t^8-42790 t^6-256355 t^4\nn\\
&\hspace{8cm}-712536 t^2-994488-712536t^{-2}+\cdots+t^{-12})\nn\\
&-26(\chi^f_{(11)_{-6}}+\chi^f_{(11)_6})(224 t^8+4774 t^6+32700 t^4+96877 t^2+137830+96877t^{-2}+\cdots +224t^{-8})\nn\\
&-13(\chi^f_{(00)_{-6}}+\chi^f_{(00)_6})(231 t^8+6182 t^6+48796 t^4+156338 t^2+228074+156338 t^{-2}+\cdots+231t^{-8})\nn\\
&+351(\chi^f_{(40)_{-5}}+\chi^f_{(04)_5})(t^9+28 t^7+253 t^5+1001 t^3+1947t+(t\rightarrow t^{-1}))\nn\\
&-27(\chi^f_{(21)_{-5}}+\chi^f_{(12)_5})(t^{11}-22 t^9-1694 t^7-19965 t^5-89298 t^3-182952t+(t\rightarrow t^{-1}))\nn\\
&+702(\chi^f_{(02)_{-5}}+\chi^f_{(20)_5})(49 t^7+707 t^5+3399 t^3+7150t+(t\rightarrow t^{-1}))\nn\\
&+702(\chi^f_{(10)_{-5}}+\chi^f_{(01)_5})(77 t^7+1281 t^5+6677 t^3+14575t+(t\rightarrow t^{-1}))\nn\\
&-27(\chi^f_{(50)_{-4}}+\chi^f_{(05)_4})(t^{10}+42 t^8+539 t^6+2948 t^4+7854 t^2+10824+7854t^{-2}+\cdots t^{-10})\nn\\
&-351(\chi^f_{(31)_{-4}}+\chi^f_{(13)_4})(21 t^8+434 t^6+2926 t^4+8602 t^2+12210+8602t^{-2}+\cdots+21t^{-8})\nn\\
&+351(\chi^f_{(12)_{-4}}+\chi^f_{(21)_4})(t^{10}-22 t^8-869 t^6-6908 t^4-21714 t^2-31416-21714t^{-2}+\cdots+t^{-10})\nn\\
&+351(\chi^f_{(20)_{-4}}+\chi^f_{(02)_4})(t^{10}-22 t^8-1177 t^6-10500 t^4-34936 t^2-51436-34936 t^{-2}+\cdots+t^{-10})\nn\\
&-1404(\chi^f_{(01)_{-4}}+\chi^f_{(10)_4})(294 t^6+3132 t^4+10989 t^2+16390+10989 t^{-2}+3132t^{-4}+294t^{-6})\nn\\
&+(\chi^f_{(60)_{-3}}+\chi^f_{(06)_3})(t^{11}+56 t^9+945 t^7+6776 t^5+23815 t^3+43989t+(t\rightarrow t^{-1}))\nn\\
&+13(\chi^f_{(41)_{-3}}+\chi^f_{(14)_3})(50 t^9+1573 t^7+15219 t^5+62623 t^3+124025t+(t\rightarrow t^{-1}))\nn\\
&+(\chi^f_{(22)_{-3}}+\chi^f_{(22)_3})(t^{13}-22 t^{11}+231 t^9+68530 t^7+919589 t^5\nn\\
&\hspace{8cm}+4310670 t^3+8985999t+(t\rightarrow t^{-1}))\nn\\
&-13(\chi^f_{(03)_{-3}}+\chi^f_{(30)_3})(6 t^{11}+93 t^9-3564 t^7-60115 t^5-303171 t^3-650699t+(t\rightarrow t^{-1}))\nn\\
&+13(\chi^f_{(30)_{-3}}+\chi^f_{(03)_3})(6075 t^7+95425 t^5+483483 t^3+1042937t+(t\rightarrow t^{-1}))\nn\\
&-832(\chi^f_{(11)_{-3}}+\chi^f_{(11)_3}(7 t^9-154 t^7-4095 t^5-23683 t^3-53471t+(t\rightarrow t^{-1}))\nn\\
&+(\chi^f_{(00)_{-3}}+\chi^f_{(00)_3})(t^{15}-22 t^{13}+231 t^{11}-1540 t^9+7315 t^7\nn\\
&\hspace{8cm}+1533042 t^5+10536141 t^3+24960012t+(t\rightarrow t^{-1}))\nn\\
&-27(\chi^f_{(51)_{-2}}+\chi^f_{(15)_2})(t^{10}+42 t^8+539 t^6+2948 t^4+7854 t^2+10824+7854t^{-2}+\cdots+t^{-10})\nn\\
&-351(\chi^f_{(32)_{-2}}+\chi^f_{(23)_2})(21 t^8+434 t^6+2926 t^4+8602 t^2+12210+8602t^{-2}+\cdots+21t^{-8})\nn\\
&+351(\chi^f_{(13)_{-2}}+\chi^f_{(31)_2})(t^{10}-22 t^8-869 t^6-6908 t^4-21714 t^2-31416-21714t^{-2}+\cdots+t^{-10})\nn\\
&-702(\chi^f_{(40)_{-2}}+\chi^f_{(04)_2})(11 t^8+270 t^6+2002 t^4+6182 t^2+8910+6182t^{2-}+\cdots+11t^{-8})\nn\\
&-27(\chi^f_{(21)_{-2}}+\chi^f_{(12)_2})(t^{12}-22 t^{10}+231 t^8+34300 t^6+334235 t^4\nn\\
&\hspace{8cm}+1139314 t^2+1686762+1139314 t^{-2}+\cdots+t^{-12})\nn\\
&+27(\chi^f_{(02)_{-2}}+\chi^f_{(20)_2})(64 t^{10}+517 t^8-27566 t^6-317548 t^4-1145354 t^2\nn\\
&\hspace{8cm}-1723106-1145354 t^{-2}+\cdots+64t^{-10})\nn\\
&+4914(\chi^f_{(10)_{-2}}+\chi^f_{(01)_2})(7 t^8-154 t^6-2310 t^4-8866 t^2-13530-8866t^{-2}+\cdots+7t^{-8})\nn\\
&+351(\chi^f_{(42)_{-1}}+\chi^f_{(24)_1})(t^9+28 t^7+253 t^5+1001 t^3+1947t+(t\rightarrow t^{-1}))\nn\\
&+351(\chi^f_{(50)_{-1}}+\chi^f_{(05)_1})(t^9+33 t^7+330 t^5+1386 t^3+2772t+(t\rightarrow t^{-1}))\nn\\
&-27(\chi^f_{(23)_{-1}}+\chi^f_{(32)_1})(t^{11}-22 t^9-1694 t^7-19965 t^5-89298 t^3-182952t+(t\rightarrow t^{-1}))\nn\\
&+22464(\chi^f_{(31)_{-1}}+\chi^f_{(13)_1})(5 t^7+77 t^5+385 t^3+825t+(t\rightarrow t^{-1}))\nn\\
&+702(\chi^f_{(04)_{-1}}+\chi^f_{(40)_1})(49 t^7+707 t^5+3399 t^3+7150t+(t\rightarrow t^{-1}))\nn\\
&-351(\chi^f_{(12)_{-1}}+\chi^f_{(21)_1})(21 t^9-462 t^7-11605 t^5-65983 t^3-148071t+(t\rightarrow t^{-1}))\nn\\
&+351(\chi^f_{(20)_{-1}}+\chi^f_{(02)_1})(t^{11}-22 t^9+231 t^7+11516 t^5+72799 t^3+168707t+(t\rightarrow t^{-1}))\nn\\
&-702(\chi^f_{(01)_{-1}}+\chi^f_{(10)_1})(25 t^9-6325 t^5-44583 t^3-107387t+(t\rightarrow t^{-1}))\nn\\
&+\chi^f_{(33)_0}(t^{12}-22 t^{10}-2694 t^8-42790 t^6-256355 t^4-712536 t^2-994488-712536 t^{-2}+\cdots+t^{-12})\nn\\
&-26(\chi^f_{(41)_0}+\chi^f_{(14)_0})(224 t^8+4774 t^6+32700 t^4+96877 t^2+137830+96877 t^{-2}+\cdots+t^{-8})\nn\\
&+26\,\chi^f_{(22)_0}(25 t^{10}-550 t^8-27030 t^6-231990 t^4-756657 t^2-1107156-756657 t^{-2}+\cdots+25t^{-10})\nn\\
&+(\chi^f_{(03)_0}+\chi^f_{(30)_0})(t^{14}-22 t^{12}+231 t^{10}-1540 t^8-593285 t^6-5973198 t^4\nn\\
&\hspace{8cm}-20531379 t^2-30453456-20531379 t^{-2}+\cdots+t^{-14})\nn\\
&-26\,\chi^f_{(11)_0}(3 t^{12}-66 t^{10}-2002 t^8+54670 t^6+741975 t^4+2786872 t^2\nn\\
&\hspace{8cm}+4232536+2786872 t^{-2}+\cdots+3t^{-12})\nn\\
&+2(1215 t^{10}+26070 t^8-212410 t^6-4381850 t^4-18219943 t^2\nn\\
&\hspace{8cm}-28496524-18219943 t^{-2}+\cdots+1215t^{-10})\Big),\nn
\end{align}
where the $U(3)$ character $\chi^f_{(mn)_c}$ is defined as
  \begin{align}
    \chi^f_{(mn)_c} =
      (w_1 w_2w_3)^{\frac{c-m+n}{3}}\bigg( \sum_{\substack{1\leq i_1\leq\cdots\leq i_m\leq3\\1\leq j_1\leq\cdots\leq j_n\leq 3}}\frac{w_{i_1}\cdots w_{i_m}}{w_{j_1}\cdots w_{j_n}}-\sum_{\substack{1\leq i_1\leq\cdots\leq i_{m-1}\leq3\\1\leq j_1\leq\cdots\leq j_{n-1}\leq 3}}\frac{w_{i_1}\cdots w_{i_{m-1}}}{w_{j_1}\cdots w_{j_{n-1}}}\bigg),
  \end{align}
  with
  \begin{align}
  (w_1,w_2,w_3)=
  \begin{cases}
  (y_{f,1},y_{f,2},y_{f,3}) & \text{for}\quad (N_f, N_{\bar{f}})=(3,0),\\
  (y_{f,1},y_{f,2},y_{\bar{f},1}^{-1}) & \text{for}\quad (N_f, N_{\bar{f}})=(2,1),\\
  (y_{f,1},y_{\bar{f},1}^{-1},y_{\bar{f},2}^{-1}) & \text{for}\quad (N_f, N_{\bar{f}})=(1,2),\\
  (y_{\bar{f},1}^{-1},y_{\bar{f},2}^{-1},y_{\bar{f},3}^{-1}) & \text{for}\quad (N_f, N_{\bar{f}})=(0,3).
  \end{cases}
  \end{align}
  Again, \eqref{eq:E6F3} was tested against our general 1-instanton expression \eqref{eq:1inst-formula} up to $t^{180}$ order.

\paragraph{$\bf E_7$} The flavor symmetry acting on $N_f$ hypermultiplets is $SO(2N_f)_f$. The character expansion of $\tilde{Z}_1$ at $N_f=2$ is given as follows:
  \begin{align}
    \label{eq:E7F2}
    \tilde{Z}_1&=\frac{t^{34}}{(1-t^2)^{34}}\Big((\chi^f_{(0,12)}+\chi^f_{(12,0)})(t^{17}+99 t^{15}+3410 t^{13}+56617 t^{11}+521917 t^9\\
    &\hspace{2.5cm}+2889898 t^7+10086066 t^5+22867856t^3+34289476t+(t\rightarrow t^{-1})\nn\\
    &-8(\chi^f_{(1,11)}+\chi^f_{(11,1)})(7 t^{16}+572 t^{14}+16401 t^{12}+227766 t^{10}+1759296 t^8\nn\\
    &\hspace{2.5cm}+8155308 t^6+23747878 t^4+44652608t^2+55026348+44652608t^{-2}+\cdots+7t^{-16})\nn\\
    &+19(\chi^f_{(2,10)}+\chi^f_{(10,2)})(81 t^{15}+5254 t^{13}+121550 t^{11}+1376580 t^9\nn\\
    &\hspace{2.5cm}+8725369 t^7+33273284 t^5+79629972t^3+122510670t+(t\rightarrow t^{-1}))\nn\\
    &+133(\chi^f_{(0,10)}+\chi^f_{(10,0)})(11 t^{15}+760 t^{13}+18445 t^{11}+216580 t^9\nn\\
    &\hspace{2.5cm}+1409980 t^7+5479474 t^5+13273260t^3+20541950t+(t\rightarrow t^{-1}))\nn\\
    &-152(\chi^f_{(39)}+\chi^f_{(93)})(182 t^{14}+8827 t^{12}+158592 t^{10}+1426827 t^8+7281032 t^6\nn\\
    &\hspace{2.5cm}+22506946 t^4+43735356t^2+54466776+43735356t^{-2}+\cdots+182t^{-14})\nn\\
    &-2128(\chi^f_{(19)}+\chi^f_{(91)})(24 t^{14}+1309 t^{12}+25454 t^{10}+241859 t^8\nn\\
    &\hspace{2.5cm}+1281324 t^6+4059022 t^4+7997752t^2+10005112+7997752t^{-2}+\cdots+24t^{-14})\nn\\
    &+(\chi^f_{(48)}+\chi^f_{(84)})(t^{19}-34 t^{17}+561 t^{15}+359766 t^{13}+11997546 t^{11}+161435604 t^9\nn\\
    &\hspace{2.5cm}+1130192844t^7+4579505424 t^5+11356618494t^3+17763983094t+(t\rightarrow t^{-1}))\nn\\
    &+10773(\chi^f_{(28)}+\chi^f_{(82)})(91 t^{13}+3668 t^{11}+54893 t^9\nn\\
    &\hspace{2.5cm}+411026 t^7+1739100 t^5+4427038 t^3+7011004t+(t\rightarrow t^{-1}))\nn\\
    &+5187(\chi^f_{(08)}+\chi^f_{(80)})(119 t^{13}+5269 t^{11}+83499 t^9\nn\\
    &\hspace{2.5cm}+648329 t^7+2806870 t^5+7243122 t^3+11543952t+(t\rightarrow t^{-1}))\nn\\
    &+912(\chi^f_{(57)}+\chi^f_{(75)})(t^{16}-34 t^{14}-3597 t^{12}-78540 t^{10}-776832 t^8-4186896 t^6\nn\\
    &\hspace{2.5cm}-13370126 t^4-26439556t^2-33110220-26439556t^{-2}+\cdots+t^{-16})\nn\\
    &-56(\chi^f_{(37)}+\chi^f_{(73)})(t^{18}-34 t^{16}+561 t^{14}+227392 t^{12}+6213449 t^{10}+69122350 t^8+400174169 t^6\nn\\
    &\hspace{2.5cm}+1335305664t^4+2705039932 t^2+3413732872+2705039932 t^{-2}+\cdots+t^{-18})\nn\\
    &-27664(\chi^f_{(17)}+\chi^f_{(71)})(539 t^{12}+17314 t^{10}+208879 t^8+1267860 t^6\nn\\
    &\hspace{2.5cm}+4351490 t^4+8949752 t^2+11348792+8949752 t^{-2}+\cdots+539t^{-12})\nn\\
    &-19\,\chi^f_{(66)}(7 t^{17}+217 t^{15}-24908 t^{13}-1021757 t^{11}-14769022 t^9\nn\\
    &\hspace{2.5cm}-107322042 t^7-444417927t^5-1115908152 t^3-1755535056 t+(r\rightarrow t^{-1}))\nn\\
    &-95(\chi^f_{(46)}+\chi^f_{(64)})(429 t^{15}-14586 t^{13}-1157156 t^{11}-20646010 t^9\nn\\
    &\hspace{2.5cm}-168250530 t^7-746606798 t^5-1952106107t^3-3129466862t+(t\rightarrow t^{-1}))\nn\\
    &+57(\chi^f_{(26)}+\chi^f_{(62)})(27 t^{17}-918 t^{15}+15147 t^{13}+3520341 t^{11}+75470346 t^9\nn\\
    &\hspace{2.5cm}+672625723 t^7+3141068903t^5+8453641548 t^3+13732731903t+(t\rightarrow t^{-1}))\nn\\
    &+(\chi^f_{(06)}+\chi^f_{(60)})(t^{21}-34 t^{19}+561 t^{17}-5984 t^{15}+46376 t^{13}+108842392 t^{11}+2613712872t^9\nn\\
    &\hspace{2.5cm}+24490191704 t^7+117519798814 t^5+321089011759 t^3+525183176299t+(t\rightarrow t^{-1}))\nn\\
    &+1296\,\chi^f_{(55)}(5 t^{16}+110 t^{14}-14030 t^{12}-458914 t^{10}-5440765 t^8-32547180 t^6\nn\\
    &\hspace{2.5cm}-110625460t^4-226279180 t^2-286427492-226279180 t^{-2}+\cdots+5t^{-16})\nn\\
    &+1064(\chi^f_{(35)}+\chi^f_{(53)})(810 t^{14}-27540 t^{12}-1538126 t^{10}-21575635 t^8-140780490 t^6\nn\\
    &\hspace{2.5cm}-502663905 t^4-1055162460t^2-1346539128-1055162460t^{-2}+\cdots+810t^{-14})\nn\\
    &-27664(\chi^f_{(15)}+\chi^f_{(51)})(t^{16}-34 t^{14}+561 t^{12}+62832 t^{10}+1000416 t^8\nn\\
    &\hspace{2.5cm}+6920904 t^6+25507174 t^4+54425228t^2+69808596+54425228t^{-2}+\cdots+t^{-16})\nn\\
    &+\chi^f_{(44)}(t^{21}-34 t^{19}+561 t^{17}-158136 t^{15}-1922955 t^{13}+320810876 t^{11}+7970822266t^9\nn\\
    &\hspace{2.5cm}+74975208858 t^7+359889450611 t^5+983025661861 t^3+1607508212091t+(t\rightarrow t^{-1}))\nn\\
    &-133(\chi^f_{(24)}+\chi^f_{(42)})(t^{19}-34 t^{17}+561 t^{15}+79101 t^{13}-2846514 t^{11}-102197931 t^9\nn\\
    &\hspace{2.5cm}-1080814746t^7-5500823076 t^5-15503708076 t^3-25710027486t+(t\rightarrow t^{-1}))\nn\\
    &+665(\chi^f_{(04)}+\chi^f_{(40)})(13 t^{17}+108 t^{15}-11407 t^{13}+230758 t^{11}+11122199 t^9\nn\\
    &\hspace{2.5cm}+125832753 t^7+660902603t^5+1893530023 t^3+3162878730t+(t\rightarrow t^{-1}))\nn\\
    &+152\,\chi^f_{(33)}(6 t^{18}-204 t^{16}+18381 t^{14}+20306 t^{12}-21755987 t^{10}-387061196 t^8-2796155121t^6\nn\\
    &\hspace{2.5cm}-10534894066 t^4-22728127951 t^2-29251476496-22728127951 t^{-2}+\cdots+6t^{-18})\nn\\
    &-1064(\chi^f_{(13)}+\chi^f_{(31)})(81 t^{16}-2754 t^{14}-49181 t^{12}+2732444 t^{10}+59237424 t^8+458851114 t^6\nn\\
    &\hspace{2.5cm}+1789977134 t^4+3929114222 t^2+5083736372+3929114222 t^{-2}+\cdots+t^{-16})\nn\\
    &+81\,\chi^f_{(22)}(91 t^{17}+41 t^{15}-356609 t^{13}+2951795 t^{11}+247685515 t^9\nn\\
    &\hspace{2.5cm}+3029637009 t^7+16451185429 t^5+47931732849 t^3+80650803640 t+(t\rightarrow t^{-1}))\nn\\
    &-1312311(\chi^f_{(02)}+\chi^f_{(20)})(14 t^{13}-17 t^{11}-7752 t^9\nn\\
    &\hspace{2.5cm}-103411 t^7-581570 t^5-1724208 t^3-2923116 t+(t\rightarrow t^{-1}))\nn\\
    &+304\,\chi^f_{(11)}(11960 t^{14}+343681 t^{12}-7234554 t^{10}-208524209 t^8-1747615980 t^6\nn\\
    &\hspace{2.5cm}-7073563915 t^4-15807799502 t^2-20565064322-15807799502 t^{-2}+\cdots+11960t^{-14})\nn\\
    &+(t^{23}-34 t^{21}+561 t^{19}-5984 t^{17}-192226 t^{15}-11212452 t^{13}-46556642 t^{11}+4966300623 t^9\nn\\
    &\hspace{2.5cm}+73315010528 t^7+427928422856 t^5+1291626014327 t^3+2206690491962 t+(t\rightarrow t^{-1}))\Big). \nn
\end{align}
This was tested against the closed-form expression \eqref{eq:1inst-formula} up to $t^{280}$ order.

\bibliographystyle{JHEP}
\bibliography{ref}

\providecommand{\href}[2]{#2}\begingroup\raggedright\begin{thebibliography}{10}

\bibitem{Seiberg:1994rs}
N.~Seiberg and E.~Witten, \emph{{Electric--magnetic Duality, Monopole
  Condensation, and Confinement in ${\mathcal{N}}\!=2$ Supersymmetric
  Yang-Mills Theory}}, \href{http://dx.doi.org/10.1016/0550-3213(94)90124-4,
  10.1016/0550-3213(94)00449-8}{\emph{Nucl. Phys.} {\bf B426} (1994) 19--52},
  [\href{http://arxiv.org/abs/hep-th/9407087}{{\tt hep-th/9407087}}].

\bibitem{Seiberg:1994aj}
N.~Seiberg and E.~Witten, \emph{{Monopoles, Duality and Chiral Symmetry
  Breaking in ${\mathcal{N}}\!=2$ Supersymmetric QCD}},
  \href{http://dx.doi.org/10.1016/0550-3213(94)90214-3}{\emph{Nucl. Phys.} {\bf
  B431} (1994) 484--550}, [\href{http://arxiv.org/abs/hep-th/9408099}{{\tt
  hep-th/9408099}}].

\bibitem{Nekrasov:2002qd}
N.~A. Nekrasov, \emph{{Seiberg-Witten Prepotential from Instanton Counting}},
  \href{http://dx.doi.org/10.4310/ATMP.2003.v7.n5.a4}{\emph{Adv. Theor. Math.
  Phys.} {\bf 7} (2003) 831--864},
  [\href{http://arxiv.org/abs/hep-th/0206161}{{\tt hep-th/0206161}}].

\bibitem{Nekrasov:2003rj}
N.~Nekrasov and A.~Okounkov, \emph{{Seiberg-Witten Theory and Random
  Partitions}}, \href{http://dx.doi.org/10.1007/0-8176-4467-9_15}{\emph{Prog.
  Math.} {\bf 244} (2006) 525--596},
  [\href{http://arxiv.org/abs/hep-th/0306238}{{\tt hep-th/0306238}}].

\bibitem{Nakajima:2003pg}
H.~Nakajima and K.~Yoshioka, \emph{{Instanton counting on blowup. 1.}},
  \href{http://dx.doi.org/10.1007/s00222-005-0444-1}{\emph{Invent. Math.} {\bf
  162} (2005) 313--355}, [\href{http://arxiv.org/abs/math/0306198}{{\tt
  math/0306198}}].

\bibitem{Braverman:2004cr}
A.~Braverman and P.~Etingof, \emph{{Instanton Counting via Affine Lie Algebras
  Ii: from Whittaker Vectors to the Seiberg-Witten Prepotential}},
  \href{http://arxiv.org/abs/math/0409441}{{\tt math/0409441}}.

\bibitem{Atiyah:1978ri}
M.~F. Atiyah, N.~J. Hitchin, V.~G. Drinfeld and {\relax Yu}.~I. Manin,
  \emph{{Construction of Instantons}},
  \href{http://dx.doi.org/10.1016/0375-9601(78)90141-X}{\emph{Phys. Lett.} {\bf
  A65} (1978) 185--187}.

\bibitem{Moore:1997dj}
G.~W. Moore, N.~Nekrasov and S.~Shatashvili, \emph{{Integrating over Higgs
  Branches}}, \href{http://dx.doi.org/10.1007/PL00005525}{\emph{Commun. Math.
  Phys.} {\bf 209} (2000) 97--121},
  [\href{http://arxiv.org/abs/hep-th/9712241}{{\tt hep-th/9712241}}].

\bibitem{Bruzzo:2002xf}
U.~Bruzzo, F.~Fucito, J.~F. Morales and A.~Tanzini, \emph{{Multiinstanton
  Calculus and Equivariant Cohomology}},
  \href{http://dx.doi.org/10.1088/1126-6708/2003/05/054}{\emph{JHEP} {\bf 05}
  (2003) 054}, [\href{http://arxiv.org/abs/hep-th/0211108}{{\tt
  hep-th/0211108}}].

\bibitem{Nekrasov:2004vw}
N.~Nekrasov and S.~Shadchin, \emph{{ABCD of Instantons}},
  \href{http://dx.doi.org/10.1007/s00220-004-1189-1}{\emph{Commun. Math. Phys.}
  {\bf 252} (2004) 359--391}, [\href{http://arxiv.org/abs/hep-th/0404225}{{\tt
  hep-th/0404225}}].

\bibitem{Marino:2004cn}
M.~Mari\~no and N.~Wyllard, \emph{{A Note on Instanton Counting for
  ${\mathcal{N}}\!=2$ Gauge Theories with Classical Gauge Groups}},
  \href{http://dx.doi.org/10.1088/1126-6708/2004/05/021}{\emph{JHEP} {\bf 05}
  (2004) 021}, [\href{http://arxiv.org/abs/hep-th/0404125}{{\tt
  hep-th/0404125}}].

\bibitem{Fucito:2004gi}
F.~Fucito, J.~F. Morales and R.~Poghossian, \emph{{Instantons on Quivers and
  Orientifolds}},
  \href{http://dx.doi.org/10.1088/1126-6708/2004/10/037}{\emph{JHEP} {\bf 10}
  (2004) 037}, [\href{http://arxiv.org/abs/hep-th/0408090}{{\tt
  hep-th/0408090}}].

\bibitem{Hollands:2010xa}
L.~Hollands, C.~A. Keller and J.~Song, \emph{{From SO/Sp Instantons to
  W-Algebra Blocks}},
  \href{http://dx.doi.org/10.1007/JHEP03(2011)053}{\emph{JHEP} {\bf 03} (2011)
  053}, [\href{http://arxiv.org/abs/1012.4468}{{\tt 1012.4468}}].

\bibitem{Hollands:2011zc}
L.~Hollands, C.~A. Keller and J.~Song, \emph{{Towards a 4D/2D Correspondence
  for Sicilian Quivers}},
  \href{http://dx.doi.org/10.1007/JHEP10(2011)100}{\emph{JHEP} {\bf 10} (2011)
  100}, [\href{http://arxiv.org/abs/1107.0973}{{\tt 1107.0973}}].

\bibitem{Hwang:2014uwa}
C.~Hwang, J.~Kim, S.~Kim and J.~Park, \emph{{General Instanton Counting and 5D
  SCFT}}, \href{http://dx.doi.org/10.1007/JHEP07(2015)063,
  10.1007/JHEP04(2016)094}{\emph{JHEP} {\bf 07} (2015) 063},
  [\href{http://arxiv.org/abs/1406.6793}{{\tt 1406.6793}}].

\bibitem{Cordova:2014oxa}
C.~Cordova and S.-H. Shao, \emph{{An Index Formula for Supersymmetric Quantum
  Mechanics}},  \href{http://arxiv.org/abs/1406.7853}{{\tt 1406.7853}}.

\bibitem{Hori:2014tda}
K.~Hori, H.~Kim and P.~Yi, \emph{{Witten Index and Wall Crossing}},
  \href{http://dx.doi.org/10.1007/JHEP01(2015)124}{\emph{JHEP} {\bf 01} (2015)
  124}, [\href{http://arxiv.org/abs/1407.2567}{{\tt 1407.2567}}].

\bibitem{Benini:2013xpa}
F.~Benini, R.~Eager, K.~Hori and Y.~Tachikawa, \emph{{Elliptic Genera of 2d
  ${\mathcal{N}}$ = 2 Gauge Theories}},
  \href{http://dx.doi.org/10.1007/s00220-014-2210-y}{\emph{Commun. Math. Phys.}
  {\bf 333} (2015) 1241--1286}, [\href{http://arxiv.org/abs/1308.4896}{{\tt
  1308.4896}}].

\bibitem{Benini:2013nda}
F.~Benini, R.~Eager, K.~Hori and Y.~Tachikawa, \emph{{Elliptic Genera of
  Two-Dimensional ${\mathcal{N}}\!=2$ Gauge Theories with Rank-One Gauge
  Groups}}, \href{http://dx.doi.org/10.1007/s11005-013-0673-y}{\emph{Lett.
  Math. Phys.} {\bf 104} (2014) 465--493},
  [\href{http://arxiv.org/abs/1305.0533}{{\tt 1305.0533}}].

\bibitem{Nakajima:2003uh}
H.~Nakajima and K.~Yoshioka, \emph{{Lectures on instanton counting}},  in
  \emph{{CRM Workshop on Algebraic Structures and Moduli Spaces Montreal,
  Canada, July 14-20, 2003}}, 2003.
\newblock \href{http://arxiv.org/abs/math/0311058}{{\tt math/0311058}}.

\bibitem{Nakajima:2005fg}
H.~Nakajima and K.~Yoshioka, \emph{{Instanton Counting on Blowup. II.
  K-Theoretic Partition Function}},
  \href{http://arxiv.org/abs/math/0505553}{{\tt math/0505553}}.

\bibitem{Gottsche:2006bm}
L.~Gottsche, H.~Nakajima and K.~Yoshioka, \emph{{K-theoretic Donaldson
  invariants via instanton counting}},
  \href{http://dx.doi.org/10.4310/PAMQ.2009.v5.n3.a5}{\emph{Pure Appl. Math.
  Quart.} {\bf 5} (2009) 1029--1111},
  [\href{http://arxiv.org/abs/math/0611945}{{\tt math/0611945}}].

\bibitem{Nakajima:2009qjc}
H.~Nakajima and K.~Yoshioka, \emph{{Perverse coherent sheaves on blowup, III:
  Blow-up formula from wall-crossing}},
  \href{http://dx.doi.org/10.1215/21562261-1214366}{\emph{Kyoto J. Math.} {\bf
  51} (2011) 263--335}, [\href{http://arxiv.org/abs/0911.1773}{{\tt
  0911.1773}}].

\bibitem{Gottsche:2010ig}
L.~Gottsche, H.~Nakajima and K.~Yoshioka, \emph{{Donaldson = Seiberg-Witten
  from Mochizuki's Formula and Instanton Counting}}, {\emph{Publ. Res. Inst.
  Math. Sci. Kyoto} {\bf 47} (2011) 307--359},
  [\href{http://arxiv.org/abs/1001.5024}{{\tt 1001.5024}}].

\bibitem{FintushelStern}
R.~Fintushel and R.~J. Stern, \emph{The blowup formula for donaldson
  invariants}, {\emph{Annals of Mathematics} {\bf 143} (1996) 529--546}.

\bibitem{Moore:1997pc}
G.~W. Moore and E.~Witten, \emph{{Integration over the U Plane in Donaldson
  Theory}}, \href{http://dx.doi.org/10.4310/ATMP.1997.v1.n2.a7}{\emph{Adv.
  Theor. Math. Phys.} {\bf 1} (1997) 298--387},
  [\href{http://arxiv.org/abs/hep-th/9709193}{{\tt hep-th/9709193}}].

\bibitem{Marino:1998bm}
M.~Mari\~no and G.~W. Moore, \emph{{The Donaldson-Witten Function for Gauge
  Groups of Rank Larger Than One}},
  \href{http://dx.doi.org/10.1007/s002200050494}{\emph{Commun. Math. Phys.}
  {\bf 199} (1998) 25--69}, [\href{http://arxiv.org/abs/hep-th/9802185}{{\tt
  hep-th/9802185}}].

\bibitem{Keller:2012da}
C.~A. Keller and J.~Song, \emph{{Counting Exceptional Instantons}},
  \href{http://dx.doi.org/10.1007/JHEP07(2012)085}{\emph{JHEP} {\bf 07} (2012)
  085}, [\href{http://arxiv.org/abs/1205.4722}{{\tt 1205.4722}}].

\bibitem{Gaiotto:2012uq}
D.~Gaiotto and S.~S. Razamat, \emph{{Exceptional Indices}},
  \href{http://dx.doi.org/10.1007/JHEP05(2012)145}{\emph{JHEP} {\bf 05} (2012)
  145}, [\href{http://arxiv.org/abs/1203.5517}{{\tt 1203.5517}}].

\bibitem{Grassi:2016nnt}
A.~Grassi and J.~Gu, \emph{{BPS Relations from Spectral Problems and Blowup
  Equations}}, \href{http://dx.doi.org/10.1007/s11005-019-01163-1}{\emph{Lett.
  Math. Phys.} {\bf 109} (2019) 1271--1302},
  [\href{http://arxiv.org/abs/1609.05914}{{\tt 1609.05914}}].

\bibitem{Gu:2017ccq}
J.~Gu, M.-x. Huang, A.-K. Kashani-Poor and A.~Klemm, \emph{{Refined BPS
  invariants of 6d SCFTs from anomalies and modularity}},
  \href{http://dx.doi.org/10.1007/JHEP05(2017)130}{\emph{JHEP} {\bf 05} (2017)
  130}, [\href{http://arxiv.org/abs/1701.00764}{{\tt 1701.00764}}].

\bibitem{Huang:2017mis}
M.-x. Huang, K.~Sun and X.~Wang, \emph{{Blowup Equations for Refined
  Topological Strings}},
  \href{http://dx.doi.org/10.1007/JHEP10(2018)196}{\emph{JHEP} {\bf 10} (2018)
  196}, [\href{http://arxiv.org/abs/1711.09884}{{\tt 1711.09884}}].

\bibitem{Gu:2018gmy}
J.~Gu, B.~Haghighat, K.~Sun and X.~Wang, \emph{{Blowup Equations for 6D SCFTs.
  I}}, \href{http://dx.doi.org/10.1007/JHEP03(2019)002}{\emph{JHEP} {\bf 03}
  (2019) 002}, [\href{http://arxiv.org/abs/1811.02577}{{\tt 1811.02577}}].

\bibitem{Gu:2019dan}
J.~Gu, A.~Klemm, K.~Sun and X.~Wang, \emph{{Elliptic Blowup Equations for 6D
  SCFTs. II: Exceptional Cases}},  \href{http://arxiv.org/abs/1905.00864}{{\tt
  1905.00864}}.

\bibitem{Honda:2016mvg}
M.~Honda, \emph{{Borel Summability of Perturbative Series in 4D $N=2$ and 5D
  $N$=1 Supersymmetric Theories}},
  \href{http://dx.doi.org/10.1103/PhysRevLett.116.211601}{\emph{Phys. Rev.
  Lett.} {\bf 116} (2016) 211601}, [\href{http://arxiv.org/abs/1603.06207}{{\tt
  1603.06207}}].

\bibitem{Edelstein:1998sp}
J.~D. Edelstein, M.~Mari\~no and J.~Mas, \emph{{Whitham Hierarchies, Instanton
  Corrections and Soft Supersymmetry Breaking in ${\mathcal{N}}\!=2$ $SU(N)$
  Superyang-Mills Theory}},
  \href{http://dx.doi.org/10.1016/S0550-3213(98)00798-6}{\emph{Nucl. Phys.}
  {\bf B541} (1999) 671--697}, [\href{http://arxiv.org/abs/hep-th/9805172}{{\tt
  hep-th/9805172}}].

\bibitem{Edelstein:1999xk}
J.~D. Edelstein, M.~G\'omez-Reino, M.~Mari\~no and J.~Mas,
  \emph{{${\mathcal{N}}\!=2$ Supersymmetric Gauge Theories with Massive
  Hypermultiplets and the Whitham Hierarchy}},
  \href{http://dx.doi.org/10.1016/S0550-3213(00)00034-1}{\emph{Nucl. Phys.}
  {\bf B574} (2000) 587--619}, [\href{http://arxiv.org/abs/hep-th/9911115}{{\tt
  hep-th/9911115}}].

\bibitem{Nekrasov:2003vi}
N.~A. Nekrasov, \emph{{Localizing Gauge Theories}},  in \emph{{Mathematical
  Physics. Proceedings, 14Th International Congress, Icmp 2003, Lisbon,
  Portugal, July 28-August 2, 2003}}, pp.~645--654, 2003.

\bibitem{Gottsche:2006tn}
L.~Gottsche, H.~Nakajima and K.~Yoshioka, \emph{{Instanton Counting and
  Donaldson Invariants}}, {\emph{J. Diff. Geom.} {\bf 80} (2008) 343--390},
  [\href{http://arxiv.org/abs/math/0606180}{{\tt math/0606180}}].

\bibitem{Gasparim:2008ri}
E.~Gasparim and C.-C.~M. Liu, \emph{{The Nekrasov Conjecture for Toric
  Surfaces}}, \href{http://dx.doi.org/10.1007/s00220-009-0948-4}{\emph{Commun.
  Math. Phys.} {\bf 293} (2010) 661--700},
  [\href{http://arxiv.org/abs/0808.0884}{{\tt 0808.0884}}].

\bibitem{Bonelli:2012ny}
G.~Bonelli, K.~Maruyoshi, A.~Tanzini and F.~Yagi, \emph{{${\mathcal{N}}\!=2$
  Gauge Theories on Toric Singularities, Blow-Up Formulae and W-Algebrae}},
  \href{http://dx.doi.org/10.1007/JHEP01(2013)014}{\emph{JHEP} {\bf 01} (2013)
  014}, [\href{http://arxiv.org/abs/1208.0790}{{\tt 1208.0790}}].

\bibitem{Sasaki:2006vq}
T.~Sasaki, \emph{{O(-2) Blow-Up Formula via Instanton Calculus on Affine $
  C^2$/$Z_2$ and Weil Conjecture}},
  \href{http://arxiv.org/abs/hep-th/0603162}{{\tt hep-th/0603162}}.

\bibitem{Ito:2013kpa}
Y.~Ito, K.~Maruyoshi and T.~Okuda, \emph{{Scheme Dependence of Instanton
  Counting in ALE Spaces}},
  \href{http://dx.doi.org/10.1007/JHEP05(2013)045}{\emph{JHEP} {\bf 05} (2013)
  045}, [\href{http://arxiv.org/abs/1303.5765}{{\tt 1303.5765}}].

\bibitem{Bruzzo:2013daa}
U.~Bruzzo, M.~Pedrini, F.~Sala and R.~J. Szabo, \emph{{Framed Sheaves on Root
  Stacks and Supersymmetric Gauge Theories on ALE Spaces}},
  \href{http://dx.doi.org/10.1016/j.aim.2015.11.005}{\emph{Adv. Math.} {\bf
  288} (2016) 1175--1308}, [\href{http://arxiv.org/abs/1312.5554}{{\tt
  1312.5554}}].

\bibitem{Bruzzo:2014jza}
U.~Bruzzo, F.~Sala and R.~J. Szabo, \emph{{${\mathcal{N} = 2}$ Quiver Gauge
  Theories on A-type ALE Spaces}},
  \href{http://dx.doi.org/10.1007/s11005-014-0734-x}{\emph{Lett. Math. Phys.}
  {\bf 105} (2015) 401--445}, [\href{http://arxiv.org/abs/1410.2742}{{\tt
  1410.2742}}].

\bibitem{Witten:1988ze}
E.~Witten, \emph{{Topological Quantum Field Theory}},
  \href{http://dx.doi.org/10.1007/BF01223371}{\emph{Commun. Math. Phys.} {\bf
  117} (1988) 353}.

\bibitem{Bershtein:2015xfa}
M.~Bershtein, G.~Bonelli, M.~Ronzani and A.~Tanzini, \emph{{Exact results for $
  \mathcal{N} $ = 2 supersymmetric gauge theories on compact toric manifolds
  and equivariant Donaldson invariants}},
  \href{http://dx.doi.org/10.1007/JHEP07(2016)023}{\emph{JHEP} {\bf 07} (2016)
  023}, [\href{http://arxiv.org/abs/1509.00267}{{\tt 1509.00267}}].

\bibitem{Baulieu:1997nj}
L.~Baulieu, A.~Losev and N.~Nekrasov, \emph{{Chern-Simons and Twisted
  Supersymmetry in Various Dimensions}},
  \href{http://dx.doi.org/10.1016/S0550-3213(98)00096-0}{\emph{Nucl. Phys.}
  {\bf B522} (1998) 82--104}, [\href{http://arxiv.org/abs/hep-th/9707174}{{\tt
  hep-th/9707174}}].

\bibitem{Losev:1995cr}
A.~Losev, G.~W. Moore, N.~Nekrasov and S.~Shatashvili, \emph{{Four-Dimensional
  Avatars of Two-Dimensional RCFT}},
  \href{http://dx.doi.org/10.1016/0920-5632(96)00015-1}{\emph{Nucl. Phys. Proc.
  Suppl.} {\bf 46} (1996) 130--145},
  [\href{http://arxiv.org/abs/hep-th/9509151}{{\tt hep-th/9509151}}].

\bibitem{Intriligator:1997pq}
K.~A. Intriligator, D.~R. Morrison and N.~Seiberg, \emph{{Five-Dimensional
  Supersymmetric Gauge Theories and Degenerations of Calabi-Yau Spaces}},
  \href{http://dx.doi.org/10.1016/S0550-3213(97)00279-4}{\emph{Nucl. Phys.}
  {\bf B497} (1997) 56--100}, [\href{http://arxiv.org/abs/hep-th/9702198}{{\tt
  hep-th/9702198}}].

\bibitem{Shadchin:2005mx}
S.~Shadchin, \emph{{On Certain Aspects of String Theory/Gauge Theory
  Correspondence}}.
\newblock PhD thesis, Orsay, LPT, 2005.
\newblock \href{http://arxiv.org/abs/hep-th/0502180}{{\tt hep-th/0502180}}.

\bibitem{Keller:2011ek}
C.~A. Keller, N.~Mekareeya, J.~Song and Y.~Tachikawa, \emph{{The ABCDEFG of
  Instantons and W-Algebras}},
  \href{http://dx.doi.org/10.1007/JHEP03(2012)045}{\emph{JHEP} {\bf 03} (2012)
  045}, [\href{http://arxiv.org/abs/1111.5624}{{\tt 1111.5624}}].

\bibitem{Bhardwaj:2013qia}
L.~Bhardwaj and Y.~Tachikawa, \emph{{Classification of 4D ${\mathcal{N}}\!=2$
  Gauge Theories}},
  \href{http://dx.doi.org/10.1007/JHEP12(2013)100}{\emph{JHEP} {\bf 12} (2013)
  100}, [\href{http://arxiv.org/abs/1309.5160}{{\tt 1309.5160}}].

\bibitem{Jefferson:2017ahm}
P.~Jefferson, H.-C. Kim, C.~Vafa and G.~Zafrir, \emph{{Towards Classification
  of 5D SCFTs: Single Gauge Node}},
  \href{http://arxiv.org/abs/1705.05836}{{\tt 1705.05836}}.

\bibitem{Hayashi:2019yxj}
H.~Hayashi, S.-S. Kim, K.~Lee and F.~Yagi, \emph{{Rank-3 Antisymmetric Matter
  on 5-Brane Webs}},
  \href{http://dx.doi.org/10.1007/JHEP05(2019)133}{\emph{JHEP} {\bf 05} (2019)
  133}, [\href{http://arxiv.org/abs/1902.04754}{{\tt 1902.04754}}].

\bibitem{Aganagic:2003db}
M.~Aganagic, A.~Klemm, M.~Mari\~no and C.~Vafa, \emph{{The Topological
  Vertex}}, \href{http://dx.doi.org/10.1007/s00220-004-1162-z}{\emph{Commun.
  Math. Phys.} {\bf 254} (2005) 425--478},
  [\href{http://arxiv.org/abs/hep-th/0305132}{{\tt hep-th/0305132}}].

\bibitem{Iqbal:2007ii}
A.~Iqbal, C.~Koz\c{c}az and C.~Vafa, \emph{{The Refined Topological Vertex}},
  \href{http://dx.doi.org/10.1088/1126-6708/2009/10/069}{\emph{JHEP} {\bf 10}
  (2009) 069}, [\href{http://arxiv.org/abs/hep-th/0701156}{{\tt
  hep-th/0701156}}].

\bibitem{Seiberg:1996bd}
N.~Seiberg, \emph{{Five-dimensional SUSY field theories, nontrivial fixed
  points and string dynamics}},
  \href{http://dx.doi.org/10.1016/S0370-2693(96)01215-4}{\emph{Phys. Lett.}
  {\bf B388} (1996) 753--760}, [\href{http://arxiv.org/abs/hep-th/9608111}{{\tt
  hep-th/9608111}}].

\bibitem{Minahan:1996fg}
J.~A. Minahan and D.~Nemeschansky, \emph{{An ${\mathcal{N}}\!=2$ Superconformal
  Fixed Point with $E_{6}$ Global Symmetry}},
  \href{http://dx.doi.org/10.1016/S0550-3213(96)00552-4}{\emph{Nucl. Phys.}
  {\bf B482} (1996) 142--152}, [\href{http://arxiv.org/abs/hep-th/9608047}{{\tt
  hep-th/9608047}}].

\bibitem{Minahan:1996cj}
J.~A. Minahan and D.~Nemeschansky, \emph{{Superconformal Fixed Points with
  $E_{N}$ Global Symmetry}},
  \href{http://dx.doi.org/10.1016/S0550-3213(97)00039-4}{\emph{Nucl. Phys.}
  {\bf B489} (1997) 24--46}, [\href{http://arxiv.org/abs/hep-th/9610076}{{\tt
  hep-th/9610076}}].

\bibitem{Aharony:1997ju}
O.~Aharony and A.~Hanany, \emph{{Branes, superpotentials and superconformal
  fixed points}},
  \href{http://dx.doi.org/10.1016/S0550-3213(97)00472-0}{\emph{Nucl. Phys.}
  {\bf B504} (1997) 239--271}, [\href{http://arxiv.org/abs/hep-th/9704170}{{\tt
  hep-th/9704170}}].

\bibitem{Diaconescu:1998cn}
D.-E. Diaconescu and R.~Entin, \emph{{Calabi-Yau Spaces and Five-Dimensional
  Field Theories with Exceptional Gauge Symmetry}},
  \href{http://dx.doi.org/10.1016/S0550-3213(98)00689-0}{\emph{Nucl. Phys.}
  {\bf B538} (1999) 451--484}, [\href{http://arxiv.org/abs/hep-th/9807170}{{\tt
  hep-th/9807170}}].

\bibitem{Jefferson:2018irk}
P.~Jefferson, S.~Katz, H.-C. Kim and C.~Vafa, \emph{{On Geometric
  Classification of 5D SCFTs}},
  \href{http://dx.doi.org/10.1007/JHEP04(2018)103}{\emph{JHEP} {\bf 04} (2018)
  103}, [\href{http://arxiv.org/abs/1801.04036}{{\tt 1801.04036}}].

\bibitem{Bhardwaj:2018yhy}
L.~Bhardwaj and P.~Jefferson, \emph{{Classifying 5D SCFTs via 6D SCFTs: Rank
  One}}, \href{http://dx.doi.org/10.1007/JHEP07(2019)178}{\emph{JHEP} {\bf 07}
  (2019) 178}, [\href{http://arxiv.org/abs/1809.01650}{{\tt 1809.01650}}].

\bibitem{Apruzzi:2019opn}
F.~Apruzzi, C.~Lawrie, L.~Lin, S.~Schäfer-Nameki and Y.-N. Wang, \emph{{Fibers
  add Flavor, Part I: Classification of 5d SCFTs, Flavor Symmetries and BPS
  States}},  \href{http://arxiv.org/abs/1907.05404}{{\tt 1907.05404}}.

\bibitem{Kim:2018gjo}
H.-C. Kim, J.~Kim, S.~Kim, K.-H. Lee and J.~Park, \emph{{6D Strings and
  Exceptional Instantons}},  \href{http://arxiv.org/abs/1801.03579}{{\tt
  1801.03579}}.

\bibitem{Zafrir:2015ftn}
G.~Zafrir, \emph{{Brane webs and $O5$-planes}},
  \href{http://dx.doi.org/10.1007/JHEP03(2016)109}{\emph{JHEP} {\bf 03} (2016)
  109}, [\href{http://arxiv.org/abs/1512.08114}{{\tt 1512.08114}}].

\bibitem{Hayashi:2018bkd}
H.~Hayashi, S.-S. Kim, K.~Lee and F.~Yagi, \emph{{5-brane webs for 5d $
  \mathcal{N} $ = 1 G$_{2}$ gauge theories}},
  \href{http://dx.doi.org/10.1007/JHEP03(2018)125}{\emph{JHEP} {\bf 03} (2018)
  125}, [\href{http://arxiv.org/abs/1801.03916}{{\tt 1801.03916}}].

\bibitem{Hollowood:2003cv}
T.~J. Hollowood, A.~Iqbal and C.~Vafa, \emph{{Matrix Models, Geometric
  Engineering and Elliptic Genera}},
  \href{http://dx.doi.org/10.1088/1126-6708/2008/03/069}{\emph{JHEP} {\bf 03}
  (2008) 069}, [\href{http://arxiv.org/abs/hep-th/0310272}{{\tt
  hep-th/0310272}}].

\bibitem{DelZotto:2016pvm}
M.~Del~Zotto and G.~Lockhart, \emph{{On Exceptional Instanton Strings}},
  \href{http://dx.doi.org/10.1007/JHEP09(2017)081}{\emph{JHEP} {\bf 09} (2017)
  081}, [\href{http://arxiv.org/abs/1609.00310}{{\tt 1609.00310}}].

\bibitem{DelZotto:2018tcj}
M.~Del~Zotto and G.~Lockhart, \emph{{Universal Features of BPS Strings in
  Six-Dimensional SCFTs}},
  \href{http://dx.doi.org/10.1007/JHEP08(2018)173}{\emph{JHEP} {\bf 08} (2018)
  173}, [\href{http://arxiv.org/abs/1804.09694}{{\tt 1804.09694}}].

\bibitem{Benini:2009gi}
F.~Benini, S.~Benvenuti and Y.~Tachikawa, \emph{{Webs of Five-Branes and
  ${\mathcal{N}}\!=2$ Superconformal Field Theories}},
  \href{http://dx.doi.org/10.1088/1126-6708/2009/09/052}{\emph{JHEP} {\bf 09}
  (2009) 052}, [\href{http://arxiv.org/abs/0906.0359}{{\tt 0906.0359}}].

\bibitem{Putrov:2015jpa}
P.~Putrov, J.~Song and W.~Yan, \emph{{(0,4) Dualities}},
  \href{http://dx.doi.org/10.1007/JHEP03(2016)185}{\emph{JHEP} {\bf 03} (2016)
  185}, [\href{http://arxiv.org/abs/1505.07110}{{\tt 1505.07110}}].

\bibitem{Gadde:2010te}
A.~Gadde, L.~Rastelli, S.~S. Razamat and W.~Yan, \emph{{The Superconformal
  Index of the $E_{6}$ SCFT}},
  \href{http://dx.doi.org/10.1007/JHEP08(2010)107}{\emph{JHEP} {\bf 08} (2010)
  107}, [\href{http://arxiv.org/abs/1003.4244}{{\tt 1003.4244}}].

\bibitem{Gadde:2011uv}
A.~Gadde, L.~Rastelli, S.~S. Razamat and W.~Yan, \emph{{Gauge Theories and
  Macdonald Polynomials}},
  \href{http://dx.doi.org/10.1007/s00220-012-1607-8}{\emph{Commun. Math. Phys.}
  {\bf 319} (2013) 147--193}, [\href{http://arxiv.org/abs/1110.3740}{{\tt
  1110.3740}}].

\bibitem{Gadde:2015xta}
A.~Gadde, S.~S. Razamat and B.~Willett, \emph{{"Lagrangian" for a
  Non-Lagrangian Field Theory with $\mathcal N=2$ Supersymmetry}},
  \href{http://dx.doi.org/10.1103/PhysRevLett.115.171604}{\emph{Phys. Rev.
  Lett.} {\bf 115} (2015) 171604}, [\href{http://arxiv.org/abs/1505.05834}{{\tt
  1505.05834}}].

\bibitem{Agarwal:2018ejn}
P.~Agarwal, K.~Maruyoshi and J.~Song, \emph{{A “Lagrangian” for the E$_{7}$
  superconformal theory}},
  \href{http://dx.doi.org/10.1007/JHEP05(2018)193}{\emph{JHEP} {\bf 05} (2018)
  193}, [\href{http://arxiv.org/abs/1802.05268}{{\tt 1802.05268}}].

\bibitem{Intriligator:1996ex}
K.~A. Intriligator and N.~Seiberg, \emph{{Mirror Symmetry in Three-Dimensional
  Gauge Theories}},
  \href{http://dx.doi.org/10.1016/0370-2693(96)01088-X}{\emph{Phys. Lett.} {\bf
  B387} (1996) 513--519}, [\href{http://arxiv.org/abs/hep-th/9607207}{{\tt
  hep-th/9607207}}].

\bibitem{Cremonesi:2013lqa}
S.~Cremonesi, A.~Hanany and A.~Zaffaroni, \emph{{Monopole operators and Hilbert
  series of Coulomb branches of $3d$ $\mathcal{N} = 4$ gauge theories}},
  \href{http://dx.doi.org/10.1007/JHEP01(2014)005}{\emph{JHEP} {\bf 01} (2014)
  005}, [\href{http://arxiv.org/abs/1309.2657}{{\tt 1309.2657}}].

\bibitem{Cremonesi:2014xha}
S.~Cremonesi, G.~Ferlito, A.~Hanany and N.~Mekareeya, \emph{{Coulomb Branch and
  the Moduli Space of Instantons}},
  \href{http://dx.doi.org/10.1007/JHEP12(2014)103}{\emph{JHEP} {\bf 12} (2014)
  103}, [\href{http://arxiv.org/abs/1408.6835}{{\tt 1408.6835}}].

\bibitem{Kim:2011mv}
H.-C. Kim, S.~Kim, E.~Koh, K.~Lee and S.~Lee, \emph{{On Instantons as
  Kaluza-Klein Modes of M5-Branes}},
  \href{http://dx.doi.org/10.1007/JHEP12(2011)031}{\emph{JHEP} {\bf 12} (2011)
  031}, [\href{http://arxiv.org/abs/1110.2175}{{\tt 1110.2175}}].

\bibitem{Hwang:2016gfw}
Y.~Hwang, J.~Kim and S.~Kim, \emph{{M5-Branes, Orientifolds, and S-Duality}},
  \href{http://dx.doi.org/10.1007/JHEP12(2016)148}{\emph{JHEP} {\bf 12} (2016)
  148}, [\href{http://arxiv.org/abs/1607.08557}{{\tt 1607.08557}}].

\bibitem{Lee:2017lfw}
S.-J. Lee and P.~Yi, \emph{{D-Particles on Orientifolds and Rational
  Invariants}}, \href{http://dx.doi.org/10.1007/JHEP07(2017)046}{\emph{JHEP}
  {\bf 07} (2017) 046}, [\href{http://arxiv.org/abs/1702.01749}{{\tt
  1702.01749}}].

\bibitem{Kim:2012gu}
H.-C. Kim, S.-S. Kim and K.~Lee, \emph{{5-Dim Superconformal Index with
  Enhanced En Global Symmetry}},
  \href{http://dx.doi.org/10.1007/JHEP10(2012)142}{\emph{JHEP} {\bf 10} (2012)
  142}, [\href{http://arxiv.org/abs/1206.6781}{{\tt 1206.6781}}].

\bibitem{VinbergPopov}
E.~B. Vinberg and V.~L. Popov, \emph{{On a class of quasihomogeneous affine
  varieties}},
  \href{http://dx.doi.org/10.1070/IM1972v006n04ABEH001898}{\emph{Math.
  USSR-Izv.} {\bf 6} (1972) 743}.

\bibitem{Garfinkle}
quoted in Chap. III~in D.~Garfinkle, \emph{{A new construction of the Joseph
  ideal}},  1982.

\bibitem{Benvenuti:2010pq}
S.~Benvenuti, A.~Hanany and N.~Mekareeya, \emph{{The Hilbert Series of the One
  Instanton Moduli Space}},
  \href{http://dx.doi.org/10.1007/JHEP06(2010)100}{\emph{JHEP} {\bf 06} (2010)
  100}, [\href{http://arxiv.org/abs/1005.3026}{{\tt 1005.3026}}].

\bibitem{Gaiotto:2009we}
D.~Gaiotto, \emph{{${\mathcal{N}}\!=2$ Dualities}},
  \href{http://dx.doi.org/10.1007/JHEP08(2012)034}{\emph{JHEP} {\bf 08} (2012)
  034}, [\href{http://arxiv.org/abs/0904.2715}{{\tt 0904.2715}}].

\bibitem{Alday:2009aq}
L.~F. Alday, D.~Gaiotto and Y.~Tachikawa, \emph{{Liouville Correlation
  Functions from Four-Dimensional Gauge Theories}},
  \href{http://dx.doi.org/10.1007/s11005-010-0369-5}{\emph{Lett. Math. Phys.}
  {\bf 91} (2010) 167--197}, [\href{http://arxiv.org/abs/0906.3219}{{\tt
  0906.3219}}].

\bibitem{Coman:2019eex}
I.~Coman, E.~Pomoni and J.~Teschner, \emph{{Trinion Conformal Blocks from
  Topological Strings}},  \href{http://arxiv.org/abs/1906.06351}{{\tt
  1906.06351}}.

\bibitem{Feger:2012bs}
R.~Feger and T.~W. Kephart, \emph{{LieART—A Mathematica application for Lie
  algebras and representation theory}},
  \href{http://dx.doi.org/10.1016/j.cpc.2014.12.023}{\emph{Comput. Phys.
  Commun.} {\bf 192} (2015) 166--195},
  [\href{http://arxiv.org/abs/1206.6379}{{\tt 1206.6379}}].

\end{thebibliography}\endgroup

\end{document}